\documentclass[aps,prd,twocolumn,notitlepage,superscriptaddress,nofootinbib]{revtex4-1}
\usepackage[utf8]{inputenc}
\usepackage[english]{babel}
\usepackage{amsmath}
\usepackage{amsfonts}
\usepackage{amssymb}
\usepackage{listings}
\usepackage{lipsum}
\usepackage{multirow}
\usepackage{datetime}
\usepackage{graphicx}
\usepackage{mathtools}
\usepackage{mathrsfs}
\usepackage{dcolumn}
\usepackage{multirow}
\usepackage{color}   
\usepackage{hyperref}
\hypersetup{
    colorlinks=true, 
    pdfborder = {0 0 0.5 [3 3]},
    anchorcolor=black,
    citecolor=blue,
    linktoc=all,    
    linktocpage=true,
    linkcolor=red,
	urlcolor=blue
}

\begin{document}

\title{Stability of rotating scalar boson stars with nonlinear interactions}
\author{Nils Siemonsen}
\email[]{nsiemonsen@perimeterinstitute.ca}
\affiliation{Perimeter Institute for Theoretical Physics, Waterloo, Ontario N2L 2Y5, Canada}
\affiliation{Department of Physics \& Astronomy, University of Waterloo, Waterloo, ON N2L 3G1, Canada}
\author{William E.\ East}
\affiliation{Perimeter Institute for Theoretical Physics, Waterloo, Ontario N2L 2Y5, Canada}

\date{\today}

\begin{abstract} 
    We study the stability of rotating scalar boson stars, comparing those made
    from a simple massive complex scalar (referred to as mini boson stars), to those
    with several different types of nonlinear interactions.  To that end, we numerically
    evolve the nonlinear Einstein-Klein-Gordon equations in 3D, beginning with
    stationary boson star solutions. We show that the linear, non-axisymmetric
    instability found in mini boson stars with azimuthal number $m=1$ persists
    across the entire parameter space for these stars, though the timescale
    diverges in the Newtonian limit.  Therefore, any boson star with $m=1$ 
    that is sufficiently far into the non-relativistic regime, where the leading order
    mass term dominates, will be unstable, independent of the nonlinear scalar self-interactions.
    However, we do find regions of $m=1$ boson star parameter space where adding nonlinear
    interactions to the scalar potential quenches the non-axisymmetric instability,
    both on the non-relativistic, and the relativistic branches of solutions.  
    We also consider select boson stars with $m=2$, finding instability in all cases.
    For the cases exhibiting instability, we follow the nonlinear development,
    finding a range of dynamics including fragmentation into multiple unbound
    non-rotating stars, and formation of binary black holes. Finally, we comment on the relationship between stability and
    criteria based on the rotating boson star's frequency in relation to that
    of a spherical boson star or the existence of a co-rotation point.
    The boson stars that we find not to exhibit instability when evolved for many dynamical
    times include rapidly rotating cases where the compactness is comparable to that of a black hole
    or neutron star.
\end{abstract}

\maketitle

\section{Introduction} \label{introduction}

Gravitational wave (GW) astronomy \cite{Abbott:2016blz, LIGOScientific:2018mvr,Abbott:2020niy}
and electromagnetic observations \cite{1997MNRAS.291..219G,Ghez:1998ph,Akiyama:2019cqa}
have shown that the black hole (BH) paradigm can be used to explain
phenomena ranging from the merger and ringdown of compact object binaries,
to active galactic nuclei. 
However, while this paradigm explains physics across several orders of
magnitude extremely successfully, the BH's defining feature---the
event horizon---cannot be directly probed experimentally. 
Additionally, some quantum gravity models predict strong deviations from general relativity approaching the boundaries of
BHs, and
extensions to the Standard Model can lead to the formation of highly compact
BH-like objects. This has motivated the construction of a large class of exotic
compact objects (ECOs), with various physical and mathematical motivations,
mimicking many of the gravitational properties attributed to classical BH
spacetimes, and providing an alternative to the BH paradigm.

Beside BHs and neutron stars, a large class of ECOs has been conceived
\cite{Cardoso:2019rvt}. The compactness of ordinary fluid stars, like neutron
stars, is restricted by the Buchdahl limit \cite{Buchdahl:1959zz} for
spherically symmetric configurations, and more stringently by the requirement
that the sound speed be less than the speed of light. However, these objects could capture and accrete
dark matter from their environment, forming a composite object 
\cite{Goldman:1989nd, Gould:1989gw,
Henriques:1989ez, Henriques:1989ar, Leung:2011zz, DiGiovanni:2020frc}, which could exceed these compactness limitations. Various
models of ordinary matter also predict the existence of anisotropic stars with
compactness levels arbitrarily close to those of BHs \cite{Bowers:1974tgi,
Letelier:1980mxb, Herrera:2004xc}. String theory inspired solutions, called
fuzzballs, emerge as averages over microstates, generating horizonless, but
highly compact, BH-like models \cite{David:2002wn, Bena:2007kg, Myers:1997qi,
Balasubramanian:2008da, Bena:2013dka}. There is speculation that quantum effects of a collapsing
horizonless spacetime could halt the complete classical collapse to a BH and
yield a highly compact configuration just outside the BH-limit
\cite{Visser:2009pw, Chen:2017pkl, Berthiere:2017tms, Baccetti:2016lsb}.
Furthermore, ultralight scalar or vector particles arise in compactifications of
string theory, solutions to the strong CP-problem \cite{Guerra:2019srj,
diCortona:2015ldu}, as well as as ultralight and fuzzy dark matter
\cite{Hu:2000ke, Hui:2016ltb, Li:2013nal, Robles:2012uy, Bar:2018acw, Annulli:2020lyc}. If 
such particles exist in the universe, stationary boson stars (BSs)
\cite{Kaup:1968zz, Ruffini:1969qy, Seidel:1993zk, Narain:2006kx,
Raidal:2018eoo, Deliyergiyev:2019vti, Liebling:2012fv, Giudice:2016zpa} can be
formed through a gravitational cooling mechanisms out of a diffuse distribution
of bosonic matter \cite{Seidel:1993zk, Sanchis-Gual:2019ljs}. Lastly, there is
an extended class of other ECOs like gravastars, wormholes, or
firewalls \cite{Cardoso:2019rvt}. The common feature of all these approaches is
a certain ``closeness," measured by the compactness and related features, to classical BHs. 

Accurate predictions of the properties of BH mimickers, especially of their dynamics
in the nonlinear regime in the case of mergers, is needed to confirm or disfavor the existence of BHs using GW or
electromagnetic observations \cite{Kesden:2004qx, Berti:2015itd,
Bambi:2015kza, Sennett:2017etc, Macedo:2013jja, Yoshida:1994xi}. However, for many ECO models, determining their 
nonlinear evolution is challenging or ill-posed, due to their vastly different physical and
mathematical origins. BSs, on the other hand, obeying standard energy conditions (as long as their
potential is non-negative), 
evolve according to well behaved wave-like equations, and can be treated numerically
using the same techniques as the Einstein equations. Thus they provide a simple
and tractable setting to explore dynamical properties of ECOs and BH mimickers
\cite{Guzman:2009zz, Liebling:2012fv}. Stationary rotating BSs can have
compactnesses approaching that of BHs, and therefore, capture the main
gravitational features of a large set of ECOs. BSs can exhibit an innermost
stable circular orbit, and unstable and stable photon orbits
while being horizonless and regular everywhere. Therefore, these
solutions provide an ideal test bed to study the nonlinear dynamics of a class
of ultracompact objects. 

However, while there is an extensive literature constructing stationary BS
solutions in general relativity, the number of studies that have looked at the
dynamics or stability of these objects is more limited, in particular for BSs
with angular momentum (see Ref.~\cite{Liebling:2012fv} for a recent review.)
Recently, there has been increasing interest in studying the nonlinear
dynamics of scalar and vector BSs in various scenarios
(e.g.~Refs.~\cite{Palenzuela:2007dm, Palenzuela:2006wp,Choptuik:2009ww,
Palenzuela:2017kcg,Helfer:2016ljl,Dietrich:2018bvi,Olivares:2018abq,
Sanchis-Gual:2020mzb} and see Ref.~\cite{Liebling:2012fv}).
And in the past few years, several studies have presented evidence that
rotating scalar BSs might be inherently unstable.  In
Refs.~\cite{Palenzuela:2017kcg,Bezares:2017mzk}, the inspiral and merger of
binary scalar BSs settled, above a certain critical BH-threshold, into a
\textit{non}-rotating scalar BSs in the final state, shedding all angular
momentum in the process. Similar results were found for the collapse of
rotating clouds of scalar field~\cite{DiGiovanni:2020ror}. Furthermore,
Ref.~\cite{Sanchis-Gual:2019ljs} considered a number of rotating scalar BSs
made of massive bosons without interactions, and found them all to be subject
to a non-axisymmetric instability (NAI), rendering them little use, e.g., for
studying the dynamics of a merger. 

However, in this work, we show that this problem can be cured by considering
nonlinear interactions for the scalar field, and present evidence for the
stability of a large class of rotating BSs both in the relativistic (high
compactness) \textit{and} in the non-relativistic (dilute) regime.  In
particular, we show that when considering parameterized families of BSs for several
different choices of the scalar field potential, the growth rate for the NAI
approaches zero at certain critical values (e.g., of the frequency of the BS).
Nonlinearly evolving select cases beyond these critical values for many
dynamical times, we find no evidence for instability. (Though our methods do
not allow us to rule out some instability operating on even longer timescales.)
Hence, such rotating scalar BS solutions are promising candidates for studying
the dynamics of isolated and binary ECOs in a nonlinear scenario, and
comparing to BHs.

In Sec.~\ref{sec:statsolutions}, we outline the scalar field models we
consider, describe the numerical techniques we use to construct rotating BS
solutions in these models, and review the linear stability results in the
literature. Following this, we present our numerical results in
Sec.~\ref{sec:results}. We first identify the form and nature of the NAI in the
linear regime, then measure the growth rates of the NAI for a set of BSs and
potentials, and present select isolated rotating BSs that show no sign of
instability.  We also analyze the final state of the instability, and discuss
various physical explanations for the onset of instability.  Finally, we
conclude in Sec.~\ref{sec:disscussion}.  Additional details on the numerical
methods and error estimates from convergence studies are given in the appendices.  In the
following, we use units with $G=c=1$ and the $(- + + +)$ metric signature.


\section{Stationary Scalar Boson stars} \label{sec:statsolutions}

\begin{figure}[t]
\includegraphics[width=0.455\textwidth]{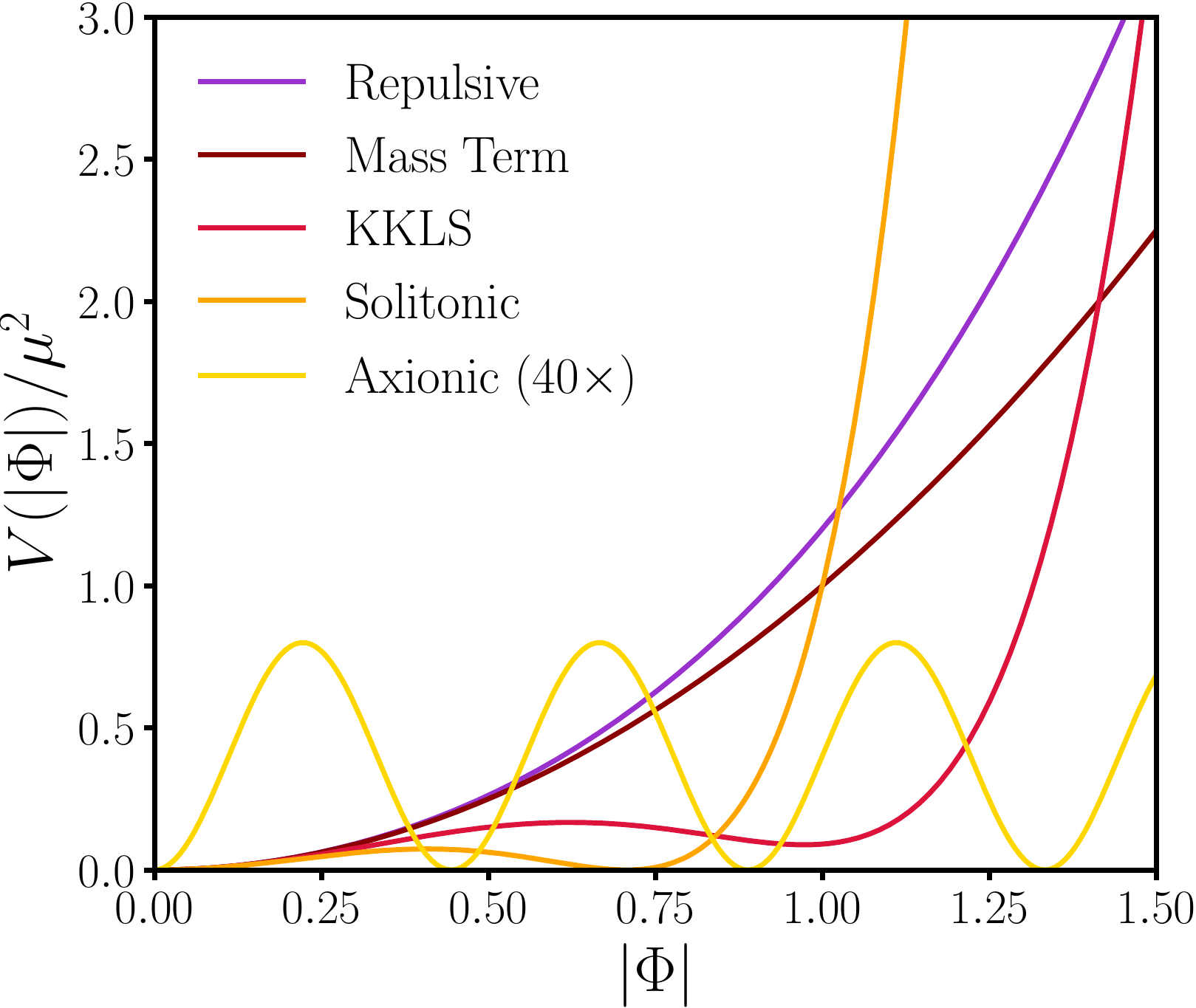}
\caption{Examples of the nonlinear self-interactions, specified by $V(|\Phi|)$, considered here. We emphasize the characteristic features of each: The solitonic potential has a non-trivial vacuum state, the KKLS self-interaction has a metastable state, the axionic potential is periodic, and finally, the repulsive potential has a positive first correction to the mass term.}
\label{fig:BSpotentials}
\end{figure}

\subsection{Models considered}
\label{sec:models}

In this work, we consider rotating scalar BSs in minimally coupled massive
complex scalar models, comparing stars where the scalar potential only has a
mass term (referred to as mini BSs) to stars where the potential has additional
higher order terms, corresponding to non-trivial scalar self-interactions. The
latter are motivated by axion-like particles, effective field theory
descriptions of light scalar degrees of freedom, or simply by the finding that
such models can produce highly compact BSs. As such, we consider a complex
scalar field $\Phi$, minimally coupled to the Einstein-Hilbert action,
exhibiting a global U(1) symmetry:
\begin{align}
S=\int d^4x\sqrt{-g}\left[ \frac{R}{16\pi}-g^{\alpha\beta}\nabla_{(\alpha} \bar{\Phi}\nabla_{\beta)} \Phi-V(|\Phi|) \right].
\end{align}
Here $R$ is the curvature scalar, the overbar denotes complex conjugation, and the potential $V(|\Phi|)$ contains both the mass term and nonlinear self-interactions of the complex scalar field. All such self-interactions considered here (see also \figurename{ \ref{fig:BSpotentials}}) reduce to the mass term\footnote{The scalar mass $m_\Phi$ and the mass parameter $\mu$ are related by $\mu=m_\Phi/\hbar$}, $V(|\Phi|) \rightarrow \mu^2 |\Phi|^2$, in a small-coupling limit.

\textit{Solitonic potential:} This potential is characterized by a single additional coupling parameter $\sigma$, such that \cite{Friedberg:1986tq}
\begin{align}
V(|\Phi|)=\mu^2|\Phi|^2\left(1-\frac{2|\Phi|^2}{\sigma^2}\right)^2,
\label{eq:solitonic}
\end{align}
reduces to the mass-term if $\sigma\rightarrow \infty$. As indicated in \figurename{ \ref{fig:BSpotentials}}, this potential has a negative first correction in the expansion in $|\Phi|$, and features a non-trivial vacuum state at $|\Phi|=\sigma/\sqrt{2}$. With these self-interactions, localized scalar field configurations (i.e., non-topological solitons) exist even in the absence of gravity \cite{Friedberg:1986tq}, and spherically symmetric BSs of this type are sufficiently compact to support stable trapped null geodesics \cite{Palenzuela:2017kcg}.

\textit{KKLS potential:} The potential \cite{Kleihaus:2005me, Kleihaus:2007vk},
\begin{align}
V(|\Phi|)=\mu^2|\Phi|^2\left[1 -\frac{16\pi}{1.1 \kappa}|\Phi|^2+ \frac{64\pi^2}{1.1\kappa^2} |\Phi|^4\right],
\label{eq:metastable}
\end{align}
is parameterized by $\kappa$ (see also Ref.~\cite{Kleihaus:2011sx}). It also exhibits a negative first correction beyond the mass-term, and a local minimum at a non-zero value of $|\Phi|$ (see also \figurename{ \ref{fig:BSpotentials}}). The KKLS potential simplifies to the mass term if $\kappa\rightarrow \infty$ \cite{Kleihaus:2005me}. After rescaling $|\Phi|\rightarrow (\kappa/(8\pi))^{1/2}|\Phi|$, BSs in this model reduce to non-gravitating non-topological Q-balls \cite{Coleman:1985ki} in the $\kappa=0$ limit.

\textit{Axionic potential:} Inspired by ultralight particles predicted, for instance, by string theory compactifications, or to solve the QCD CP-problem, we consider a generic axion-like potential of the form
\begin{align}
V(|\Phi|)=\mu^2f^2\big\{1-\cos[\sqrt{2 |\Phi|^2}f^{-1}]\big\}.
\label{eq:axionpot}
\end{align}
This periodic potential is parameterized by the coupling $f$, reduces to the mass term if $f\rightarrow \infty$, and has a negative first correction, when expanded in small $|\Phi|$.

\textit{Repulsive potential:} While all the above scalar self-interactions have attractive (i.e., negative) first corrections to the mass term, we also study the effects of a non-trivial \textit{repulsive} first correction of the form
\begin{align}
V(|\Phi|)=\mu^2|\Phi|^2+\lambda |\Phi|^4,
\label{eq:repulsive}
\end{align}
with $\lambda>0$. In the following, we also briefly comment on the properties of BSs in a scalar model with a Liouville potential \cite{Schunck:1999zu, Choi:2019mva}, 
\begin{align}
V(|\Phi|)=\mu^2\alpha^2[e^{|\Phi|^2/\alpha^2}-1],
\label{eq:Liouville}
\end{align}
which has the same form as Eq.~\eqref{eq:repulsive} when expanded to quadratic order in $|\Phi|^2$.

\subsection{Stationary solutions and classical observables} \label{sec:classical_observables}

\begin{figure*}[t!]
\includegraphics[width=1\textwidth]{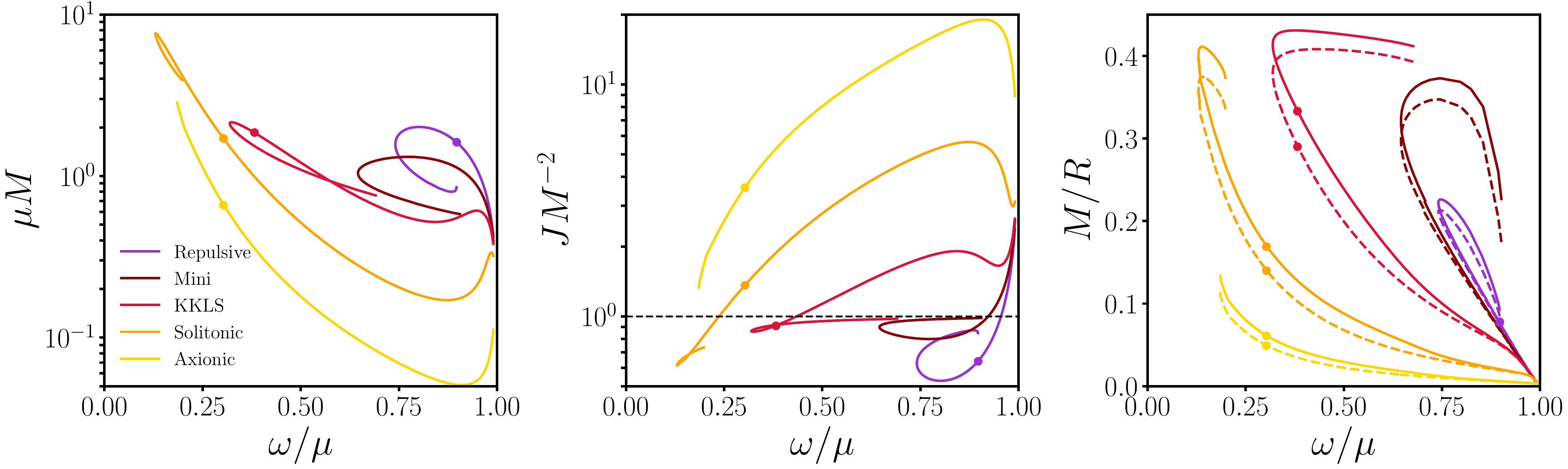}
\caption{We present the main properties of five families of $m=1$ rotating BSs
in the above scalar models, for choices of coupling focused on in this work.
The solitonic BS family has coupling $\sigma=0.05$, the axionic family has
coupling $f=0.005$, the BSs in the KKLS model have $\kappa=0.1$, the family of
BSs with repulsive self-interactions have coupling $\lambda/\mu^2=5\times
10^2$, while the mini BS family is given for reference. 
From left to right, the three plots show the total mass $M$ in units of $\mu^{-1}$, 
the dimensionless spin $J/M^2$, and the  
compactness $M/R$ of each
of these families of BSs as a function of the boson's frequency $\omega/\mu$.
For determining the radius of the BSs in the last plot, we use
either the surface containing $99\%$ of the mass $C^K$ (solid), or $99\%$ of the U(1)-charge density $C^Q$
(dashed), as discussed in Sec.~\ref{sec:classical_observables}.
Based on the turning
point argument described in the text, BSs past the maximum of the total mass 
are expected to be unstable.
Notice that in
the non-relativistic limit, i.e., when $\omega/\mu\rightarrow 1$, the behavior of
all families of BSs with non-trivial self-interactions reduce to that of mini
BSs \cite{Hertzberg:2018lmt, Davidson:2016uok}. The individual data points indicate BSs that we evolve and where we find no sign of an instability (see Sec.~\ref{sec:NAIgrowthrate}). Note that, although
not evident in the plot, the axionic family reaches a global maximum of $\mu M$ at
$\omega/\mu \approx 0.187$.}
\label{fig:BSproperties}
\end{figure*}

We consider BS solutions where the spacetime is regular, stationary, axisymmetric, and asymptotically flat.
In Lewis-Papapetrou coordinates, the metric takes the general form
\begin{align}
\begin{aligned}
ds^2=-f dt^2+l f^{-1}\big\{ & g(dr^2+ r^2 d\theta^2) \\
+ & r^2\sin^2\theta \left( d\varphi - \Omega r^{-1} dt\right)^2 \big\}.
\label{eq:metricansatz}
\end{aligned}
\end{align}
Here all metric functions $\{f,l,g,\Omega\}$ depend only on $(r,\theta)$, in
accordance with our assumptions. The complex scalar field is assumed to have
the form $\Phi=e^{i\omega t+im\varphi}\phi(r,\theta)$, with
$\phi\in\mathbb{R}$. This time and azimuthal dependency ensures that the
scalar field stress-energy tensor is stationary and axisymmetric (maintaining the symmetry of the metric).  In addition
to the above five physical fields, we introduce two auxiliary fields
$\rho(r,\theta)$ and $\omega_s(r,\theta)$, following
Ref.~\cite{Kleihaus:2005me}, which aid in imposing certain conditions on the BSs (see appendix \ref{sec:NumMeth} for details). With
this ansatz, in conjunction with regularity conditions at the origin and
asymptotic flatness conditions at infinity, the coupled system of
Einstein-complex-Klein-Gordon equations 
\begin{align}
G_{\mu\nu}=8\pi T_{\mu\nu}, & & \nabla_\alpha\nabla^\alpha\Phi + \Phi\partial_{|\Phi|^2} V(|\Phi|)=0
\label{eq:EinsteinKleinGordon}
\end{align}
[together with Eq.~\eqref{eq:auxiliaryeq}]
reduce to an elliptic boundary value problem. 
Given a sufficiently accurate initial seed for a Newton-Raphson type relaxation scheme, the parameter space is explored by marching along the respective parameters. We give more details on how we numerically solve these equations to construct BS solutions in Appendix~\ref{sec:NumMeth}. 

In asymptotically flat, stationary spacetimes, the Komar mass integrated over a sphere at spatial infinity, and the ADM mass of a spatial slice $\Sigma_t$ coincide. Both can be written as
\begin{align}
M = \int_D drd\theta d\varphi\sqrt{-g}(2 T^t{}_t-T^\alpha{}_\alpha),
\label{eq:komarmass}
\end{align}
where the integral is over the Lewis-Papapetrou spatial coordinate domain $D$. The axisymmetry and regularity of Eq.~\eqref{eq:metricansatz} ensures that the total angular momentum of the spacetime is due
to the scalar field through
\begin{align}
J=-\int_D drd\theta d\varphi\sqrt{-g} T^t{}_\varphi.
\end{align}
Finally, the global U(1) symmetry of the complex scalar theory 
gives rise to a conserved U(1)-Noether charge $Q$ that measures the particle number (i.e., the occupation number of the Bose-Einstein condensate\footnote{When quantizing the complex scalar field theory, the charge $Q$ counts the (anti)-particles of a given state, making the relation manifest.}). 
The associated Noether current is
\begin{align}
j^\mu = -i(\bar\Phi \partial^\mu \Phi-\Phi \partial^\mu \bar\Phi), & & \nabla_\mu j^\mu =0.
\label{eq:noethercurrent}
\end{align}
Following immediately from noticing that $T^t{}_\varphi=m j^t$ (in terms of the
Noether current Eq.~\eqref{eq:noethercurrent}), the angular momentum of the
system is ``quantized" into integer increments of the scalar charge, $J=mQ$,
dictated by the azimuthal number $m$. Of course, this relation is a purely
classical constraint, but hints towards the quantum interpretation of
the system as a coherent condensate of a set of bosons.

At sufficiently large radial coordinate, the scalar BS solution exhibits 
an exponential tail, $\phi\sim\exp(-\beta r)$, for some $\beta>0$, making 
any notion of a radius for the star ambiguous. In the non-rotating case, the
radius of a BS is typically defined by the \textit{areal radius}, $R_{99}$, at
which $99 \%$ of the BS's Komar mass Eq.~\eqref{eq:komarmass} is contained within a
coordinate 2-sphere of radius $R_{99}$. For rotating BSs, we introduce two
distinct notions of size: \textit{(i)} We define the $R^K_{99}$ as the
\textit{circular radius}\footnote{The radius $\tilde{r}$ for which the proper
length of a circle $C$ in the equatorial plane is $C=2\pi \tilde{r}$.}
$\tilde{r}$, for which $M(\tilde{r})=0.99 M(\tilde{r}\rightarrow\infty)$,
and \textit{(ii)} we define $R^Q_{99}$ as the circular radius, for which
$Q(\tilde{r})=0.99 Q(\tilde{r}\rightarrow\infty)$. Note that, while $M$
contains both the scalar and the gravitational binding energy, $Q$ measures
only the scalar rest mass. Based on this notion of size, the compactness of
a rotating BS is given either by $C^K=M/R^K_{99}$ or $C^Q=M/R^Q_{99}$.
For comparison, for a non-rotating BH
$C^K_\text{BH}=1/2$, while for a typical neutron star $C^K_\text{NS}\sim 0.1$. 
For any given BS, the difference between $C^Q$ and $C^K$ is indicative of the 
ambiguity in defining its radius. 

In \figurename{ \ref{fig:BSproperties}}, we present these observables for
several sets of families of solutions. The Newtonian limit is approached as
the gravitational binding energy, of order $M/R$, is small. In that limit, the
bosons' frequency $\omega$ approaches $\mu$, the marginally bound value, since $\omega/\mu-1\sim M/R$.

\subsection{Stability arguments} 
\label{sec:stable_arg}

The stability of non-rotating BSs has been investigated by means of
\textit{(i)} analyzing the temporal dependence of individual modes or more
general linear perturbations \cite{Lee:1988av, Gleiser:1988rq, Gleiser:1988ih},
\textit{(ii)} applying catastrophe theory or thermodynamic stability to families of BS solutions
\cite{Kusmartsev:1990cr,
Tamaki:2011zza, Kleihaus:2011sx,Sorkin:1981jc, Schiffrin:2013zta}, and \textit{(iii)} evolving the
Einstein-Klein-Gordon system of equations numerically to study the nonlinear
stability of BSs \cite{Seidel:1990jh, Balakrishna:1997ej, Guzman:2004jw, ValdezAlvarado:2012xc,Liebling:2012fv}.  The consensus of these methods is that spherically symmetric BSs
switch their radial stability properties whenever the BS's mass reaches an
extremum as a function of the central scalar field value $\phi(0)$:
$dM/d\phi(0)=0$. As such, non-rotating BSs have at least one stable branch
reaching from the non-relativistic (i.e., dilute) limit, to the first maximum of
$M$. Depending on the potential, non-rotating BSs can have a number of stable
branches, analogous to the white dwarf and neutron star branches of fluid stars.

However, the stability analysis of \textit{rotating} BSs
is considerably
more complex\footnote{Here we will not consider BSs with ergoregions,
which would be subject to the ergoregion instability~\cite{friedman1978}.}. 
Even at the linear level, there is not expected to be a clean decoupling of the
scalar and the gravitational modes, making a linear
stability analysis difficult. Based on turning point arguments in
Ref.~\cite{Kleihaus:2011sx} (see also Ref.~\cite{Collodel:2017biu} for a brief
analysis of the stability of excited BSs with the KKLS potential), the stability of
rotating BSs should switch, analogous to non-rotating BSs, at the extrema of
$M(\omega)$. In \figurename{ \ref{fig:BSproperties}}, we present the total
mass, angular momentum, and compactness of five different potentials and
families of $m=1$ fundamental BSs. Applying the arguments above to these families of BSs, we
see that mini BSs and those in the repulsive scalar model Eq.~\eqref{eq:repulsive}
exhibit a single (potentially) stable branch reaching from the non-relativistic limit,
$\omega/\mu \lesssim 1$, to the frequency where $M$ (or equivalently $Q$)
reaches the global maximum, while all solutions past that maximum are unstable.
Similarly, the family of BSs in the axionic, solitonic, and KKLS scalar models
exhibit two distinct potentially stable branches. Since all these potentials reduce to the
mass term in the non-relativistic limit, it is not surprising that BSs with
these self-interactions are stable in the non-relativistic limit based on the
turning point arguments. However, these BSs have another potentially stable
branch in the relativistic (high-compactness) regime, between the first local
minimum of $M$ and its global maximum, where $\partial M/\partial \omega <0$.

However, as seen, for instance, in Refs.~\cite{Sorkin:1981jc, Schiffrin:2013zta}, turning points are only a \textit{sufficient} condition for the existence of a thermodynamic instability, \textit{not} a necessary condition. 
In fact, a thermodynamic instability could appear without the presence of a
turning point. On top of this, thermodynamically unstable systems are not
necessarily dynamically unstable: While there exists a preferred solution,
there may not be a path through the solutions space that is dynamically
achievable. Given these arguments, and the fact that some BSs are dynamically
unstable in a regime not indicated by turning point arguments
\cite{Sanchis-Gual:2019ljs}, a more detailed analysis of the dynamical
nonlinear stability is necessary.
 
\section{Results} \label{sec:results}


In contrast to the stability analysis outlined in the last section, in Ref.
\cite{Sanchis-Gual:2019ljs}, it was discovered that $m=1$ scalar mini BSs are
subject to a NAI in a regime where they are expected to be stable based on
turning point arguments. In those cases, the evolution of the perturbed
stationary scalar BSs revealed exponentially growing non-axisymmetric modes
whose nonlinear evolution ultimately lead to the gravitational collapse of the
BS to a BH. In the following, expanding on the work presented in
Ref.~\cite{Sanchis-Gual:2019ljs} (see also Ref.~\cite{DiGiovanni:2020ror} for
the same instability present in dilute bosonic clouds), we study the nonlinear
dynamical behavior of a large class of $m=1$ (and some $m=2$) mini BSs and BSs
in the models presented in \ref{sec:models}. To that end, we numerically evolve
the Einstein-Klein-Gordon equations, Eq.~\eqref{eq:EinsteinKleinGordon}, in 3D from
stationary BS initial data that is perturbed only by numerical truncation error
(see Appendix~\ref{sec:NumEvo} for details of the numerical evolution). In many
cases, we find a similar NAI and determine its growth rate as a function of the
BS parameters. However, we identify two distinct regions of the parameter space, where the NAI growth rate approaches
zero. Past this point, rotating scalar BSs
appear to be stable under nonlinear evolution. This is true for parts of the 
\textit{relativistic branch}, identified in Sec.~\ref{sec:stable_arg}, as well as 
portions of the \textit{non-relativistic branch}. Furthermore, we show that the 
instability growth rate for BSs in \textit{any} scalar model, approaches zero (both 
in units of the BS mass and BS radius)
in the dilute/Newtonian limit, where the
nonlinear interactions can be neglected, and all BSs reduce to mini BSs. 

In Sec.~\ref{sec:NAIcharac}, we illustrate the linear characteristics of the
NAI and explain how we extract the identifying features. In
Sec.~\ref{sec:NAIgrowthrate}, we present the growth rates and other results on
the unstable linear modes for a number of types of rotating BSs across the
parameter space of the BSs'
frequency and couplings, as well as explicitly demonstrate the stable evolution of a few example cases. In Sec.~\ref{sec:endstate}, we describe the ultimate fate of the unstable BSs, and finally, in Sec.~\ref{sec:nai_origin}, we examine several criteria for
characterizing the onset of the NAI in BSs.

\subsection{Characterization of the linear non-axisymmetric instability} \label{sec:NAIcharac}

In the linear regime, the NAI manifests as an exponentially growing,
non-axisymmetric perturbation to both the scalar field and metric.  We study this by
evolving stationary BSs solutions in time, letting numerical truncation
error seed the instability at a small amplitude that grows by several orders of
magnitude before becoming nonlinear. This allows us to characterize the NAI
during this extended linear phase.

We will be interested in monitoring perturbations to quantities
that are stationary in the background BS solution, which has
an axisymmetric and stationary spacetime.
For example, linear perturbations to the magnitude squared of the BS complex scalar field $\Phi_\text{BS}$
can be written as 
\begin{align} 
|\Phi|^2-|\Phi_\text{BS}|^2 = e^{-i\tilde{\omega} t}e^{i \tilde{m}\varphi}\delta\Phi(r,\theta), 
\end{align}
where $\tilde{\omega}=\tilde{\omega}_R+i\tilde{\omega}_I$ captures both the
harmonic time dependence, $\tilde{\omega}_R>0$, and possible (un)stable
dynamical behavior with $\tilde{\omega}_I<0$ ($>0$), while $\tilde{m}$ is the
azimuthal mode number. In the time domain setting, we are assuming\footnote{We
confirm explicitly that the $\varphi$-dependence of the linear perturbation has
large support over only a single azimuthal mode $\tilde{m}$.} that a
\textit{single} perturbative mode (the most unstable) dominates the dynamics
during the linear instability phase. We characterize the NAI by its
azimuthal mode number, complex frequency, as well as the radial and polar
dependency encoded in $\delta\Phi(r,\theta)$. We can extract the growth rates
$\tilde{\omega}_I$ by fitting an exponential to the global maxima 
\begin{align}
\Phi_m := \max_{x\in\Sigma_t}| \partial_t |\Phi|^2 | & & \text{and} & & g_m := \max_{x\in\Sigma_t}|\partial_t g_{tt}|
\label{eq:phimgm}
\end{align}
of a given time slice $\Sigma_t$, which quantify the divergence from a stationary solution.
\begin{figure}[t]
\includegraphics[width=0.45\textwidth]{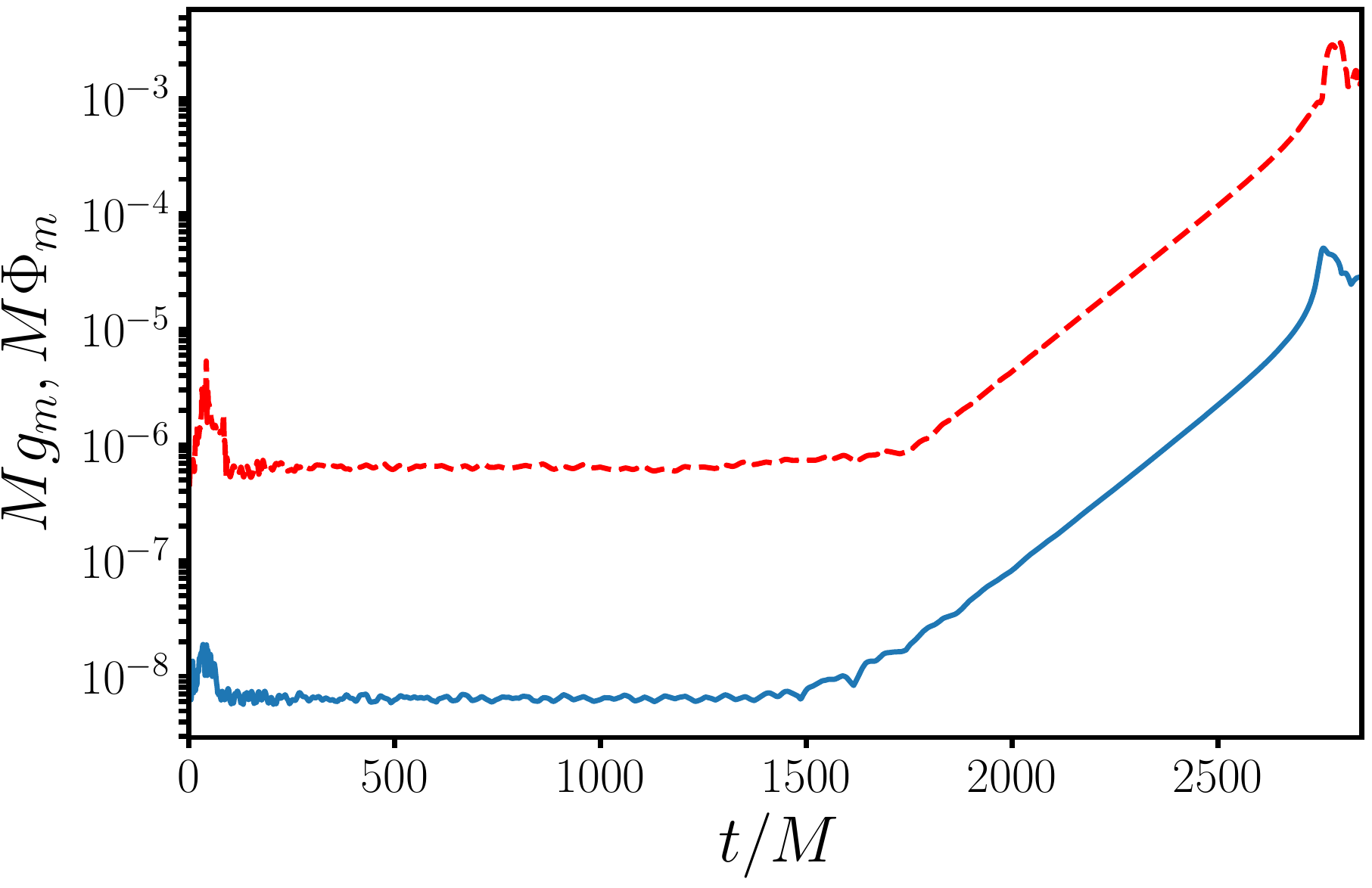}
\caption{The evolution of the global maxima $g_m=\max|\partial_t g_{tt}|$ (red dashed) and $\Phi_m=\max|\partial_t|\Phi|^2|$
(blue) [see Eq.~\eqref{eq:phimgm}] 
for an axionic $m=1$ BS with $f=5\times 10^{-2}$ and $\omega/\mu=0.425$. The
NAI is evident as a linear instability beginning around $t/M\approx 1.75\times
10^3$, and enters the nonlinear regime at roughly $t/M\approx 2.75\times
10^3$. The perturbations present at early times originate from lower order
interpolations at the mesh refinement boundaries (see Appendix~\ref{sec:NumEvo} for
details).}
\label{fig:omegaiextract}
\end{figure}
In \figurename{ \ref{fig:omegaiextract}}, we depict the typical dynamical behavior of $\Phi_m$ and $g_m$ for an $m=1$ scalar BS in the linear regime.

\begin{figure}[t]
\includegraphics[width=0.49\textwidth]{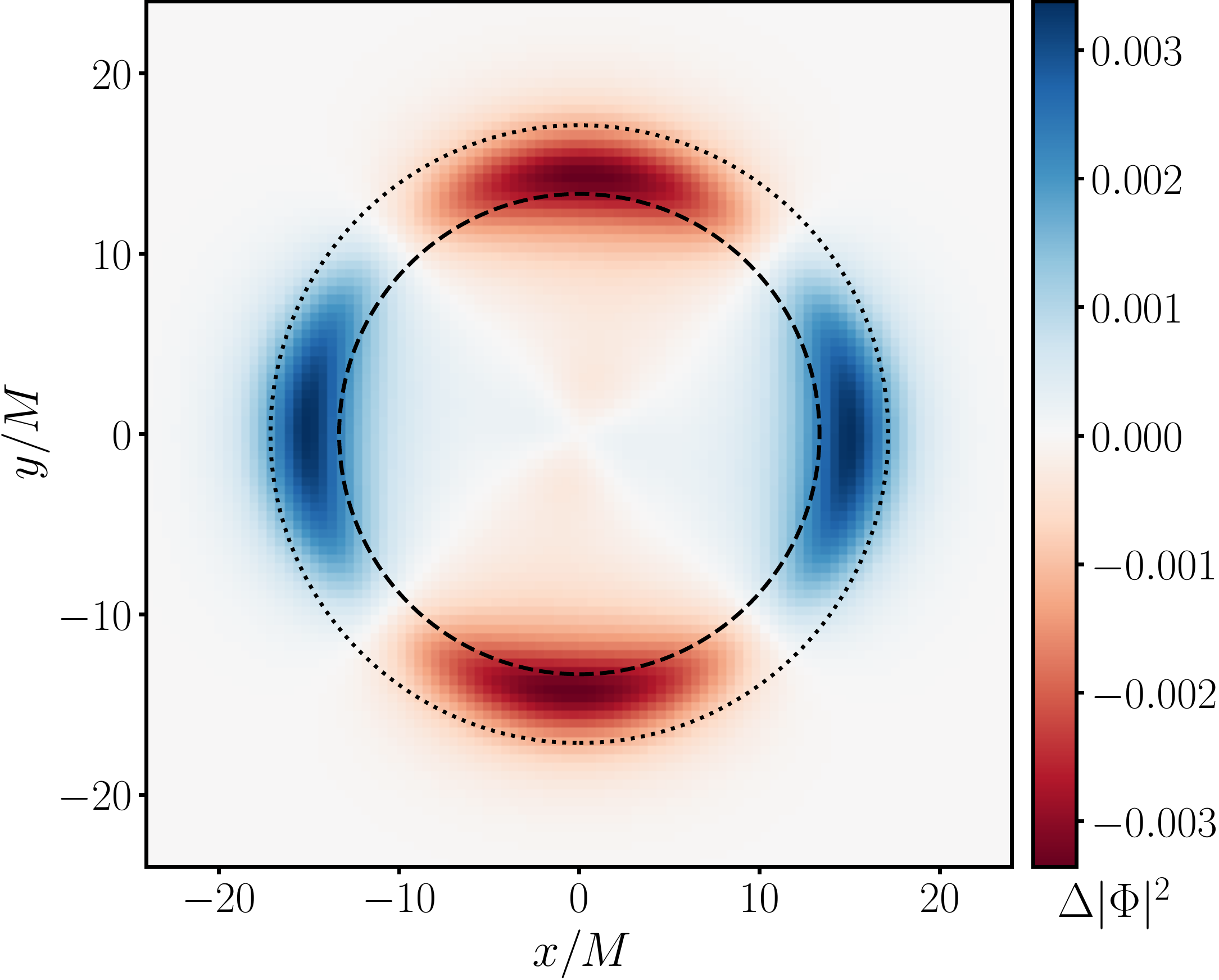}
\caption{The radial dependence of the unstable mode 
from a typical case, in particular the axionic BS considered in \figurename{
\ref{fig:omegaiextract}}. We show $\Delta|\Phi|^2$ [defined in Eq.~\eqref{eq:deltaphi}] in the equatorial plane (parameterized by the Cartesian coordinates $x$ and $y$), where we recall that $|\Phi(t=0)|^2$ is axisymmetric. 
This quantity is extracted at $t_\mathrm{NAI}=2.5\times 10^3 M$ which, 
as can be seen from \figurename{ \ref{fig:omegaiextract}}, is
solidly in the linear instability regime. The depicted structure rotates
about the center of the equatorial plane at an angular frequency $\tilde{\omega}_R
M=2.44\times 10^{-2}$, while its magnitude grows exponentially with
$\tilde{\omega}_I M = 6.4\times 10^{-3}$. For reference, we also add the radii
$R_{99}^K$ (dotted circle) and $R^Q_{99}$ (dashed circle) defined in
Sec.~\ref{sec:classical_observables}.} 
\label{fig:omegarextract}
\end{figure}

Furthermore, in \figurename{ \ref{fig:omegarextract}}, we present an example of
the difference in magnitude of the scalar field, $\Delta|\Phi|^2$, during the linear phase (at $t=t_\mathrm{NAI}$) of the
NAI, compared to the initial data (at $t=0$), in the equatorial plane of the BS: 
\begin{align}
\Delta|\Phi|^2:=|\Phi(t=0)|^2-|\Phi(t=t_\mathrm{NAI})|^2
\label{eq:deltaphi}
\end{align}
From this, we can extract the azimuthal mode number $\tilde{m}$, the radial
dependence, and the harmonic part of the frequency $\tilde{\omega}_R$. The
last-named is extracted by finding the radius $R_m$ where the perturbation is largest, 
$\max_{x\in\Sigma_t}\Delta|\Phi|^2=\Delta|\Phi(r=R_m,\theta=\pi/2)|^2$; we then
fit $\cos(\tilde{m}\varphi + \phi(t))$ to $\Delta|\Phi(R_m, \theta=\pi/2,
\varphi)|^2$ as a function of time during the exponential growth phase of the mode. The
essentially constant time derivative of the phase $\dot{\phi}$
gives $\tilde{\omega}_R$ for the unstable mode. As can be seen in 
\figurename{ \ref{fig:omegarextract}}, the azimuthal mode number $\tilde{m}$
is evident from $\Delta|\Phi|^2$, and shows large support
only over a single mode $\tilde{m}$. These modes could equivalently be extracted 
from perturbations to the energy density, or other projections of the stress-energy tensor, due to the global 
U(1) symmetry. In the following section, we present the
characteristics of the NAI extracted in this way from our time-domain
evolutions for several different families of rotating BSs.

\subsection{Results: Instability growth rates} \label{sec:NAIgrowthrate}

\begin{figure}[t]
\includegraphics[width=0.45\textwidth]{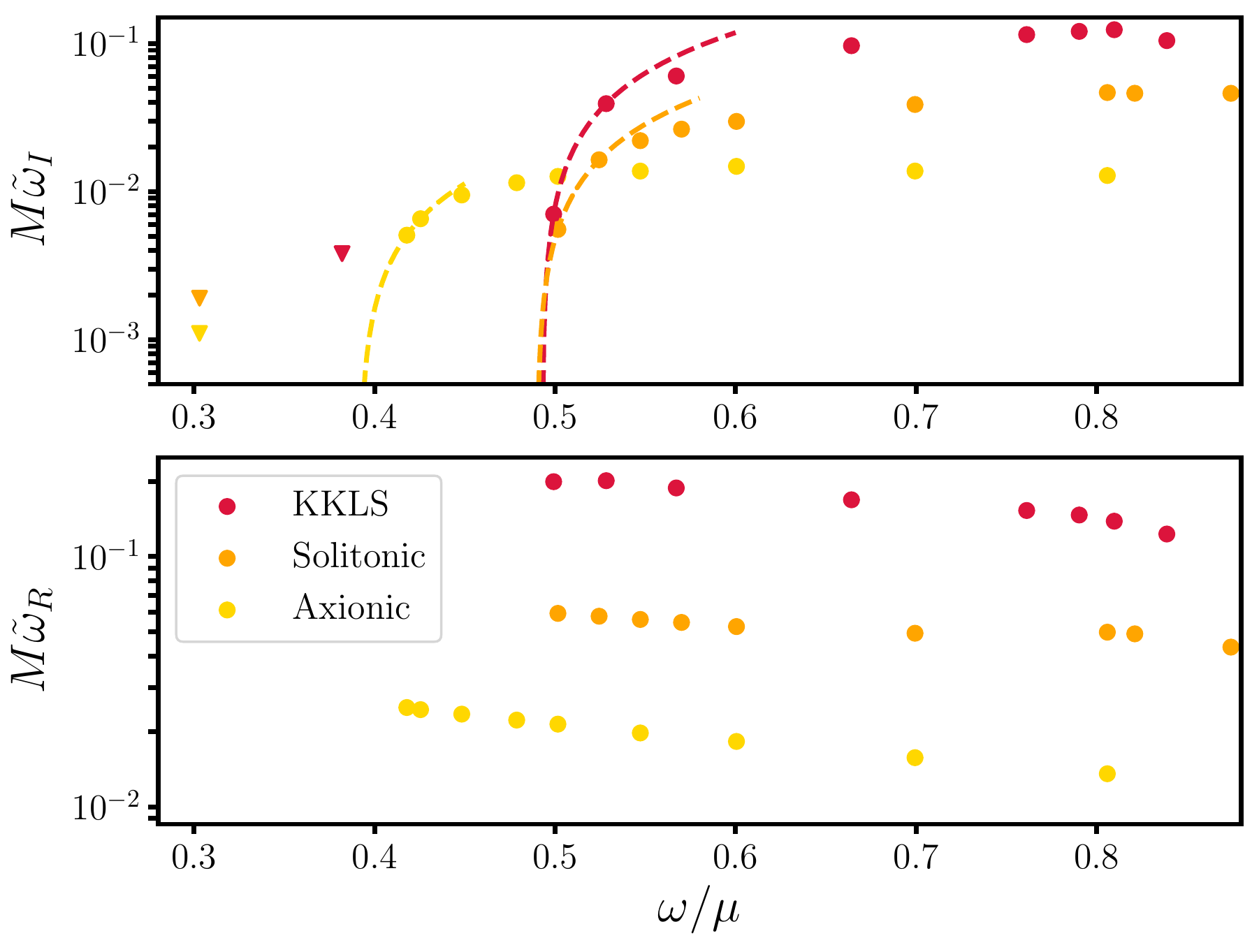}
\caption{The imaginary (top) and real (bottom) components of the frequency
of the $\tilde{m}=2$ unstable mode as a function of the BS 
frequency $\omega/\mu$ for three families of solutions shown in 
\figurename{ \ref{fig:BSproperties}}. The
axionic model has coupling $f=5\times 10^{-2}$, the solitonic potential has
$\sigma=0.05$, and the KKLS self-interactions are characterized by
$\kappa=0.1$. All three families are sequences of $m=1$ rotating BSs. The
dashed curves are linear interpolations considering only the two points with
smallest $\omega/\mu$ in each respective case. We also indicate the upper bounds on 
the growth rate of the cases considered in \tablename{ \ref{tab:properties}} by down-arrows.
}
\label{fig:NAI}
\end{figure}

In the following, we focus on finding the complex frequency $\tilde{\omega}$
for the unstable modes of the NAI. For all the families of BSs presented in
\figurename{ \ref{fig:BSproperties}}, we investigate how $\tilde{\omega}_I$
changes with the BS frequency or coupling. For self-interactions with an attractive first
correction, we focus on the second potentially stable branch (based on turning point arguments,
as discussed in Sec.~\ref{sec:stable_arg}), 
while for mini BSs and those
in models with repulsive potentials, we focus on the only branch that is
connected to the non-relativistic limit. 
In this limit, which corresponds to $\omega \to \mu$, the scalar field amplitude becomes small,
and only the lowest-order term in the potential, i.e. the mass term, will be important.
Ultimately, we identify critical points
$\omega^c/\mu$ and $\lambda^c/\mu^2$ where $\tilde{\omega}_I$ tends toward zero, i.e. the NAI shuts off.

In \figurename{ \ref{fig:NAI}}, we show the real and imaginary frequency of
the unstable mode for axionic,
solitonic, and KKLS scalar rotating BSs with $m=1$, as a function of the BS's frequency
$\omega/\mu<0.875$, for fixed coupling constants. All the cases shown
exhibit an $\tilde{m}=2$ NAI with a positive $\tilde{\omega}_I$ in the
frequency ranges with $\omega/\mu$ larger than the critical values:
\begin{align}
\omega^c_\mathrm{A}/\mu = 0.392, & & \omega^c_\mathrm{S}/\mu = 0.493, & & \omega^c_\mathrm{K}/\mu = 0.490.
\label{eq:critfreq}
\end{align}
Here the subscripts stand for axionic, solitonic, and KKLS, respectively. From
\figurename{ \ref{fig:BSproperties}}, we see that all the BSs shown in
\figurename{ \ref{fig:NAI}} are on the second branch that is nominally stable
based on turning point arguments. 
Approaching these critical values from above, we find that the instability rate $\tilde{\omega}_I$ tends toward zero.
Though it becomes more and more computationally expensive to measure longer and longer
instability timescales, $\tilde{\omega}_I$ appears to be approaching zero roughly linearly in $\omega$.
This suggests that below this critical value, $\omega < \omega^c$ (and above
the global maximum of the BS's mass), there is a range of BS solutions in the
relativistic regime of the nonlinear interactions models we consider, that are
free of the NAI.
Again, consulting \figurename{ \ref{fig:BSproperties}}, we see that such solutions have
large compactness.

\begin{figure}[t]
\includegraphics[width=0.48\textwidth]{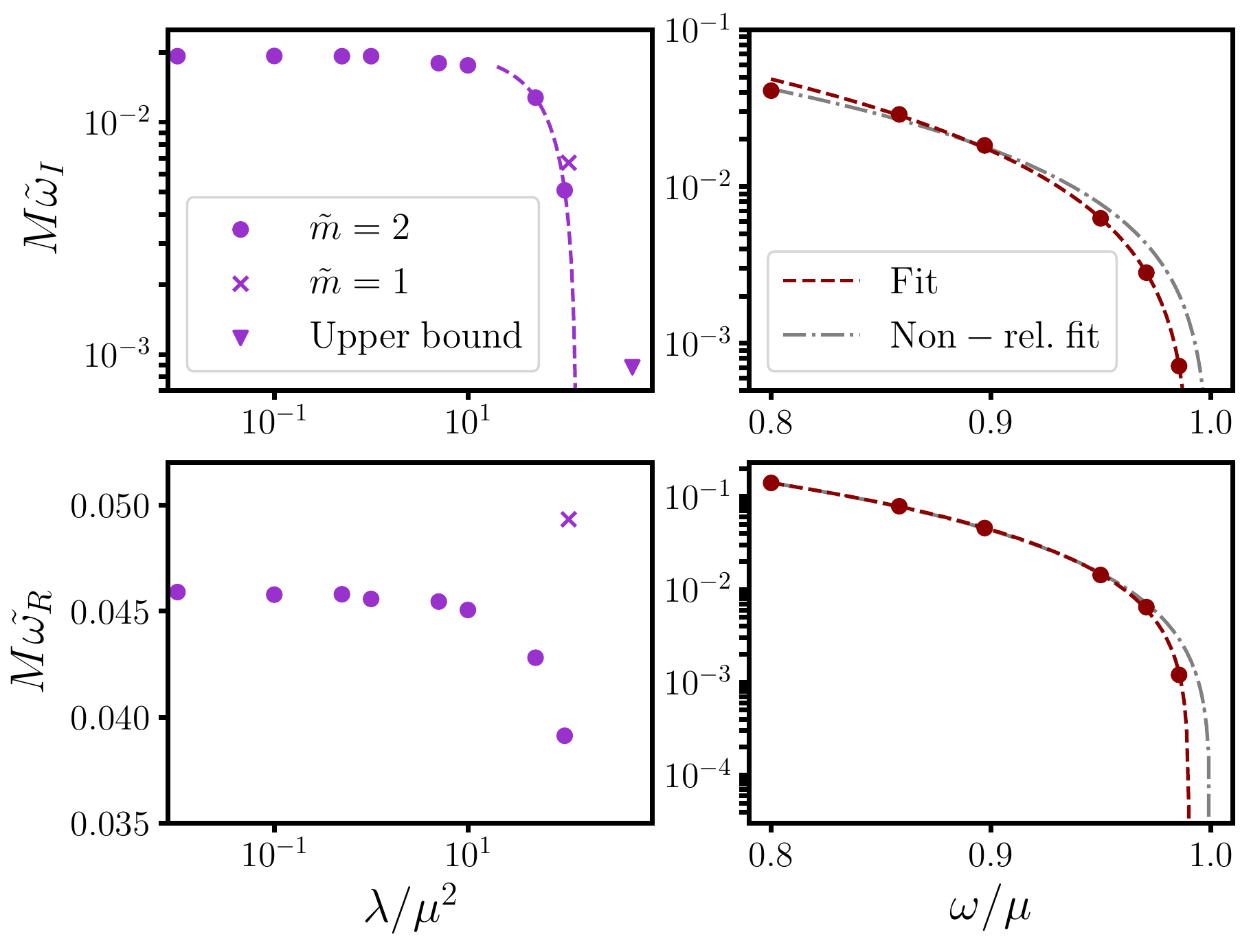}
\caption{The top and bottom panel on the left show the complex frequency of the
unstable modes of a family of repulsive potential BSs with fixed
$\omega/\mu=0.897$, but varying coupling constant $\lambda$. In all but one
case, the unstable mode has the azimuthal number $\tilde{m}=2$; for
$\lambda/\mu^2=110$, the $\tilde{m}=1$ mode is the (most) unstable. The dashed
curve is a linear interpolation based on the two $\tilde{m}=2$ cases with
largest $\lambda/\mu^2$ (ignoring the $\tilde{m}=1$ case). The top and bottom panel on the right show the
corresponding complex frequencies of the unstable modes of the family of mini
BSs. We fit both a generic quadratic ansatz with three degrees of freedom (red dashed)
and a quadratic ansatz (gray dash-dotted), which is fixed to
$\tilde{\omega}_{R,I}|_{\omega=\mu}=0$, to the data. Finally, we indicate the upper bounds on 
the growth rate of the case considered in \tablename{ \ref{tab:properties}} by down-arrows.
\label{fig:rmBSgrowthrates}
}
\end{figure}

We also study the stability of $m=1$ BSs with the repulsive nonlinear scalar
interactions given by Eq.~\eqref{eq:repulsive}. We fix the BS frequency to
$\omega/\mu=0.897$ and vary the coupling constant $\lambda$. In the
no-coupling limit, i.e., for $\lambda =0$, the $\omega/\mu=0.897$ mini BS
exhibits a $\tilde{m}=2$ NAI with growth rate $M\tilde{\omega}_I=1.8\times
10^{-2}$. As shown in \figurename{ \ref{fig:rmBSgrowthrates}}, as $\lambda$
increases, the $\tilde{m}=2$ NAI growth rate decreases (almost linearly), until
it approaches zero at 
\begin{align}
\lambda^c/\mu^2 = 133.2.
\label{eq:lamdaC}
\end{align}
Note, however, that there exists a small interval of $\lambda/\mu^2$, for which
the NAI is dominated by a $\tilde{m}=1$ mode [which we ignore for the purposes
of determining Eq.~\eqref{eq:lamdaC}]. This result seems to indicate that the
NAI is turned off for $\lambda/\mu^2>133.2$, suggesting that BSs residing in
this portion of the parameter space are stable. Finally, in contrast to the
trend in growth rates of the relativistic branch, the real part of the
perturbation's frequency $\tilde{\omega}_R M$ decreases \textit{together} with
$\tilde{\omega}_I M$ to zero at $\lambda^c/\mu^2$, on this non-relativistic
branch. When comparing the importance of the mass term $\mu^2|\Phi|^2$ to the
repulsive potential $\lambda|\Phi|^4$ we see that the self-interactions
dominate, i.e., $\lambda|\Phi|^2/\mu^2>1$ (around the maximum of the magnitude of
the scalar field inside the star), for $\omega/\mu=0.897$ repulsive BSs with
couplings $\lambda/\mu^2\geq 100$. This difference grows, with increasing
coupling, allowing for the possibility of scalar BSs where the effect of the
mass term is small in comparison to the quartic term, except at the outer and inner edges
of the star.

In \figurename{ \ref{fig:rmBSgrowthrates}}, we also show the NAI growth rates
of mini BSs in the regime between the global maximum of $\mu M$ and the
non-relativistic limit. The growth rates and harmonic frequencies
$\tilde{\omega}_R$ both decrease approaching $\omega=\mu$.  Fitting a generic
quadratic ansatz to the $\tilde{\omega}_IM$ and $\tilde{\omega}_RM$ data, we
find the respective critical boson frequencies where these two quantities would
go through zero: \begin{align} \omega^c_\mathrm{I, m}/\mu=0.991, & &
\omega^c_\mathrm{R, m}/\mu=0.990.  \label{eq:miniCrit} \end{align} We perform a
resolution study on the mini BS with $\omega/\mu=0.97$ (see
Appendix~\ref{sec:NumEvo} for details) to determine the numerical uncertainty
of our results, and hence, understand the stability of BSs with $\omega\approx
\mu$. The relative numerical uncertainty for the $\tilde{\omega}_{R,I}M$
estimates is $\approx 3\%$. Comparing this with the extrapolated critical
frequencies Eq.~\eqref{eq:miniCrit}, implies that our results are consistent
with $\tilde{\omega}_I$ and $\tilde{\omega}_R$ both reaching zero at
$\omega=\mu$. The second fit presented in \figurename{
 \ref{fig:rmBSgrowthrates}}, fixed to obey $\tilde{\omega}_I=0$ as
$\omega\rightarrow \mu$, is consistent with the free fit to within the numerical 
uncertainty. These results suggest that the timescales of the NAI
grow at least as $\tilde{\tau}/M\sim (1-\omega/\mu)^{-1}$ approaching the
non-relativistic limit. Note also, when normalizing the frequencies by the BS
radii, $\tilde{\omega}_{R,I}R^{K,Q}_{99}$, these quantities still tend to zero
in the dilute limit (see Sec.~\ref{sec:nai_origin} for a detailed discussion of
this). We point out that these results are consistent with Ref.~\cite{DiGiovanni:2020ror}; 
there a diffuse scalar cloud was evolved to form a
rotating BS and then undergo the NAI. They find (notice a factor of two
difference in the definition of the $|\Phi|^4$-term), a decrease of the
instability timescales for larger $\lambda$, which is to be expected based on
our results, because \textit{(a)} the non-relativistic limit is unstable
independent of the scalar self-interactions, and \textit{(b)} the repulsive
potential becomes more important the lower $\omega/\mu$. Therefore, the larger
$\omega/\mu$, the larger the critical coupling $\lambda^c/\mu^2$. In fact, one
could conjecture: $\lim_{\omega\rightarrow\mu}\lambda^c\rightarrow\infty$.

\begin{table}[b]
\begin{ruledtabular}
\begin{tabular}{c|cccccc}
$V(|\Phi|)$ & Coupl. & $\omega/\mu$ & $J/M^2$ & $C^K$ & $C^Q$ & $T_\text{max}/M$ \\ 
\hline \hline
Axionic & $f=0.005$ & $0.303$ & $3.59$ & $0.061$ & $0.049$ & $11.1\times 10^3$ \\ 
\hline 
Solitonic & $\sigma=0.05$ & $0.303$ & $1.36$ & $0.169$ & $0.140$ & $7.1\times 10^3$ \\ 
\hline 
KKLS & $\kappa=0.1$ & $0.382$ & $0.91$ & $0.333$ & $0.290$ & $3.7\times 10^3$ \\ 
\hline 
Repulsive & $\lambda/\mu^2=500$ & $0.897$ & $0.64$ & $0.077$ & $0.078$ & $13.5\times 10^3$ \\ 
\end{tabular} 
\end{ruledtabular}
\caption{We study the dynamical evolution of a set of $m=1$ BSs, one for each set of scalar self-interactions that is expected to be stable based on the results obtain in \figurename{ \ref{fig:NAI}} and \figurename{ \ref{fig:rmBSgrowthrates}}. We evolve these cases up to $T_\text{max}/M$ and find no sign of an instability arising.}
\label{tab:properties}
\end{table}

To further test all these results, we pick one BS solution for each potential
out of the regime conjectured to be free of the NAI (based on the trends in
\figurename{ \ref{fig:NAI}} and \figurename{ \ref{fig:rmBSgrowthrates}}) and
evolve it for many dynamical times. To that end, we evolve the cases summarized
in \tablename{ \ref{tab:properties}} up to times $T_\text{max}$, while
monitoring $\Phi_m$ [defined in Eq.~\eqref{eq:phimgm}] over time. In
Sec.~\ref{sec:NAIcharac}, we illustrated that tracking $\Phi_m$ captures the
exponential growth during the linear phase of the NAI. While the 
truncation error introduces perturbations, these slowly decay away, and---in
contrast to \figurename{ \ref{fig:omegaiextract}}---there is no sign of
exponential growth for the cases presented in \tablename{
    \ref{tab:properties}}. From this we conclude that any instability develops
on significantly longer timescales than the unstable cases with $\omega >
\omega^c$ and $\lambda>\lambda^c$ found above. Based on these simulations, we
place upper bounds on the growth rates of possible instabilities arising in
those BSs. To that end, we use the unstable cases presented in \figurename{ \ref{fig:NAI}},
and extrapolate the exponential growth backwards in time to $t=0$ to estimate
the initial amplitude at which truncation error seeds the unstable mode.
Assuming that this amplitude is similar in the stars considered in \tablename{ 
\ref{tab:properties}}, we extract approximate upper bounds on the possible growth rates
for an instability not to be evident during the simulated time. We 
include these in \figurename{ \ref{fig:NAI}} and \figurename{
\ref{fig:rmBSgrowthrates}}.


\begin{table}[b]
\begin{ruledtabular}
\begin{tabular}{c|cccccc}
$V(|\Phi|)$ & Coupl. & $m$ & $\omega/\mu$ & $C^K$ & $\tilde{m}$ & $M\tilde{\omega}_I$ \\ 
\hline \hline
KKLS & $\kappa=0.1$ & $2$ & $0.6$ & $0.26$ & $4$ & $7.4\times 10^{-2}$ \\ 
\hline 
Axionic & $f=0.1$ & $2$ &  $0.9$ & $0.08$ & $2$ & $1.1\times 10^{-2}$ \\ 
\hline 
Solitonic & $\sigma=0.05$ & $2$ & $0.4$ & $0.14$ & $2$ & $5.3\times 10^{-2}$ \\
\hline
Solitonic & $\sigma=0.05$ & $2$ & $0.9$ & $0.03$ & $4$ & $8.5\times 10^{-2}$ \\ \hline\hline
Solitonic & $\sigma=0.1$ & $1$ & $0.8$ & $0.08$ & $2$ & $1.3\times 10^{-1}$ \\ \hline
Solitonic & $\sigma=0.1$ & $1$ & $0.4$ & $0.28$ & -- & $< 5.4\times 10^{-3}$\\ \hline
Axionic & $f=0.01$ & $1$ & $0.8$ & $0.03$ & $2$ & $4.5\times 10^{-2}$ \\ \hline
Axionic & $f=0.01$ & $1$ & $0.3$ & $0.19$ & -- & $<5.3 \times 10^{-3}$\\ \hline
Liouville & $\alpha=0.1$ & $1$ & $0.9$ & $0.07$ & $2$ & $1.0\times 10^{-2}$ \\ \hline
Liouville & $\alpha=0.05$ & $1$ & $0.9$ & $0.07$ & -- & $<2.8\times 10^{-3}$ \\
\end{tabular} 
\end{ruledtabular}
\caption{Properties of several additional rotating BSs are considered and their
dynamical behavior studied. $\tilde{m}$ refers to the azimuthal number of the dominant
(i.e., most unstable) non-axisymmetric mode. However, as pointed out in the
text, for $m=2$ BSs, the NAI is composed of several competing modes
simultaneously. We estimate upper bounds on the growth rates for the cases, where no
sign of an instability can be found. We describe in the text how these upper bounds are obtained.}
\label{tab:m2BSs}
\end{table}

Finally, we use nonlinear evolutions to explore the stability of individual
$m=2$ rotating BSs in different scalar models, as well as $m=1$ BSs in various
models with different couplings from the ones shown in \figurename{
\ref{fig:NAI}}. The properties of the resulting evolutions are summarized in
\tablename{ \ref{tab:m2BSs}}. In our small set of $m=2$ BSs, we did not find a
solution that does not develop a NAI. Note that, as will be discussed in the next
section, the NAI for the $m=2$ BSs has non-negligible support over several azimuthal
modes. Therefore, the $\tilde{m}$ values for $m=2$ BSs in \tablename{
\ref{tab:m2BSs}} should be understood as rough estimates.

The result for $m=1$ BSs shown in \tablename{ \ref{tab:m2BSs}} solidify the 
conclusions drawn above. We show that even when changing the coupling constants
in scalar models considered in \figurename{ \ref{fig:NAI}}, there is an
unstable (high frequency) regime, but also a seemingly stable (low frequency)
regime, both in the solitonic and the axionic model. This shows that the
coupling parameters chosen in \figurename{ \ref{fig:NAI}} are not special. In
addition to this, 
we also study the stability of BSs in scalar models with a
Liouville potential \cite{Schunck:1999zu} [defined in Eq.~\eqref{eq:Liouville}], a non-perturbative extension of
the repulsive self-interactions. Analogous to
the family of BSs in the repulsive model, studied in \figurename{
\ref{fig:rmBSgrowthrates}}, we fix the BS frequency to $\omega/\mu=0.9$ and vary
the coupling $\alpha$. As seen from \tablename{ \ref{tab:m2BSs}}, there is a
low-coupling range, for which an $\tilde{m}=2$ NAI is present, but also a large
coupling regime, where the NAI is quenched, and no sign of an instability can be
found. Therefore, the addition of a $|\Phi|^4$ term in the scalar potential,
is sufficient to stabilize rotating BS in a portion of the parameter space, 
and a potential that has additional higher order terms does not affect this qualitative behavior.

\subsection{Results: Endstate of the NAI} \label{sec:endstate}

\begin{figure}[t]
\includegraphics[width=0.48\textwidth]{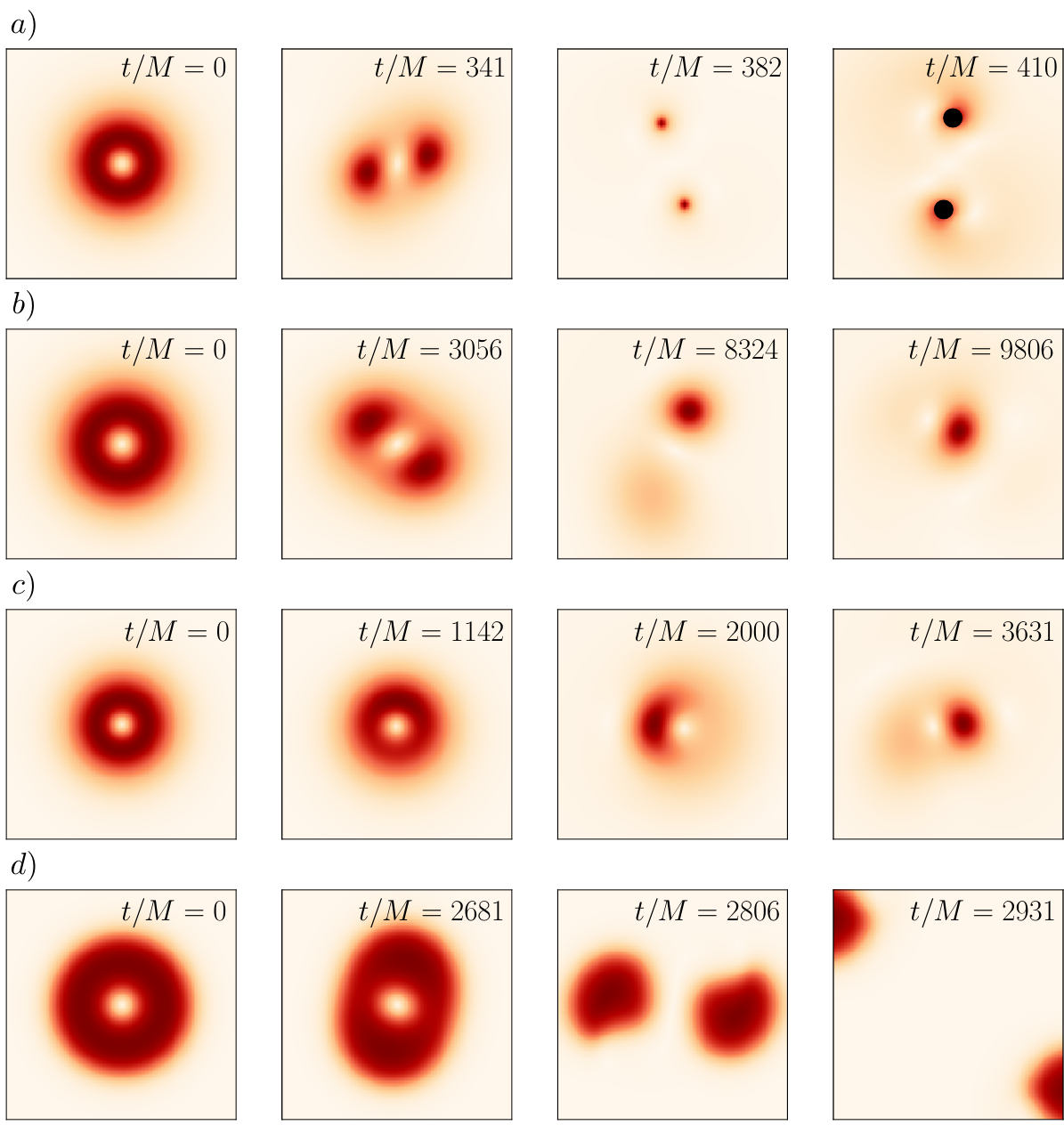}
\caption{Series of snapshots (increasing in time from left to right) showing the 
evolution of $|\Phi|^2$ in four different scenarios where a BS undergoes a NAI. 
These are representative of the possible end states of all the unstable $m=1$ BSs 
studied here. From top to bottom: $a)$
Evolution of an $m=1$ mini BS with $\omega/\mu=0.8$ that collapses to a binary BH 
(the regions inside the apparent horizons are indicated in black).
$b)$ Snapshots of an $\omega/\mu=0.95$ $m=1$ mini BS resulting in a
non-rotating BS with non-negligible linear momentum. $c)$ The NAI of an $m=1$
rotating BS, in the repulsive model with $\lambda/\mu^2=110$, and
$\omega/\mu=0.897$ (corresponding to the $\tilde{m}=1$ case in \figurename{
\ref{fig:rmBSgrowthrates}}), yielding a non-rotating BS with large linear momentum. 
$d)$ Finally, the evolution of an $m=1$ axionic BS
with $\omega/\mu=0.425$ where the NAI results in the fragmentation of the star into 
two equal-mass non-rotating BSs.}
\label{fig:Endstate1}
\end{figure}

In the previous section, we identified two distinct regimes, the
relativistic and non-relativistic branch, where the NAI is quenched and the
stars appear stable when evolved nonlinearly. In what follows, we explore the end 
state of the instability of $m=1$ and $m=2$ BSs both in the relativistic regime and 
the non-relativistic regime.

In Ref.~\cite{Sanchis-Gual:2019ljs}, it was found that a $m=1$ mini BS with
$\omega/\mu=0.83$ was subject to a $\tilde{m}=2$ NAI that lead to the formation
of a BH.  We study this family of mini BSs in more detail, following the growth
and end state of the instability for the cases shown in \figurename{
\ref{fig:rmBSgrowthrates}} with $\omega/\mu \in \{0.80, 0.86, 0.90, 0.95\}$.
These stars undergo a $\tilde{m}=2$ NAI, form two approximately non-rotating
BSs that orbit around each other, and collapse to a binary BH (for $\omega/\mu
\in \{0.80, 0.86\}$; see panel $a)$ in \figurename{ \ref{fig:Endstate1}}), or
merge into a single non-rotating BS that then collapses to an isolated BH (for
$\omega/\mu =0.90$).
\begin{figure}[t]
\includegraphics[width=0.48\textwidth]{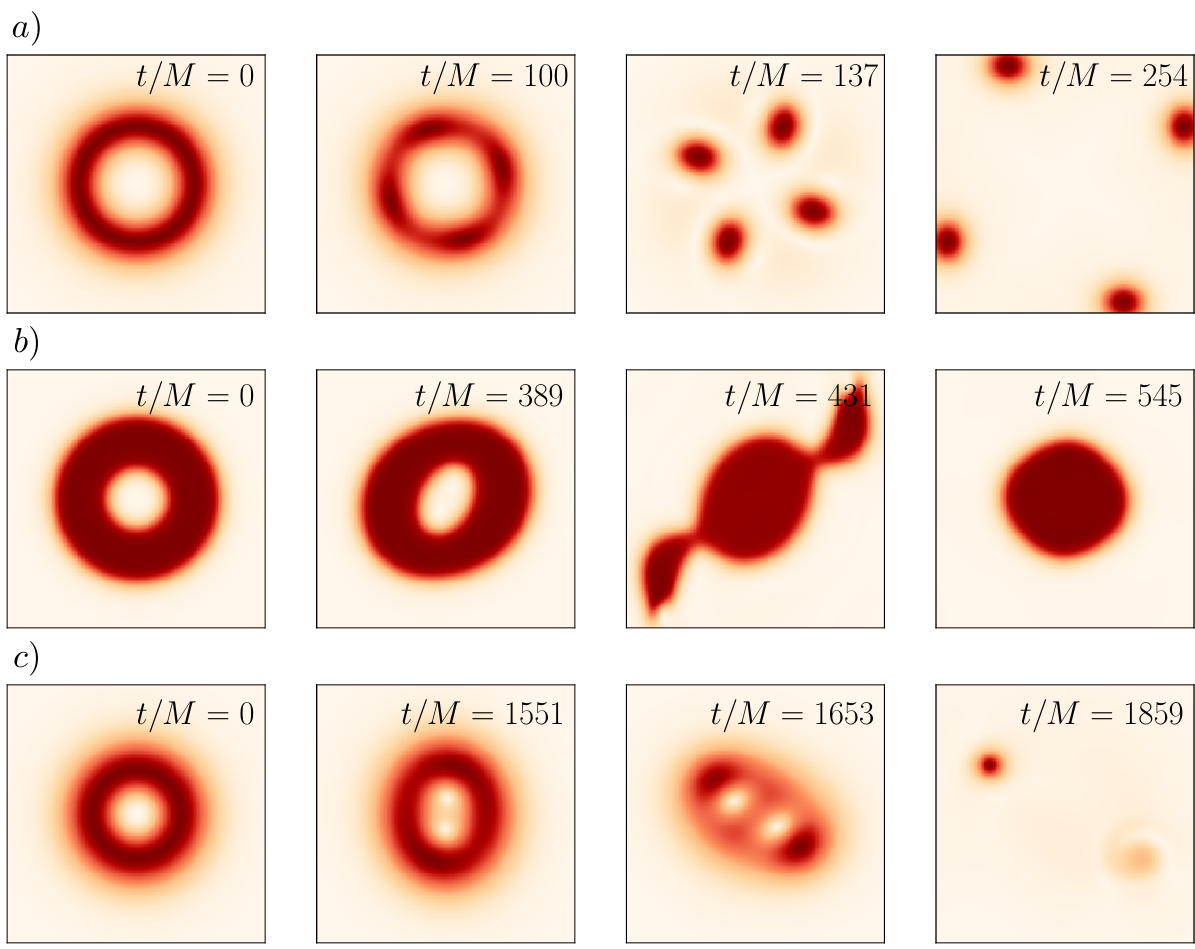}
\caption{Analogous to \figurename{ \ref{fig:Endstate1}}, here we present the
dynamics of three $m=2$ BSs undergoing a NAI (their properties can be found in
\tablename{ \ref{tab:m2BSs}}).
From top to bottom: $a)$ The evolution of a rotating BS, in the KKLS
model with $\kappa=0.1$, and $\omega/\mu=0.6$, results in four 
non-rotating equal mass BSs that are flung out from the center of mass. $b)$ The fragmentation of a solitonic BS with
$\sigma=0.05$ and $\omega/\mu=0.4$ into a large, oscillating, approximately
spherically symmetric BS at the center, and two non-rotating BSs flung out at
relativistic velocities. $c)$ Finally, the NAI of an axionic BS with $f=0.1$
and $\omega/\mu=0.9$, undergoing a complex fragmentation process resulting in
a single non-rotating BS with significant linear momentum.}
\label{fig:Endstate2}
\end{figure}
In the case of $\omega/\mu = 0.80$ mini BSs, the individual BSs 
collapse to BHs with negligible spin, and subsequently merge into a single BH $\approx 60M$ later. 
Additionally, we find that mini BSs with $\omega/\mu=0.95$
undergo an $\tilde{m}=2$ NAI and settle to a (highly perturbed) non-rotating BS
(that does not collapse further), through the emission of scalar and
gravitational waves, with non-zero linear momentum (see panel $b)$ in
\figurename{ \ref{fig:Endstate1}}). This suggests that the contributions from
the $\tilde{m}=1$ unstable mode is non-negligible in this case. Similarly,
repulsive self-interactions Eq.~\eqref{eq:repulsive} with sufficiently large
coupling can prevent the BSs from collapsing to BHs.  Instead, the $m=1$ BSs in
the repulsive model, considered in \figurename{ \ref{fig:rmBSgrowthrates}},
undergoing the $\tilde{m}=2$ NAI, collapse to BHs for couplings up to
$\lambda/\mu^2 = 50$. For $\lambda/\mu^2 \geq 100$, however, we find that the
$\tilde{m}=2$ NAI eventually results in an approximately spherically symmetric
BS, while the instability is quenched at Eq.~\eqref{eq:lamdaC}.  In \figurename{
\ref{fig:Endstate1}}, we also depict the nonlinear evolution of a BS undergoing
a $\tilde{m}=1$ NAI. 

This results in a highly perturbed, approximately
non-rotating BS, with significant linear momentum, radiating both scalar and
gravitational waves outwards. We can conclude that, the closer the BS is to the
critical frequency/coupling, the less it seems to be prone to forming a BH.
Therefore, we can also conclude that the end state of the NAI for $m=1$ BSs on
the branch directly connected to the non-relativistic limit depends both on the
nonlinear scalar self-interactions, as well as the BS frequency. In contrast to
this, we find a consistent outcome for the final fate of $m=1$ BSs undergoing the
NAI on the relativistic branch. \textit{All} BSs presented in \figurename{
\ref{fig:NAI}} have the same final state, independent of the potential or their
frequency: after the $\tilde{m}=2$ NAI, two approximately non-rotating BSs are
formed that are flung out into opposite directions at relativistic speeds. This
behavior is illustrated in panel $d)$ of \figurename{ \ref{fig:Endstate1}}.

We can conclude that the end state of the NAI in $m=1$ BSs is clearly differentiated 
in the two branches. The first non-relativistic branch exhibits quasi
stable bound non-stationary states after fragmentation of the original BS that either settle into
a BH or to a single non-rotating BS. The second, relativistic branch 
consistently results in two non-rotating BSs that are flung out after fragmentation, independent of
the frequency or the character of the nonlinear scalar self-interactions.

In \figurename{ \ref{fig:Endstate2}}, we show the evolution of $m=2$ BSs
undergoing a NAI. The dynamics of the NAI in these stars is more complex, as
more unstable non-axisymmetric modes are of non-negligible size. While the
linear phase is still mostly dominated by a single azimuthal mode $\tilde{m}$
(see also \tablename{ \ref{tab:m2BSs}}), the nonlinear evolution of these
$m=2$ BSs is substantially different from their $m=1$ counterparts.

\subsection{Results: Physical origin of the NAI} \label{sec:nai_origin}

In this section, we investigate the possible physical mechanisms leading to stable
regions in the BS parameter space. We focus on the case of $m=1$ BSs and
distinguish between stars on the relativistic branch and the non-relativistic
branch, as the NAI of each has fundamentally different properties. 

First, let us consider the non-relativistic branch, i.e., mini BSs and stars in the
repulsive scalar model with frequencies between the non-relativistic limit
$\omega/\mu =1$ and the \textit{global} maximum of the BS's mass. Since the
solitonic, axionic, and KKLS scalar models reduce to the mass term in the
Newtonian limit (for those cases, the non-relativistic branch extends from $\omega/\mu =1$
to the \textit{local} maximum of the BS's mass), the following also applies to
that part of the BS parameter space in those models. From the previous section,
we recall that the NAI timescales of mini BSs tend to infinity in the Newtonian
limit, while BSs in the repulsive scalar model with fixed frequency
$\omega/\mu=0.897$ turn stable for coupling parameters with
$\lambda>\lambda^c=133.2\mu^2$. In Ref.~\cite{DiGiovanni:2020ror}, it was suggested
that the presence of a co-rotation point (the radius at which angular velocity of
the matter and the pattern speed of the unstable modes are equal) inside the
star, is driving the NAI. There it was conjectured that, if the co-rotation
point is outside the star (or it does not exist at all), then the star is
stable, while if such a point exists within the
star's radius, the star is unstable. Here we find this conjecture to hold for BSs on the
non-relativistic branch. (Though not for the relativistic branch, as we shall discuss below.) 
The angular velocity $\tilde{\Omega}$ 
is defined in terms of the energy density $\rho$ and angular momentum density $J^\mu$, with respect to slices of
constant time with unit normal $n^{\mu}$ and projector $\gamma_{\mu \nu}=g_{\mu \nu}+n_\mu n_\nu$:
\begin{align}
\tilde{\Omega}=J^\varphi/\rho & \textrm{, with } & J_i= \gamma^\mu_i n^\nu T_{\mu\nu},  & & \rho= n^\mu n^\nu T_{\mu\nu}.
\end{align}
(Note that we are only measuring this quantity with respect to our stationary and axisymmetric
solutions.) We define the co-rotation radius to be the radius $r=R_\mathrm{cor}$
in the equatorial plane of a rotating BS where
$\tilde{\Omega}(R_\mathrm{cor})=\tilde{\omega}_R/\tilde{m}$. 

\begin{figure}[t]
\includegraphics[width=0.48\textwidth]{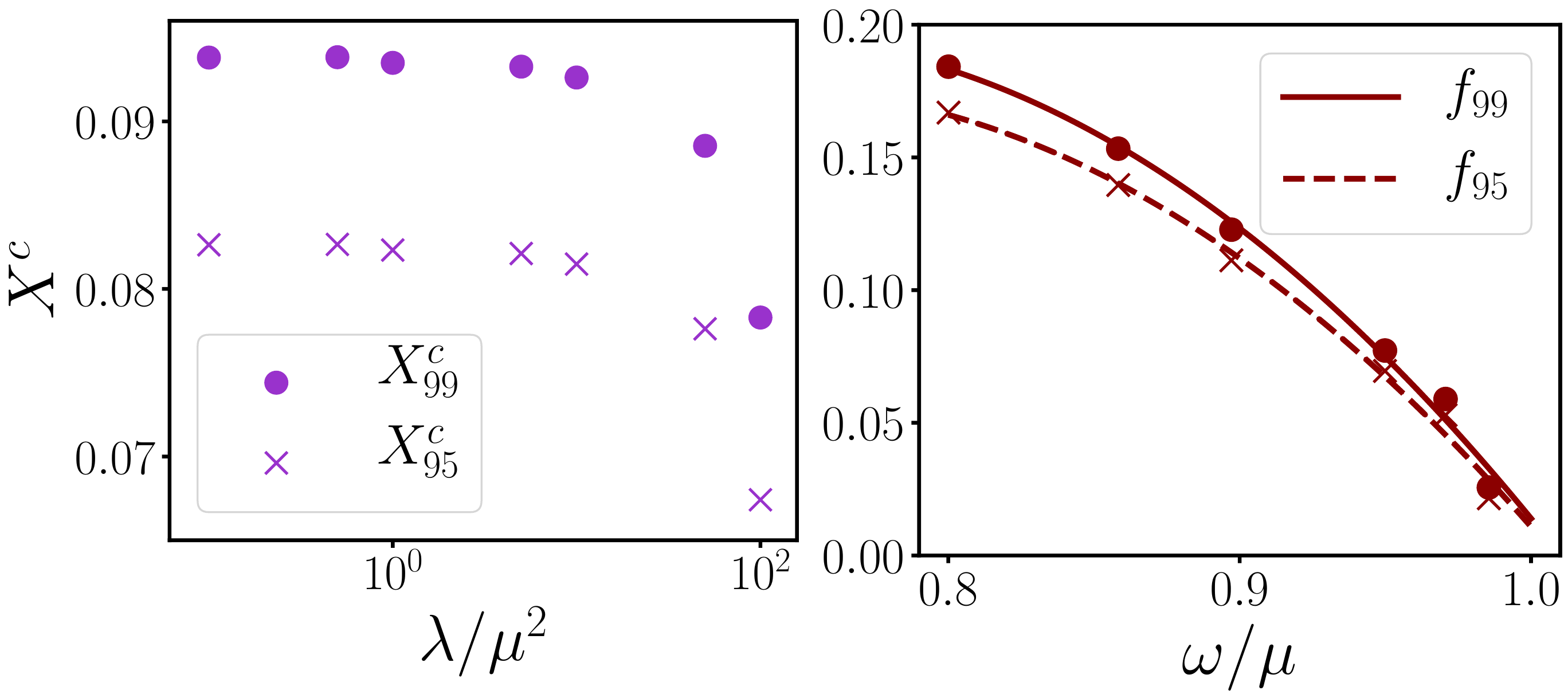}
\caption{ We plot the normalized difference $X^c_i=(R^K_i-R_\mathrm{cor})/R^K_i$,
$i\in\{99,95\}$, between the co-rotation point $R_\mathrm{cor}$ and the two
radii $R_{99}^K$ and $R_{95}^K$ for the families of BSs considered in
\figurename{ \ref{fig:rmBSgrowthrates}}. The left panel shows
the repulsive potential cases, and the right shows the mini BS cases. The
quantity $R_{95}^K$ is defined, analogously to how
$R^K_{99}$ is in Sec.~\ref{sec:classical_observables}, as the circular radius
$r=R^K_{95}$, where $95\%$ of the energy resides in $r<R^K_{95}$. In the right panel, 
the solid and dashed lines are quadratic fits to the $X^c_{99}$ and $X_{95}^c$ data points.}
\label{fig:corrot}
\end{figure}

In \figurename{ \ref{fig:corrot}}, we show that $R_\mathrm{cor}$ approaches the
radius of mini BSs, as well as BSs in the repulsive scalar model, as those
families approach the critical frequency $\omega^c$ and critical coupling $\lambda^c$,
respectively. This holds for either definition of the radius of the BS. In the
repulsive scalar model, this is not surprising, since we already saw in
Sec.~\ref{sec:NAIgrowthrate} that the real part of the unstable mode's
frequency (and therefore, the pattern speed $\tilde{\omega}_R/\tilde{m}$), tends
to zero at $\lambda^c$. While the radius remains finite passing through
$\lambda^c$, the decreasing pattern speed is pushing $R_\mathrm{cor}\rightarrow \infty$, as
$\lambda\rightarrow\lambda^c$. In the mini BS case, however, this is
\textit{not} the case. In the Newtonian limit, $\omega/\mu\rightarrow
1$, the radius of the BS diverges $R^K_{99}\rightarrow\infty$, while
simultaneously, $\tilde{\omega}_R\rightarrow 0$. Therefore, there are two
competing effects determining, whether $R^K_{99}<R_\mathrm{cor}$ or
$R^K_{99}>R_\mathrm{cor}$. Despite this, the co-rotation point approaches the
radius of the BS from inside the star, while the NAI growth rate decreases, as can be deduced from
\figurename{ \ref{fig:corrot}}. On top of this, we explicitly checked that $\lim_{\omega\rightarrow \mu}(\tilde{\omega}_{R,I}R_{99}^K)= 0$.
This suggests that the presence of the
co-rotation radius inside the star (independent of the notion of radius) is
related to the instability of BSs on the non-relativistic branch. Lastly, this also seem to indicate that there is no finite regime with stable mini BSs in
the Newtonian limit between the values in Eq.~\eqref{eq:miniCrit} and $\omega/\mu=1$.

\begin{figure}[t]
\includegraphics[width=0.48\textwidth]{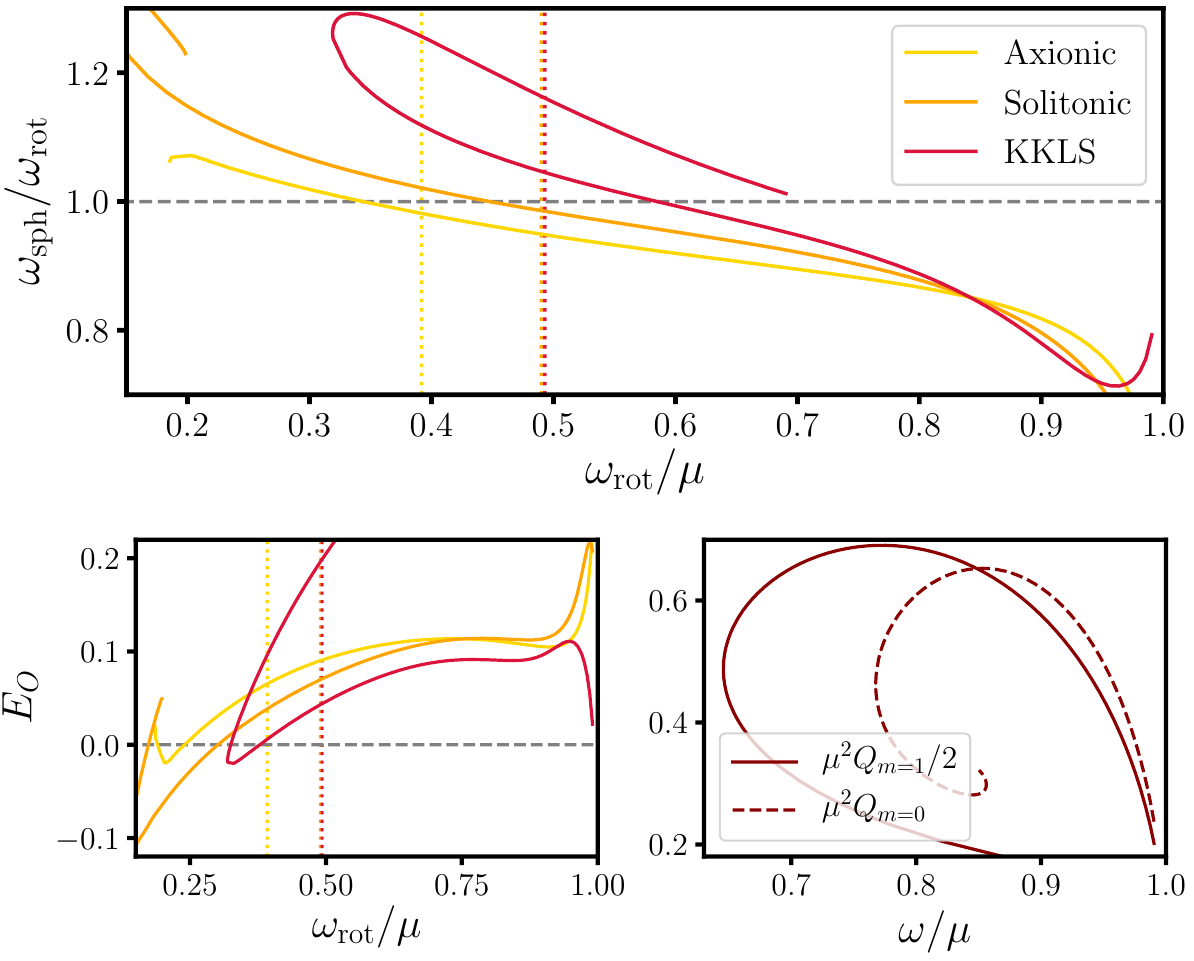}
\caption{Top panel: The ratio 
$\omega_\mathrm{sph}/\omega_\mathrm{rot}$ 
of the boson frequency in $m=1$ BSs with charge $Q$ to the frequency of $m=0$
BSs with charge $Q/2$ in several scalar models. The ratios pass through unity
at $\omega^c_A/\mu=0.34$, $\omega^c_S/\mu=0.45$ and $\omega^c_K/\mu=0.58$ for the
axionic, solitonic and KKLS models, respectively. For comparison, we indicate the 
critical frequencies, in Eq.~\eqref{eq:critfreq}, by dashed vertical lines. Bottom left: The orbital energy $E_O$, defined in
Eq.~\eqref{eq:Eorb}, of a $m=0$ binary BS system with constituent charge $Q/2$
emerging from a single $m=1$ BS with charge $Q$. (Legend from top panel also
applies here.) Bottom right: Half the scalar charge $Q/2$ for $m=1$ mini BSs
(solid line), compared with the full charge $Q$ of $m=0$ mini BSs (dashed
line), as a function of their respective frequencies.
\label{fig:Qfreq}
}
\end{figure}

Let us now discuss the relativistic branch, i.e., axionic, solitonic, and KKLS
BSs with frequencies between the local and the global maximum of $M$ (see
\figurename{ \ref{fig:BSproperties}}): $0.3\lesssim \omega/\mu \lesssim
0.9$. Recall from the discussion in Sec.~\ref{sec:endstate} that these families of BSs undergo a
$\tilde{m}=2$ NAI that results in two equal-mass non-rotating BSs being flung
out from the center of mass, for initial BS frequencies
$\omega^c<\omega\lesssim 0.9 \mu$. Fundamentally, the system posses three
distinct conserved quantities: the ADM mass $M$, the angular momentum $J$
about the symmetry axis, and the U(1)-charge $Q$. Given that the total number
of bosons is conserved, we neglect scalar radiation and assume that the
final state of the instability of a $m=1$ BS with charge $Q_\mathrm{rot}$ is a
binary $m=0$ BS of total charge $2Q_\mathrm{sph}=Q_\mathrm{rot}$. 
From the difference in the initial BS mass, and that of the putative binary,  
we can calculate the normalized orbital energy
\begin{align}
E_O=(M_\mathrm{rot}-2M_\mathrm{sph})/M_\mathrm{rot},
\label{eq:Eorb}
\end{align}
as well as the frequency $\omega_\mathrm{sph}$ of BSs in the binary, as a function of initial boson
frequency $\omega_\mathrm{rot}$. We show this in \figurename{ \ref{fig:Qfreq}}. 
Based on the values shown there,
the NAI fragments should be significantly unbound: $E_O=0.1$ corresponds
to a velocity of $\approx 0.4 c$ at infinity. 
All the unstable solutions on the relativistic branch that we find occur in the
regime where $E_O>0$, which is consistent with the fact that the instability gives 
rise to two fragments which appear unbound. However, there is some part of the 
parameter space where $E_O>0$, but we do not find the NAI.

The critical frequencies in Eq.~\eqref{eq:critfreq}, where the NAI instability
approaches zero, seem to roughly align with those, where
$\omega_\mathrm{sph}=\omega_\mathrm{rot}$ (see \figurename{ \ref{fig:Qfreq}}).
Due to the fragmentation, and ignoring radiation, the bosons can fall into lower frequency/energy
states within the stars (since there
$\omega_\mathrm{sph}<\omega_\mathrm{rot}$), while for the regime where
$\omega_\mathrm{sph}>\omega_\mathrm{rot}$, the bosons would have to climb up
into higher frequency states, if the rotating BSs were to fragment into two
non-rotating BSs. Hence, one can conjecture that the points, where
$\omega_\mathrm{sph}=\omega_\mathrm{rot}$, indicate the switching of the
stability properties. However, there is a discrepancy between the critical
frequencies where $\omega_\mathrm{sph}=\omega_\mathrm{rot}$ in \figurename{
    \ref{fig:Qfreq}}, and those derived from the exponentially growing modes in
Eq.~\eqref{eq:critfreq} (even when including the numerical uncertainty),
indicating that this criterion based on the relative frequency of the rotating
and non-rotating equilibrium BS solutions is at most approximate.
Furthermore, this approximate argument cannot be applied in the same way to the non-relativistic
branch.  Recall from Sec.~\ref{sec:endstate} that the NAI in the case of mini
BSs yields two orbiting pieces that either collapse to BHs or form a single
non-rotating BS. Assuming these two orbiting scalar field distributions are
roughly stationary isolated spherically symmetric BSs, i.e., applying the above
argument, and consulting \figurename{ \ref{fig:Qfreq}} (bottom right), we
notice two features. 
One is that $\omega_\mathrm{rot}<\omega_\mathrm{sph}$ for most of the parameter space,
at least when comparing the turning-point stable branches. This may be related
to the fact that the BS fragments collapse in some cases, which would be consistent
with them being unstable spherical stars. The second feature to note is that
for $m=1$ stars with
$\omega_\mathrm{rot}/\mu>0.846$, there exists no corresponding non-rotating
counterparts with half the U(1)-charge. Again, this may be related to the fact
that for large enough $\omega_\mathrm{rot}$, the instability gives rise to a single
BH or non-rotating BS.
Additionally, we observe no significant
qualitative change in the dynamics of the systems when moving across the point
where both curves cross.

Finally, we note that the co-rotation argument discussed above in the context
of non-relativistic BSs does not apply to the relativistic branch.  This is
because the real part of the unstable mode's frequency remain non-zero on the
entire relativistic branch (see \figurename{ \ref{fig:NAI}}), and hence the
co-rotation point remains well inside the star for all unstable BSs. Even when
extrapolating $\tilde{\omega}_RM$ deeper into the relativistic regime, the
co-rotation point seems to only exit the star at the global maximum of the BS's
mass. 


\section{Discussion and Conclusion} \label{sec:disscussion}

In this work, we study the stability of $m=1$ and 2 BSs 
in various massive complex scalar models minimally coupled to gravity.
We do this by numerically evolving the Einstein-Klein-Gordon system of equations,
starting from stationary BS initial data. We consider a number of different types of scalar interactions,
and explore the parameter space of BS solutions. We find that \textit{all}
$m=1$ mini BSs are unstable to an exponentially-growing,
non-axisymmetric mode with azimuthal number $\tilde{m}=2$, but that the growth
rate scales as $\tilde{\omega}_I \propto ( 1-\omega/\mu)$ when
$\omega\rightarrow\mu$ (see \figurename{ \ref{fig:rmBSgrowthrates}}). Since all nonlinear scalar self-interactions reduce to
the mass term for sufficiently small field values, this holds also for all
$m=1$ scalar BS that reside sufficiently deep in the Newtonian limit.
This may have some relevance for rotating distributions of dark matter in models
where they are composed of ultralight bosons.

We find that adding a nonlinear scalar potential can quench the instability in a number of cases, 
in the sense that the growth rate approaches zero as the BS frequency decreases
towards a critical value (see \figurename{ \ref{fig:NAI}} and \figurename{ \ref{fig:rmBSgrowthrates}}).  
Studying the conjecture in Ref.~\cite{DiGiovanni:2020ror},
we relate the stability of the stars on the
non-relativistic branch to the presence of the co-rotation point inside the
star, and show that, when the BSs become stable, the co-rotation point moves
outside of the star (independent of the notion of radius; see \figurename{
\ref{fig:corrot}}). 
The stability of solutions on the
relativistic branch seems unrelated to the co-rotation, though
approximately related to the BS frequency;  
the final state of the instability is always an equal-mass binary BS of
non-rotating constituents, and we find around where the rotating stars become
stable, there is a transition from this final state having lower frequency to having
higher frequency (see \figurename{ \ref{fig:Qfreq}} for the details).
It would be interesting for future work to see if this argument could be sharpened.
In Refs.~\cite{Sanchis-Gual:2019ljs, DiGiovanni:2020ror}, it was argued that the
stability of BSs is related to the morphology of surfaces of constant
scalar energy. Stars with toroidal morphology were claimed to be unstable,
while those with spherical surfaces of constant energy, stable. 
Here we find $m=1$ scalar BSs that have toroidal morphology, but do not appear
to be subject to this instability.
Though we do not explore the parameter space of $m=2$ BSs as thoroughly as the $m=1$
case, we find all the examples we consider to be linearly unstable.

We choose several examples of $m=1$ BSs with nonlinear interactions that are
past the point where the NAI growth rate approaches zero, and evolve them for
many dynamical times (in some cases $>10^4M$) and find no sign of a growing
perturbation.  These include cases with high compactness, rapid rotation, and---in the
case of the KKLS BS in \tablename{ \ref{tab:properties}}---unstable and stable photon orbits, which can be used
for future studies of mergers, and compared to the mergers of black holes and
neutron stars.  Of course, our methods can not definitely rule out instabilities
of any kind, in particular instabilities that arise on longer timescales than
considered here. For example, in Ref.~\cite{Keir:2014oka}, it was conjectured
that stable photon orbits may give rise to nonlinear instability, based on the
slow timescales associated with the decay of linear perturbations.  If such a
nonlinear instability were to exist, it would likely operate on much longer
timescales than the NAI, and so may not be evident here.

Another future direction is to see whether rotating scalar BSs could be formed
dynamically, through a non-spinning binary BS merger, or through the collapse of a rotating
scalar cloud. The parameter space identified here could be used to choose favorable
initial conditions that may dynamically give rise to a stable rotating BS.
It would also be interesting to investigate scenarios where unstable rotating
BS solutions may form, for example from the migration of a stable star to a
unstable part of the parameter space, or through rapid collapse. In such cases,
the nonlinear development of the instability will give rise to distinct
gravitational wave signatures.  For example, here we find cases where the NAI
of a rotating BS gives rise to a binary BH which will subsequently merge, similar to
the binary BH formation found to arise in the collapse of supermassive
stars~\cite{Reisswig:2013sqa}. 

\begin{acknowledgments}
We acknowledge support from an NSERC Discovery grant.
Research at Perimeter Institute is supported in part by the Government of Canada through the Department of Innovation, Science and Economic Development Canada and by
the Province of Ontario through the Ministry of Economic Development, Job
Creation and Trade.  
This research was enabled in part by support provided by SciNet
(www.scinethpc.ca) and Compute Canada (www.computecanada.ca).
Calculations were performed on the Symmetry cluster at
Perimeter Institute and the Niagara cluster at the University of Toronto.
\end{acknowledgments}

\appendix
\section{Numerical construction of stationary boson stars}	 \label{sec:NumMeth}

Here we provide the details of the numerical methods employed in order to
construct stationary BS solutions.  We assume an axisymmetric and
stationary metric $g_{\mu\nu}$, which can be written in Lewis-Papapetrou
coordinates as in Eq.~\eqref{eq:metricansatz}, where all free functions depend
only on the $r$ and $\theta$ coordinates. Additionally, we use 
a harmonic scalar field ansatz
$\Phi=e^{i\omega t+im\varphi}\phi$, where $\phi$ carries the $r$ and $\theta$
dependence, that is compatible with the metric ansatz. Plugging these expressions into the Einstein-Klein-Gordon equations,
Eq.~\eqref{eq:EinsteinKleinGordon}, results in a set of five coupled partial
differential equations in $r$ and $\theta$. Imposing regularity at the origin,
$r=0$, and asymptotic flatness at $r\rightarrow\infty$ provides the boundary
conditions for the elliptic set of equations. See, for instance,
Ref.~\cite{Kleihaus:2005me} for the explicit form of the
equations and boundary conditions for the metric ansatz Eq.~\eqref{eq:metricansatz}
chosen here. Notice a typo in their Eq. (B1): The second to last term should be $-2l(\partial_r f)^2$. Additionally, we find our axisymmetric system of equations to agree if, in the equation for $f(r,\theta)$ (i.e., Eq. (B6) in Ref.~\cite{Kleihaus:2005me}):
\begin{align}
-\frac{1}{2}\partial_\theta f & \ \rightarrow -\frac{1}{2}\frac{\partial_\theta f\partial_\theta l}{l}, \\
-2r^2\kappa l U(\phi) & \ \rightarrow -2r^2\kappa l g U(\phi).
\end{align}
As mentioned in the main text, we introduce the \textit{auxiliary} functions $\omega_s$ and $\rho$. Following the arguments in Ref.~\cite{Kleihaus:2005me}, these obey the equations
\begin{align}
\Delta_g \omega_s = 0 , & & \Delta_g \rho = j^t/\omega_s,
\label{eq:auxiliaryeq}
\end{align}
where $\Delta_g := g^{ij}\nabla_i\nabla_j$ is the Laplacian associated with the
spacetime metric $g_{\mu\nu}$, and $j^t$ is proportional to the charge density
defined by the Noether current associated with the global U(1) symmetry. Here we 
have also promoted the scalar field's frequency $\omega$ to the scalar function $\omega_s
$, which, however, is fixed to be constant by the Laplace equation (with the 
appropriate boundary conditions given below). Introducing $\omega_s$ and $\rho$ in 
such a fashion enables us to impose \textit{either} the BS's frequency $\omega$, 
\textit{or} the total
U(1)-Noether charge $Q$, as boundary conditions at spatial infinity (outlined 
below). This way, we are able to explore the full parameter space, even, when the 
family of solutions develops two different branches for a fixed frequency $\omega$ 
or charge $Q$. 

We solve these equations using a relaxation method, which requires a sufficiently
good guess for the field configurations. In order to obtain such an initial guess,
we begin by exploiting the fact that Eq.~\eqref{eq:metastable} admits solitonic
solutions even in the absence of gravity. In
the non-rotating limit, the above set of elliptic partial different equations
reduces to a single \textit{ordinary} differential equation for $\phi(r)$. We
compactify the radial coordinate to $\bar{r}\in(0,1)$ with $\bar{r}=r/(1+r)$.
Given the boundary conditions for $\phi$, we use a shooting method, starting at
the origin, and integrating outwards to $\bar{r}=1$, to generate the
non-gravitating scalar solitons (known as Q-balls) in the non-relativistic
regime (i.e., where $\omega/\mu\approx 1$). We use these Q-ball solutions as the starting 
point for solving the equations including gravity.
In anticipation of the rotating case, we use a
Newton-Raphson-type relaxation code with fifth order finite differences to
solve for these spherically symmetric BSs. In order to be able to impose either
the BS frequency $\omega/\mu$, \textit{or} its U(1)-charge $Q$ at $r\rightarrow
\infty$, we introduce the two equations Eq.~\eqref{eq:auxiliaryeq} into the
relaxation scheme. The boundary conditions for these auxiliary functions are
\begin{align}
\partial_r\rho|_{r=0}=0, & & \lim_{r\rightarrow\infty}\rho =\rho_\infty,
\end{align}
where $\rho_\infty$ is arbitrary (we set $\rho_\infty=1$). We can then impose the BS frequency $\omega$ as a boundary condition given
\begin{align}
\lim_{r\rightarrow\infty}\omega_s=\omega,
\label{eq:freqBC}
\end{align}
or the U(1)-charge by
\begin{align}
\lim_{r\rightarrow\infty}8\pi r^2\omega_s\partial_r \rho=Q.
\label{eq:chargeBC}
\end{align}
This follows directly from a volume integration of the $\rho$-equation in Eq.~\eqref{eq:auxiliaryeq}. Finally, in both cases $\lim_{r\rightarrow0}\partial_r\omega_s=0$. Gravity can be
incorporated slowly starting from the Q-ball solutions in the non-relativistic
limit by increasing $\kappa$ from $0$ incrementally. Depending on the
convergence properties of the individual solutions in the relaxation scheme,
the step size $\delta\kappa$ must be adjusted. We find that
$\delta\kappa=0.01$ is typically sufficient. 

Using these non-rotating BS solutions in the
non-relativistic limit as a starting point, the rest of the parameter space can be
explored by marching iteratively in the $\omega/\mu$ direction: We start from a
solution with $\omega=\omega_0$, take a sufficiently small step
$\omega_1=\omega_0-\delta\omega$ (with $\delta\omega>0$) towards small
frequencies, use the solution with $\omega_0$ as an initial guess, and then relax using the above
established relaxation scheme into the new solution imposing
$\omega=\omega_1$. Due to the spiral feature of the solutions in the
$(\omega,Q)$ plane, we can follow this approach only up to a global minimum,
$\omega_\text{min}$, at which point no new solutions can be found at
$\omega<\omega_\text{min}$. At these turning points, we switch from imposing the
frequency $\omega_\infty$, to imposing the corresponding BS U(1)-charge with
Eq.~\eqref{eq:chargeBC}. We then proceed by incrementally decreasing (or
increasing) the charge until we reach the next turning point, where we switch
back to imposing the frequency with Eq.~\eqref{eq:freqBC}, until the family of BS
solutions is generated across the parameter space. Spherically symmetric BSs
with different potentials can be generated using this family of KKLS BSs as an
initial seed for the relaxation into the BS solution with a different potential
in the non-relativistic limit (for a given $\kappa$). We find this works, even if the
two potentials cannot be continuously deformed into each other, as long as 
the coupling parameters are chosen such that the potentials are similar.

For rotating BSs, we proceed in much the same fashion. In this case, however, we
cannot obtain the initial rotating Q-balls using a shooting method, as the
equation is no longer an \textit{ordinary} differential equation, but a partial
differential equation in $r$ and $\theta$. We found that generating analytic
seeds based on the results of Ref.~\cite{Kleihaus:2005me} for $m>0$ rotating Q-balls
provided sufficiently good initial guesses to be able to relax into the correct
rotating Q-ball solutions in the non-relativistic regime. Once these solutions
are generated, we follow the same procedure as in the $m=0$ case.
We find that the necessary resolution varies significantly with the properties of the BS considered.
For less compact BSs, $C<0.2$, a resolution of $N_{\bar{r}}\times
N_\theta=350\times 50$ (or even lower resolution) was sufficient, while for BSs
with high compactness, resolutions up to $500\times 100$ were necessary for a
successful relaxation. 
In general, we find a resolution of $N_{\bar{r}}\times N_\theta=500\times 200$
is sufficient to generate initial data that accurate enough so that the
evolution errors are dominant, even for the highest resolution simulations we
consider.

\section{Numerical evolution of boson stars}	 \label{sec:NumEvo}

\begin{figure}[t!]
\includegraphics[width=0.48\textwidth]{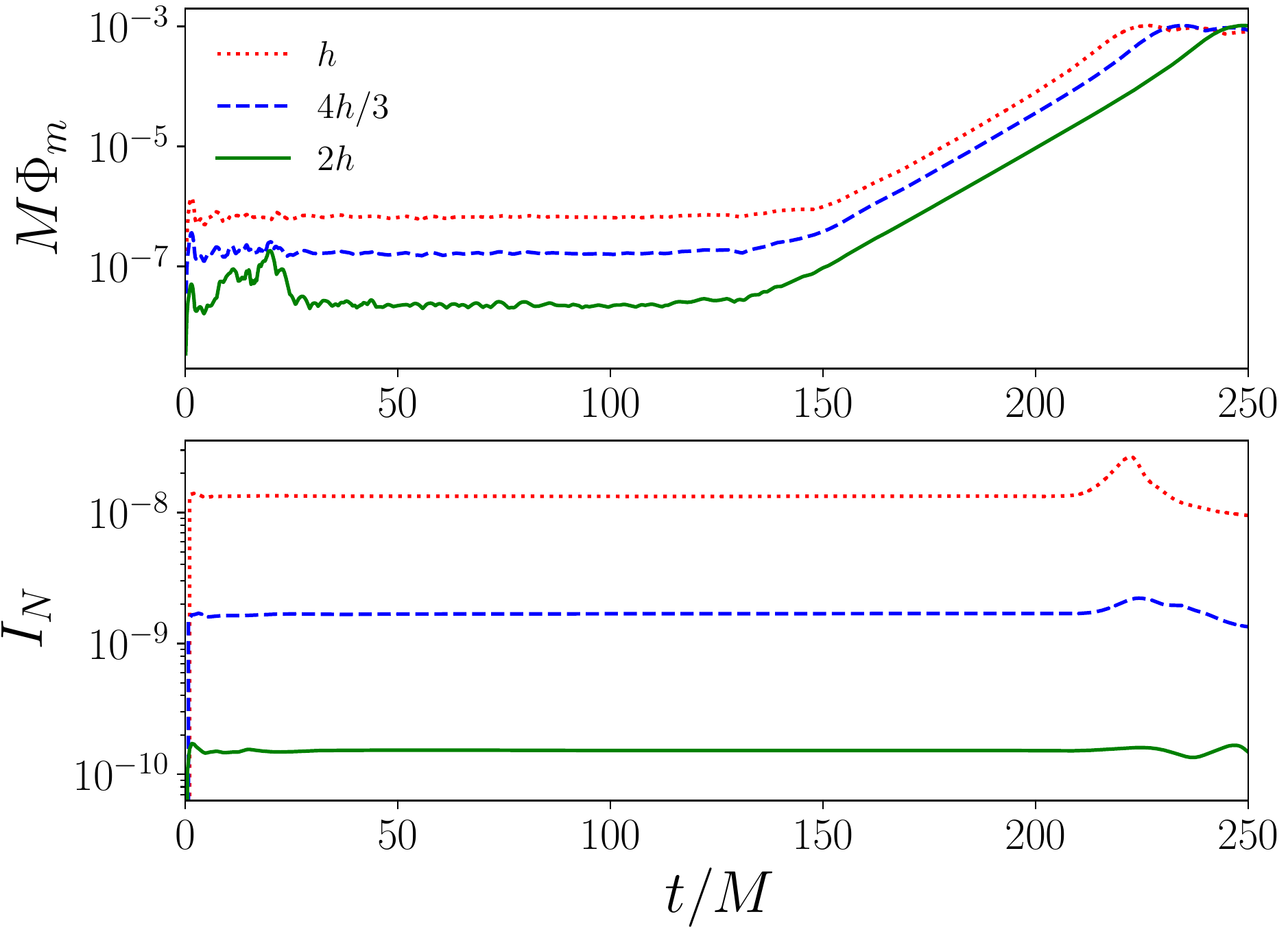}
    \caption{Time evolution of $\Phi_m$ (top panel), defined in Eq.~\eqref{eq:phimgm}, at the three
different resolutions specified in the text. We also show the integrated norm
of the generalized harmonic constraint violation $I_N:=\int_{\Sigma_t}d^3\sigma |H_\mu - \square
    x_\mu |$ (bottom panel). Both quantities converge to zero at the expected fourth order. The origin of the
perturbations of $\Phi_m$ at early times is discussed in the text. The units for $I_N$ are arbitrary since here our purpose is just to demonstrate convergence.}
\label{fig:ResStudy}
\end{figure}

Once the stationary solutions are generated, we evolve the
Einstein-Klein-Gordon equations in the generalized harmonic formulation
\cite{Pretorius:2004jg}. We utilize standard fourth order accurate finite
difference stencils and fourth order Runge-Kutta time integration, and use 
adaptive mesh refinement (AMR). Our computational grid extends to spatial infinity, where
we apply boundary conditions that the metric is flat and the scalar field is zero, through
the use of compactified coordinates. More details can be found in Ref.~\cite{East:2011aa}. 
We use a gauge where we fix the source functions $H^a$
to be equal to those calculated from the
stationary solution, and constant in time. For those mini BSs that collapse to 
binary (or single) BHs, we utilize a damped harmonic gauge~\cite{Lindblom:2009tu,Choptuik:2009ww}. 
We consider a range of cases where
the BS compactness varies significantly from $\sim 10^{-3}$ to $\approx 1/3$.
Therefore, we employ grid hierarchies with between five and eight levels of
mesh refinement (using a 2:1 refinement ratio) centered on the BS.

We perform resolution studies, using three different resolutions, on $m=1$
axionic and KKLS BSs with $\omega/\mu = 0.6$ (and $M/R_{99}^K=0.02$) and
$\omega/\mu=0.7$ (and $M/R_{99}^K=0.14$), respectively, as well as a mini BS
with $\omega/\mu = 0.97$, to check convergence and determine the numerical
accuracy of our results. In \figurename{ \ref{fig:ResStudy}}, we present the
time evolution of the magnitude of the scalar field, captured by
Eq.~\eqref{eq:phimgm}, and the integrated norm of the constraint violations of the
KKLS BS case. For this case, the lowest resolution is $49^3$ with six levels of
mesh refinement and a spatial resolution of $dx/M\approx 0.26$ on the lowest
level. The medium and high resolutions are, respectively, a
factor of $4/3$ and $2\times$ larger than the lowest resolution. Both quantities are
converging to zero, consistent with the fact that the instability is seeded
only by truncation error.  The resolution studies for the other cases showed
similar behavior.  We note that there is transient perturbation to the global
maximum $\Phi_m$ that is briefly evident in the highest resolution of
Fig.~\ref{fig:ResStudy} that is not converging at the expected rate.  This is
due to the time interpolation on the AMR 
boundaries, and in particular an
inaccuracy in how the ``past" time level used to perform this interpolation at
the initial time is set.

The medium resolution used in the resolution studies is equivalent to the resolution
we use for all the other cases studied here.
We place the mesh refinement such that both radii of the BS (defined in
Sec.~\ref{sec:classical_observables}) reside inside the finest AMR level, and set the spatial resolution to be roughly
$dx/R_{99}^K\approx 0.04$ or greater in all cases considered in the text. 
For most cases, we fix the mesh refinement to be that of the initial time slice. In 
those cases where the NAI yields binary BHs or binary BSs, however, we employ 
AMR set by the truncation error estimates between different 
refinement levels to track the fragments.
%
Using Richardson extrapolation, we are able to determine the numerical accuracy
of the NAI growth rate and harmonic frequency estimates both on the relativistic
and the non-relativistic branch of BS solutions. Based on the respective resolution
studies of the $m=1$ KKLS, axionic, and mini BSs, we estimate the relative
numerical error in $\tilde{\omega}_I$ to typically
be $2.5\%$ ($3\%$) on the relativistic (non-relativistic) branch. The error in
$\tilde{\omega}_R$ is smaller; the relative numerical error in this quantity
is $<0.5\%$ ($3\%$) on the relativistic (non-relativistic) branch.

\bibliography{bib.bib}

\begin{thebibliography}{87}%
\makeatletter
\providecommand \@ifxundefined [1]{%
 \@ifx{#1\undefined}
}%
\providecommand \@ifnum [1]{%
 \ifnum #1\expandafter \@firstoftwo
 \else \expandafter \@secondoftwo
 \fi
}%
\providecommand \@ifx [1]{%
 \ifx #1\expandafter \@firstoftwo
 \else \expandafter \@secondoftwo
 \fi
}%
\providecommand \natexlab [1]{#1}%
\providecommand \enquote  [1]{``#1''}%
\providecommand \bibnamefont  [1]{#1}%
\providecommand \bibfnamefont [1]{#1}%
\providecommand \citenamefont [1]{#1}%
\providecommand \href@noop [0]{\@secondoftwo}%
\providecommand \href [0]{\begingroup \@sanitize@url \@href}%
\providecommand \@href[1]{\@@startlink{#1}\@@href}%
\providecommand \@@href[1]{\endgroup#1\@@endlink}%
\providecommand \@sanitize@url [0]{\catcode `\\12\catcode `\$12\catcode
  `\&12\catcode `\#12\catcode `\^12\catcode `\_12\catcode `\%12\relax}%
\providecommand \@@startlink[1]{}%
\providecommand \@@endlink[0]{}%
\providecommand \url  [0]{\begingroup\@sanitize@url \@url }%
\providecommand \@url [1]{\endgroup\@href {#1}{\urlprefix }}%
\providecommand \urlprefix  [0]{URL }%
\providecommand \Eprint [0]{\href }%
\providecommand \doibase [0]{http://dx.doi.org/}%
\providecommand \selectlanguage [0]{\@gobble}%
\providecommand \bibinfo  [0]{\@secondoftwo}%
\providecommand \bibfield  [0]{\@secondoftwo}%
\providecommand \translation [1]{[#1]}%
\providecommand \BibitemOpen [0]{}%
\providecommand \bibitemStop [0]{}%
\providecommand \bibitemNoStop [0]{.\EOS\space}%
\providecommand \EOS [0]{\spacefactor3000\relax}%
\providecommand \BibitemShut  [1]{\csname bibitem#1\endcsname}%
\let\auto@bib@innerbib\@empty
\bibitem [{\citenamefont {Abbott}\ \emph {et~al.}(2016)\citenamefont {Abbott}
  \emph {et~al.}}]{Abbott:2016blz}%
  \BibitemOpen
  \bibfield  {author} {\bibinfo {author} {\bibfnamefont {B.~P.}\ \bibnamefont
  {Abbott}} \emph {et~al.} (\bibinfo {collaboration} {LIGO Scientific,
  Virgo}),\ }\href {\doibase 10.1103/PhysRevLett.116.061102} {\bibfield
  {journal} {\bibinfo  {journal} {Phys. Rev. Lett.}\ }\textbf {\bibinfo
  {volume} {116}},\ \bibinfo {pages} {061102} (\bibinfo {year} {2016})},\
  \Eprint {http://arxiv.org/abs/1602.03837} {arXiv:1602.03837 [gr-qc]}
  \BibitemShut {NoStop}%
\bibitem [{\citenamefont {Abbott}\ \emph {et~al.}(2019)\citenamefont {Abbott}
  \emph {et~al.}}]{LIGOScientific:2018mvr}%
  \BibitemOpen
  \bibfield  {author} {\bibinfo {author} {\bibfnamefont {B.}~\bibnamefont
  {Abbott}} \emph {et~al.} (\bibinfo {collaboration} {LIGO Scientific,
  Virgo}),\ }\href {\doibase 10.1103/PhysRevX.9.031040} {\bibfield  {journal}
  {\bibinfo  {journal} {Phys. Rev. X}\ }\textbf {\bibinfo {volume} {9}},\
  \bibinfo {pages} {031040} (\bibinfo {year} {2019})},\ \Eprint
  {http://arxiv.org/abs/1811.12907} {arXiv:1811.12907 [astro-ph.HE]}
  \BibitemShut {NoStop}%
\bibitem [{\citenamefont {Abbott}\ \emph {et~al.}(2020)\citenamefont {Abbott}
  \emph {et~al.}}]{Abbott:2020niy}%
  \BibitemOpen
  \bibfield  {author} {\bibinfo {author} {\bibfnamefont {R.}~\bibnamefont
  {Abbott}} \emph {et~al.} (\bibinfo {collaboration} {LIGO Scientific,
  Virgo}),\ }\href@noop {} {\  (\bibinfo {year} {2020})},\ \Eprint
  {http://arxiv.org/abs/2010.14527} {arXiv:2010.14527 [gr-qc]} \BibitemShut
  {NoStop}%
\bibitem [{\citenamefont {{Genzel}}\ \emph {et~al.}(1997)\citenamefont
  {{Genzel}}, \citenamefont {{Eckart}}, \citenamefont {{Ott}},\ and\
  \citenamefont {{Eisenhauer}}}]{1997MNRAS.291..219G}%
  \BibitemOpen
  \bibfield  {author} {\bibinfo {author} {\bibfnamefont {R.}~\bibnamefont
  {{Genzel}}}, \bibinfo {author} {\bibfnamefont {A.}~\bibnamefont {{Eckart}}},
  \bibinfo {author} {\bibfnamefont {T.}~\bibnamefont {{Ott}}}, \ and\ \bibinfo
  {author} {\bibfnamefont {F.}~\bibnamefont {{Eisenhauer}}},\ }\href {\doibase
  10.1093/mnras/291.1.219} {\bibfield  {journal} {\bibinfo  {journal} {MNRAS}\
  }\textbf {\bibinfo {volume} {291}},\ \bibinfo {pages} {219} (\bibinfo {year}
  {1997})}\BibitemShut {NoStop}%
\bibitem [{\citenamefont {Ghez}\ \emph {et~al.}(1998)\citenamefont {Ghez},
  \citenamefont {Klein}, \citenamefont {Morris},\ and\ \citenamefont
  {Becklin}}]{Ghez:1998ph}%
  \BibitemOpen
  \bibfield  {author} {\bibinfo {author} {\bibfnamefont {A.}~\bibnamefont
  {Ghez}}, \bibinfo {author} {\bibfnamefont {B.}~\bibnamefont {Klein}},
  \bibinfo {author} {\bibfnamefont {M.}~\bibnamefont {Morris}}, \ and\ \bibinfo
  {author} {\bibfnamefont {E.}~\bibnamefont {Becklin}},\ }\href {\doibase
  10.1086/306528} {\bibfield  {journal} {\bibinfo  {journal} {Astrophys. J.}\
  }\textbf {\bibinfo {volume} {509}},\ \bibinfo {pages} {678} (\bibinfo {year}
  {1998})},\ \Eprint {http://arxiv.org/abs/astro-ph/9807210}
  {arXiv:astro-ph/9807210} \BibitemShut {NoStop}%
\bibitem [{\citenamefont {Akiyama}\ \emph {et~al.}(2019)\citenamefont {Akiyama}
  \emph {et~al.}}]{Akiyama:2019cqa}%
  \BibitemOpen
  \bibfield  {author} {\bibinfo {author} {\bibfnamefont {K.}~\bibnamefont
  {Akiyama}} \emph {et~al.} (\bibinfo {collaboration} {Event Horizon
  Telescope}),\ }\href {\doibase 10.3847/2041-8213/ab0ec7} {\bibfield
  {journal} {\bibinfo  {journal} {Astrophys. J.}\ }\textbf {\bibinfo {volume}
  {875}},\ \bibinfo {pages} {L1} (\bibinfo {year} {2019})},\ \Eprint
  {http://arxiv.org/abs/1906.11238} {arXiv:1906.11238 [astro-ph.GA]}
  \BibitemShut {NoStop}%
\bibitem [{\citenamefont {Cardoso}\ and\ \citenamefont
  {Pani}(2019)}]{Cardoso:2019rvt}%
  \BibitemOpen
  \bibfield  {author} {\bibinfo {author} {\bibfnamefont {V.}~\bibnamefont
  {Cardoso}}\ and\ \bibinfo {author} {\bibfnamefont {P.}~\bibnamefont {Pani}},\
  }\href {\doibase 10.1007/s41114-019-0020-4} {\bibfield  {journal} {\bibinfo
  {journal} {Living Rev. Rel.}\ }\textbf {\bibinfo {volume} {22}},\ \bibinfo
  {pages} {4} (\bibinfo {year} {2019})},\ \Eprint
  {http://arxiv.org/abs/1904.05363} {arXiv:1904.05363 [gr-qc]} \BibitemShut
  {NoStop}%
\bibitem [{\citenamefont {Buchdahl}(1959)}]{Buchdahl:1959zz}%
  \BibitemOpen
  \bibfield  {author} {\bibinfo {author} {\bibfnamefont {H.~A.}\ \bibnamefont
  {Buchdahl}},\ }\href {\doibase 10.1103/PhysRev.116.1027} {\bibfield
  {journal} {\bibinfo  {journal} {Phys. Rev.}\ }\textbf {\bibinfo {volume}
  {116}},\ \bibinfo {pages} {1027} (\bibinfo {year} {1959})}\BibitemShut
  {NoStop}%
\bibitem [{\citenamefont {Goldman}\ and\ \citenamefont
  {Nussinov}(1989)}]{Goldman:1989nd}%
  \BibitemOpen
  \bibfield  {author} {\bibinfo {author} {\bibfnamefont {I.}~\bibnamefont
  {Goldman}}\ and\ \bibinfo {author} {\bibfnamefont {S.}~\bibnamefont
  {Nussinov}},\ }\href {\doibase 10.1103/PhysRevD.40.3221} {\bibfield
  {journal} {\bibinfo  {journal} {Phys. Rev. D}\ }\textbf {\bibinfo {volume}
  {40}},\ \bibinfo {pages} {3221} (\bibinfo {year} {1989})}\BibitemShut
  {NoStop}%
\bibitem [{\citenamefont {Gould}\ \emph {et~al.}(1990)\citenamefont {Gould},
  \citenamefont {Draine}, \citenamefont {Romani},\ and\ \citenamefont
  {Nussinov}}]{Gould:1989gw}%
  \BibitemOpen
  \bibfield  {author} {\bibinfo {author} {\bibfnamefont {A.}~\bibnamefont
  {Gould}}, \bibinfo {author} {\bibfnamefont {B.~T.}\ \bibnamefont {Draine}},
  \bibinfo {author} {\bibfnamefont {R.~W.}\ \bibnamefont {Romani}}, \ and\
  \bibinfo {author} {\bibfnamefont {S.}~\bibnamefont {Nussinov}},\ }\href
  {\doibase 10.1016/0370-2693(90)91745-W} {\bibfield  {journal} {\bibinfo
  {journal} {Phys. Lett. B}\ }\textbf {\bibinfo {volume} {238}},\ \bibinfo
  {pages} {337} (\bibinfo {year} {1990})}\BibitemShut {NoStop}%
\bibitem [{\citenamefont {Henriques}\ \emph {et~al.}(1990)\citenamefont
  {Henriques}, \citenamefont {Liddle},\ and\ \citenamefont
  {Moorhouse}}]{Henriques:1989ez}%
  \BibitemOpen
  \bibfield  {author} {\bibinfo {author} {\bibfnamefont {A.}~\bibnamefont
  {Henriques}}, \bibinfo {author} {\bibfnamefont {A.~R.}\ \bibnamefont
  {Liddle}}, \ and\ \bibinfo {author} {\bibfnamefont {R.}~\bibnamefont
  {Moorhouse}},\ }\href {\doibase 10.1016/0550-3213(90)90514-E} {\bibfield
  {journal} {\bibinfo  {journal} {Nucl. Phys. B}\ }\textbf {\bibinfo {volume}
  {337}},\ \bibinfo {pages} {737} (\bibinfo {year} {1990})}\BibitemShut
  {NoStop}%
\bibitem [{\citenamefont {Henriques}\ \emph {et~al.}(1989)\citenamefont
  {Henriques}, \citenamefont {Liddle},\ and\ \citenamefont
  {Moorhouse}}]{Henriques:1989ar}%
  \BibitemOpen
  \bibfield  {author} {\bibinfo {author} {\bibfnamefont {A.}~\bibnamefont
  {Henriques}}, \bibinfo {author} {\bibfnamefont {A.~R.}\ \bibnamefont
  {Liddle}}, \ and\ \bibinfo {author} {\bibfnamefont {R.}~\bibnamefont
  {Moorhouse}},\ }\href {\doibase 10.1016/0370-2693(89)90623-0} {\bibfield
  {journal} {\bibinfo  {journal} {Phys. Lett. B}\ }\textbf {\bibinfo {volume}
  {233}},\ \bibinfo {pages} {99} (\bibinfo {year} {1989})}\BibitemShut
  {NoStop}%
\bibitem [{\citenamefont {Leung}\ \emph {et~al.}(2011)\citenamefont {Leung},
  \citenamefont {Chu},\ and\ \citenamefont {Lin}}]{Leung:2011zz}%
  \BibitemOpen
  \bibfield  {author} {\bibinfo {author} {\bibfnamefont {S.}~\bibnamefont
  {Leung}}, \bibinfo {author} {\bibfnamefont {M.}~\bibnamefont {Chu}}, \ and\
  \bibinfo {author} {\bibfnamefont {L.}~\bibnamefont {Lin}},\ }\href {\doibase
  10.1103/PhysRevD.84.107301} {\bibfield  {journal} {\bibinfo  {journal} {Phys.
  Rev. D}\ }\textbf {\bibinfo {volume} {84}},\ \bibinfo {pages} {107301}
  (\bibinfo {year} {2011})},\ \Eprint {http://arxiv.org/abs/1111.1787}
  {arXiv:1111.1787 [astro-ph.CO]} \BibitemShut {NoStop}%
\bibitem [{\citenamefont {Di~Giovanni}\ \emph
  {et~al.}(2020{\natexlab{a}})\citenamefont {Di~Giovanni}, \citenamefont
  {Fakhry}, \citenamefont {Sanchis-Gual}, \citenamefont {Degollado},\ and\
  \citenamefont {Font}}]{DiGiovanni:2020frc}%
  \BibitemOpen
  \bibfield  {author} {\bibinfo {author} {\bibfnamefont {F.}~\bibnamefont
  {Di~Giovanni}}, \bibinfo {author} {\bibfnamefont {S.}~\bibnamefont {Fakhry}},
  \bibinfo {author} {\bibfnamefont {N.}~\bibnamefont {Sanchis-Gual}}, \bibinfo
  {author} {\bibfnamefont {J.~C.}\ \bibnamefont {Degollado}}, \ and\ \bibinfo
  {author} {\bibfnamefont {J.~A.}\ \bibnamefont {Font}},\ }\href@noop {} {\
  (\bibinfo {year} {2020}{\natexlab{a}})},\ \Eprint
  {http://arxiv.org/abs/2006.08583} {arXiv:2006.08583 [gr-qc]} \BibitemShut
  {NoStop}%
\bibitem [{\citenamefont {Bowers}\ and\ \citenamefont
  {Liang}(1974)}]{Bowers:1974tgi}%
  \BibitemOpen
  \bibfield  {author} {\bibinfo {author} {\bibfnamefont {R.~L.}\ \bibnamefont
  {Bowers}}\ and\ \bibinfo {author} {\bibfnamefont {E.}~\bibnamefont {Liang}},\
  }\href {\doibase 10.1086/152760} {\bibfield  {journal} {\bibinfo  {journal}
  {Astrophys. J.}\ }\textbf {\bibinfo {volume} {188}},\ \bibinfo {pages} {657}
  (\bibinfo {year} {1974})}\BibitemShut {NoStop}%
\bibitem [{\citenamefont {Letelier}(1980)}]{Letelier:1980mxb}%
  \BibitemOpen
  \bibfield  {author} {\bibinfo {author} {\bibfnamefont {P.~S.}\ \bibnamefont
  {Letelier}},\ }\href {\doibase 10.1103/PhysRevD.22.807} {\bibfield  {journal}
  {\bibinfo  {journal} {Phys. Rev. D}\ }\textbf {\bibinfo {volume} {22}},\
  \bibinfo {pages} {807} (\bibinfo {year} {1980})}\BibitemShut {NoStop}%
\bibitem [{\citenamefont {Herrera}\ \emph {et~al.}(2004)\citenamefont
  {Herrera}, \citenamefont {Di~Prisco}, \citenamefont {Martin}, \citenamefont
  {Ospino}, \citenamefont {Santos},\ and\ \citenamefont
  {Troconis}}]{Herrera:2004xc}%
  \BibitemOpen
  \bibfield  {author} {\bibinfo {author} {\bibfnamefont {L.}~\bibnamefont
  {Herrera}}, \bibinfo {author} {\bibfnamefont {A.}~\bibnamefont {Di~Prisco}},
  \bibinfo {author} {\bibfnamefont {J.}~\bibnamefont {Martin}}, \bibinfo
  {author} {\bibfnamefont {J.}~\bibnamefont {Ospino}}, \bibinfo {author}
  {\bibfnamefont {N.}~\bibnamefont {Santos}}, \ and\ \bibinfo {author}
  {\bibfnamefont {O.}~\bibnamefont {Troconis}},\ }\href {\doibase
  10.1103/PhysRevD.69.084026} {\bibfield  {journal} {\bibinfo  {journal} {Phys.
  Rev. D}\ }\textbf {\bibinfo {volume} {69}},\ \bibinfo {pages} {084026}
  (\bibinfo {year} {2004})},\ \Eprint {http://arxiv.org/abs/gr-qc/0403006}
  {arXiv:gr-qc/0403006} \BibitemShut {NoStop}%
\bibitem [{\citenamefont {David}\ \emph {et~al.}(2002)\citenamefont {David},
  \citenamefont {Mandal},\ and\ \citenamefont {Wadia}}]{David:2002wn}%
  \BibitemOpen
  \bibfield  {author} {\bibinfo {author} {\bibfnamefont {J.~R.}\ \bibnamefont
  {David}}, \bibinfo {author} {\bibfnamefont {G.}~\bibnamefont {Mandal}}, \
  and\ \bibinfo {author} {\bibfnamefont {S.~R.}\ \bibnamefont {Wadia}},\ }\href
  {\doibase 10.1016/S0370-1573(02)00271-5} {\bibfield  {journal} {\bibinfo
  {journal} {Phys. Rept.}\ }\textbf {\bibinfo {volume} {369}},\ \bibinfo
  {pages} {549} (\bibinfo {year} {2002})},\ \Eprint
  {http://arxiv.org/abs/hep-th/0203048} {arXiv:hep-th/0203048} \BibitemShut
  {NoStop}%
\bibitem [{\citenamefont {Bena}\ and\ \citenamefont
  {Warner}(2008)}]{Bena:2007kg}%
  \BibitemOpen
  \bibfield  {author} {\bibinfo {author} {\bibfnamefont {I.}~\bibnamefont
  {Bena}}\ and\ \bibinfo {author} {\bibfnamefont {N.~P.}\ \bibnamefont
  {Warner}},\ }\href {\doibase 10.1007/978-3-540-79523-0_1} {\bibfield
  {journal} {\bibinfo  {journal} {Lect. Notes Phys.}\ }\textbf {\bibinfo
  {volume} {755}},\ \bibinfo {pages} {1} (\bibinfo {year} {2008})},\ \Eprint
  {http://arxiv.org/abs/hep-th/0701216} {arXiv:hep-th/0701216} \BibitemShut
  {NoStop}%
\bibitem [{\citenamefont {Myers}(1997)}]{Myers:1997qi}%
  \BibitemOpen
  \bibfield  {author} {\bibinfo {author} {\bibfnamefont {R.~C.}\ \bibnamefont
  {Myers}},\ }\href {\doibase 10.1023/A:1018855611972} {\bibfield  {journal}
  {\bibinfo  {journal} {Gen. Rel. Grav.}\ }\textbf {\bibinfo {volume} {29}},\
  \bibinfo {pages} {1217} (\bibinfo {year} {1997})},\ \Eprint
  {http://arxiv.org/abs/gr-qc/9705065} {arXiv:gr-qc/9705065} \BibitemShut
  {NoStop}%
\bibitem [{\citenamefont {Balasubramanian}\ \emph {et~al.}(2008)\citenamefont
  {Balasubramanian}, \citenamefont {de~Boer}, \citenamefont {El-Showk},\ and\
  \citenamefont {Messamah}}]{Balasubramanian:2008da}%
  \BibitemOpen
  \bibfield  {author} {\bibinfo {author} {\bibfnamefont {V.}~\bibnamefont
  {Balasubramanian}}, \bibinfo {author} {\bibfnamefont {J.}~\bibnamefont
  {de~Boer}}, \bibinfo {author} {\bibfnamefont {S.}~\bibnamefont {El-Showk}}, \
  and\ \bibinfo {author} {\bibfnamefont {I.}~\bibnamefont {Messamah}},\ }\href
  {\doibase 10.1088/0264-9381/25/21/214004} {\bibfield  {journal} {\bibinfo
  {journal} {Class. Quant. Grav.}\ }\textbf {\bibinfo {volume} {25}},\ \bibinfo
  {pages} {214004} (\bibinfo {year} {2008})},\ \Eprint
  {http://arxiv.org/abs/0811.0263} {arXiv:0811.0263 [hep-th]} \BibitemShut
  {NoStop}%
\bibitem [{\citenamefont {Bena}\ and\ \citenamefont
  {Warner}(2013)}]{Bena:2013dka}%
  \BibitemOpen
  \bibfield  {author} {\bibinfo {author} {\bibfnamefont {I.}~\bibnamefont
  {Bena}}\ and\ \bibinfo {author} {\bibfnamefont {N.~P.}\ \bibnamefont
  {Warner}},\ }\href@noop {} {\  (\bibinfo {year} {2013})},\ \Eprint
  {http://arxiv.org/abs/1311.4538} {arXiv:1311.4538 [hep-th]} \BibitemShut
  {NoStop}%
\bibitem [{\citenamefont {Visser}\ \emph {et~al.}(2008)\citenamefont {Visser},
  \citenamefont {Barcelo}, \citenamefont {Liberati},\ and\ \citenamefont
  {Sonego}}]{Visser:2009pw}%
  \BibitemOpen
  \bibfield  {author} {\bibinfo {author} {\bibfnamefont {M.}~\bibnamefont
  {Visser}}, \bibinfo {author} {\bibfnamefont {C.}~\bibnamefont {Barcelo}},
  \bibinfo {author} {\bibfnamefont {S.}~\bibnamefont {Liberati}}, \ and\
  \bibinfo {author} {\bibfnamefont {S.}~\bibnamefont {Sonego}},\ }\href
  {\doibase 10.22323/1.075.0010} {\bibfield  {journal} {\bibinfo  {journal}
  {PoS}\ }\textbf {\bibinfo {volume} {BHGRS}},\ \bibinfo {pages} {010}
  (\bibinfo {year} {2008})},\ \Eprint {http://arxiv.org/abs/0902.0346}
  {arXiv:0902.0346 [gr-qc]} \BibitemShut {NoStop}%
\bibitem [{\citenamefont {Chen}\ \emph {et~al.}(2018)\citenamefont {Chen},
  \citenamefont {Unruh}, \citenamefont {Wu},\ and\ \citenamefont
  {Yeom}}]{Chen:2017pkl}%
  \BibitemOpen
  \bibfield  {author} {\bibinfo {author} {\bibfnamefont {P.}~\bibnamefont
  {Chen}}, \bibinfo {author} {\bibfnamefont {W.~G.}\ \bibnamefont {Unruh}},
  \bibinfo {author} {\bibfnamefont {C.-H.}\ \bibnamefont {Wu}}, \ and\ \bibinfo
  {author} {\bibfnamefont {D.-H.}\ \bibnamefont {Yeom}},\ }\href {\doibase
  10.1103/PhysRevD.97.064045} {\bibfield  {journal} {\bibinfo  {journal} {Phys.
  Rev. D}\ }\textbf {\bibinfo {volume} {97}},\ \bibinfo {pages} {064045}
  (\bibinfo {year} {2018})},\ \Eprint {http://arxiv.org/abs/1710.01533}
  {arXiv:1710.01533 [gr-qc]} \BibitemShut {NoStop}%
\bibitem [{\citenamefont {Berthiere}\ \emph {et~al.}(2018)\citenamefont
  {Berthiere}, \citenamefont {Sarkar},\ and\ \citenamefont
  {Solodukhin}}]{Berthiere:2017tms}%
  \BibitemOpen
  \bibfield  {author} {\bibinfo {author} {\bibfnamefont {C.}~\bibnamefont
  {Berthiere}}, \bibinfo {author} {\bibfnamefont {D.}~\bibnamefont {Sarkar}}, \
  and\ \bibinfo {author} {\bibfnamefont {S.~N.}\ \bibnamefont {Solodukhin}},\
  }\href {\doibase 10.1016/j.physletb.2018.09.027} {\bibfield  {journal}
  {\bibinfo  {journal} {Phys. Lett. B}\ }\textbf {\bibinfo {volume} {786}},\
  \bibinfo {pages} {21} (\bibinfo {year} {2018})},\ \Eprint
  {http://arxiv.org/abs/1712.09914} {arXiv:1712.09914 [hep-th]} \BibitemShut
  {NoStop}%
\bibitem [{\citenamefont {Baccetti}\ \emph {et~al.}(2018)\citenamefont
  {Baccetti}, \citenamefont {Mann},\ and\ \citenamefont
  {Terno}}]{Baccetti:2016lsb}%
  \BibitemOpen
  \bibfield  {author} {\bibinfo {author} {\bibfnamefont {V.}~\bibnamefont
  {Baccetti}}, \bibinfo {author} {\bibfnamefont {R.~B.}\ \bibnamefont {Mann}},
  \ and\ \bibinfo {author} {\bibfnamefont {D.~R.}\ \bibnamefont {Terno}},\
  }\href {\doibase 10.1088/1361-6382/aad70e} {\bibfield  {journal} {\bibinfo
  {journal} {Class. Quant. Grav.}\ }\textbf {\bibinfo {volume} {35}},\ \bibinfo
  {pages} {185005} (\bibinfo {year} {2018})},\ \Eprint
  {http://arxiv.org/abs/1610.07839} {arXiv:1610.07839 [gr-qc]} \BibitemShut
  {NoStop}%
\bibitem [{\citenamefont {Guerra}\ \emph {et~al.}(2019)\citenamefont {Guerra},
  \citenamefont {Macedo},\ and\ \citenamefont {Pani}}]{Guerra:2019srj}%
  \BibitemOpen
  \bibfield  {author} {\bibinfo {author} {\bibfnamefont {D.}~\bibnamefont
  {Guerra}}, \bibinfo {author} {\bibfnamefont {C.~F.}\ \bibnamefont {Macedo}},
  \ and\ \bibinfo {author} {\bibfnamefont {P.}~\bibnamefont {Pani}},\ }\href
  {\doibase 10.1088/1475-7516/2019/09/061} {\bibfield  {journal} {\bibinfo
  {journal} {JCAP}\ }\textbf {\bibinfo {volume} {09}},\ \bibinfo {pages} {061}
  (\bibinfo {year} {2019})},\ \Eprint {http://arxiv.org/abs/1909.05515}
  {arXiv:1909.05515 [gr-qc]} \BibitemShut {NoStop}%
\bibitem [{\citenamefont {Grilli~di Cortona}\ \emph {et~al.}(2016)\citenamefont
  {Grilli~di Cortona}, \citenamefont {Hardy}, \citenamefont {Pardo~Vega},\ and\
  \citenamefont {Villadoro}}]{diCortona:2015ldu}%
  \BibitemOpen
  \bibfield  {author} {\bibinfo {author} {\bibfnamefont {G.}~\bibnamefont
  {Grilli~di Cortona}}, \bibinfo {author} {\bibfnamefont {E.}~\bibnamefont
  {Hardy}}, \bibinfo {author} {\bibfnamefont {J.}~\bibnamefont {Pardo~Vega}}, \
  and\ \bibinfo {author} {\bibfnamefont {G.}~\bibnamefont {Villadoro}},\ }\href
  {\doibase 10.1007/JHEP01(2016)034} {\bibfield  {journal} {\bibinfo  {journal}
  {JHEP}\ }\textbf {\bibinfo {volume} {01}},\ \bibinfo {pages} {034} (\bibinfo
  {year} {2016})},\ \Eprint {http://arxiv.org/abs/1511.02867} {arXiv:1511.02867
  [hep-ph]} \BibitemShut {NoStop}%
\bibitem [{\citenamefont {Hu}\ \emph {et~al.}(2000)\citenamefont {Hu},
  \citenamefont {Barkana},\ and\ \citenamefont {Gruzinov}}]{Hu:2000ke}%
  \BibitemOpen
  \bibfield  {author} {\bibinfo {author} {\bibfnamefont {W.}~\bibnamefont
  {Hu}}, \bibinfo {author} {\bibfnamefont {R.}~\bibnamefont {Barkana}}, \ and\
  \bibinfo {author} {\bibfnamefont {A.}~\bibnamefont {Gruzinov}},\ }\href
  {\doibase 10.1103/PhysRevLett.85.1158} {\bibfield  {journal} {\bibinfo
  {journal} {Phys. Rev. Lett.}\ }\textbf {\bibinfo {volume} {85}},\ \bibinfo
  {pages} {1158} (\bibinfo {year} {2000})},\ \Eprint
  {http://arxiv.org/abs/astro-ph/0003365} {arXiv:astro-ph/0003365} \BibitemShut
  {NoStop}%
\bibitem [{\citenamefont {Hui}\ \emph {et~al.}(2017)\citenamefont {Hui},
  \citenamefont {Ostriker}, \citenamefont {Tremaine},\ and\ \citenamefont
  {Witten}}]{Hui:2016ltb}%
  \BibitemOpen
  \bibfield  {author} {\bibinfo {author} {\bibfnamefont {L.}~\bibnamefont
  {Hui}}, \bibinfo {author} {\bibfnamefont {J.~P.}\ \bibnamefont {Ostriker}},
  \bibinfo {author} {\bibfnamefont {S.}~\bibnamefont {Tremaine}}, \ and\
  \bibinfo {author} {\bibfnamefont {E.}~\bibnamefont {Witten}},\ }\href
  {\doibase 10.1103/PhysRevD.95.043541} {\bibfield  {journal} {\bibinfo
  {journal} {Phys. Rev. D}\ }\textbf {\bibinfo {volume} {95}},\ \bibinfo
  {pages} {043541} (\bibinfo {year} {2017})},\ \Eprint
  {http://arxiv.org/abs/1610.08297} {arXiv:1610.08297 [astro-ph.CO]}
  \BibitemShut {NoStop}%
\bibitem [{\citenamefont {Li}\ \emph {et~al.}(2014)\citenamefont {Li},
  \citenamefont {Rindler-Daller},\ and\ \citenamefont {Shapiro}}]{Li:2013nal}%
  \BibitemOpen
  \bibfield  {author} {\bibinfo {author} {\bibfnamefont {B.}~\bibnamefont
  {Li}}, \bibinfo {author} {\bibfnamefont {T.}~\bibnamefont {Rindler-Daller}},
  \ and\ \bibinfo {author} {\bibfnamefont {P.~R.}\ \bibnamefont {Shapiro}},\
  }\href {\doibase 10.1103/PhysRevD.89.083536} {\bibfield  {journal} {\bibinfo
  {journal} {Phys. Rev. D}\ }\textbf {\bibinfo {volume} {89}},\ \bibinfo
  {pages} {083536} (\bibinfo {year} {2014})},\ \Eprint
  {http://arxiv.org/abs/1310.6061} {arXiv:1310.6061 [astro-ph.CO]} \BibitemShut
  {NoStop}%
\bibitem [{\citenamefont {Robles}\ and\ \citenamefont
  {Matos}(2012)}]{Robles:2012uy}%
  \BibitemOpen
  \bibfield  {author} {\bibinfo {author} {\bibfnamefont {V.~H.}\ \bibnamefont
  {Robles}}\ and\ \bibinfo {author} {\bibfnamefont {T.}~\bibnamefont {Matos}},\
  }\href {\doibase 10.1111/j.1365-2966.2012.20603.x} {\bibfield  {journal}
  {\bibinfo  {journal} {Mon. Not. Roy. Astron. Soc.}\ }\textbf {\bibinfo
  {volume} {422}},\ \bibinfo {pages} {282} (\bibinfo {year} {2012})},\ \Eprint
  {http://arxiv.org/abs/1201.3032} {arXiv:1201.3032 [astro-ph.CO]} \BibitemShut
  {NoStop}%
\bibitem [{\citenamefont {Bar}\ \emph {et~al.}(2018)\citenamefont {Bar},
  \citenamefont {Blas}, \citenamefont {Blum},\ and\ \citenamefont
  {Sibiryakov}}]{Bar:2018acw}%
  \BibitemOpen
  \bibfield  {author} {\bibinfo {author} {\bibfnamefont {N.}~\bibnamefont
  {Bar}}, \bibinfo {author} {\bibfnamefont {D.}~\bibnamefont {Blas}}, \bibinfo
  {author} {\bibfnamefont {K.}~\bibnamefont {Blum}}, \ and\ \bibinfo {author}
  {\bibfnamefont {S.}~\bibnamefont {Sibiryakov}},\ }\href {\doibase
  10.1103/PhysRevD.98.083027} {\bibfield  {journal} {\bibinfo  {journal} {Phys.
  Rev. D}\ }\textbf {\bibinfo {volume} {98}},\ \bibinfo {pages} {083027}
  (\bibinfo {year} {2018})},\ \Eprint {http://arxiv.org/abs/1805.00122}
  {arXiv:1805.00122 [astro-ph.CO]} \BibitemShut {NoStop}%
\bibitem [{\citenamefont {Annulli}\ \emph {et~al.}(2020)\citenamefont
  {Annulli}, \citenamefont {Cardoso},\ and\ \citenamefont
  {Vicente}}]{Annulli:2020lyc}%
  \BibitemOpen
  \bibfield  {author} {\bibinfo {author} {\bibfnamefont {L.}~\bibnamefont
  {Annulli}}, \bibinfo {author} {\bibfnamefont {V.}~\bibnamefont {Cardoso}}, \
  and\ \bibinfo {author} {\bibfnamefont {R.}~\bibnamefont {Vicente}},\ }\href
  {\doibase 10.1103/PhysRevD.102.063022} {\bibfield  {journal} {\bibinfo
  {journal} {Phys. Rev. D}\ }\textbf {\bibinfo {volume} {102}},\ \bibinfo
  {pages} {063022} (\bibinfo {year} {2020})},\ \Eprint
  {http://arxiv.org/abs/2009.00012} {arXiv:2009.00012 [gr-qc]} \BibitemShut
  {NoStop}%
\bibitem [{\citenamefont {Kaup}(1968)}]{Kaup:1968zz}%
  \BibitemOpen
  \bibfield  {author} {\bibinfo {author} {\bibfnamefont {D.~J.}\ \bibnamefont
  {Kaup}},\ }\href {\doibase 10.1103/PhysRev.172.1331} {\bibfield  {journal}
  {\bibinfo  {journal} {Phys. Rev.}\ }\textbf {\bibinfo {volume} {172}},\
  \bibinfo {pages} {1331} (\bibinfo {year} {1968})}\BibitemShut {NoStop}%
\bibitem [{\citenamefont {Ruffini}\ and\ \citenamefont
  {Bonazzola}(1969)}]{Ruffini:1969qy}%
  \BibitemOpen
  \bibfield  {author} {\bibinfo {author} {\bibfnamefont {R.}~\bibnamefont
  {Ruffini}}\ and\ \bibinfo {author} {\bibfnamefont {S.}~\bibnamefont
  {Bonazzola}},\ }\href {\doibase 10.1103/PhysRev.187.1767} {\bibfield
  {journal} {\bibinfo  {journal} {Phys. Rev.}\ }\textbf {\bibinfo {volume}
  {187}},\ \bibinfo {pages} {1767} (\bibinfo {year} {1969})}\BibitemShut
  {NoStop}%
\bibitem [{\citenamefont {Seidel}\ and\ \citenamefont
  {Suen}(1994)}]{Seidel:1993zk}%
  \BibitemOpen
  \bibfield  {author} {\bibinfo {author} {\bibfnamefont {E.}~\bibnamefont
  {Seidel}}\ and\ \bibinfo {author} {\bibfnamefont {W.-M.}\ \bibnamefont
  {Suen}},\ }\href {\doibase 10.1103/PhysRevLett.72.2516} {\bibfield  {journal}
  {\bibinfo  {journal} {Phys. Rev. Lett.}\ }\textbf {\bibinfo {volume} {72}},\
  \bibinfo {pages} {2516} (\bibinfo {year} {1994})},\ \Eprint
  {http://arxiv.org/abs/gr-qc/9309015} {arXiv:gr-qc/9309015} \BibitemShut
  {NoStop}%
\bibitem [{\citenamefont {Narain}\ \emph {et~al.}(2006)\citenamefont {Narain},
  \citenamefont {Schaffner-Bielich},\ and\ \citenamefont
  {Mishustin}}]{Narain:2006kx}%
  \BibitemOpen
  \bibfield  {author} {\bibinfo {author} {\bibfnamefont {G.}~\bibnamefont
  {Narain}}, \bibinfo {author} {\bibfnamefont {J.}~\bibnamefont
  {Schaffner-Bielich}}, \ and\ \bibinfo {author} {\bibfnamefont {I.~N.}\
  \bibnamefont {Mishustin}},\ }\href {\doibase 10.1103/PhysRevD.74.063003}
  {\bibfield  {journal} {\bibinfo  {journal} {Phys. Rev. D}\ }\textbf {\bibinfo
  {volume} {74}},\ \bibinfo {pages} {063003} (\bibinfo {year} {2006})},\
  \Eprint {http://arxiv.org/abs/astro-ph/0605724} {arXiv:astro-ph/0605724}
  \BibitemShut {NoStop}%
\bibitem [{\citenamefont {Raidal}\ \emph {et~al.}(2018)\citenamefont {Raidal},
  \citenamefont {Solodukhin}, \citenamefont {Vaskonen},\ and\ \citenamefont
  {Veermäe}}]{Raidal:2018eoo}%
  \BibitemOpen
  \bibfield  {author} {\bibinfo {author} {\bibfnamefont {M.}~\bibnamefont
  {Raidal}}, \bibinfo {author} {\bibfnamefont {S.}~\bibnamefont {Solodukhin}},
  \bibinfo {author} {\bibfnamefont {V.}~\bibnamefont {Vaskonen}}, \ and\
  \bibinfo {author} {\bibfnamefont {H.}~\bibnamefont {Veermäe}},\ }\href
  {\doibase 10.1103/PhysRevD.97.123520} {\bibfield  {journal} {\bibinfo
  {journal} {Phys. Rev. D}\ }\textbf {\bibinfo {volume} {97}},\ \bibinfo
  {pages} {123520} (\bibinfo {year} {2018})},\ \Eprint
  {http://arxiv.org/abs/1802.07728} {arXiv:1802.07728 [astro-ph.CO]}
  \BibitemShut {NoStop}%
\bibitem [{\citenamefont {Deliyergiyev}\ \emph {et~al.}(2019)\citenamefont
  {Deliyergiyev}, \citenamefont {Del~Popolo}, \citenamefont {Tolos},
  \citenamefont {Le~Delliou}, \citenamefont {Lee},\ and\ \citenamefont
  {Burgio}}]{Deliyergiyev:2019vti}%
  \BibitemOpen
  \bibfield  {author} {\bibinfo {author} {\bibfnamefont {M.}~\bibnamefont
  {Deliyergiyev}}, \bibinfo {author} {\bibfnamefont {A.}~\bibnamefont
  {Del~Popolo}}, \bibinfo {author} {\bibfnamefont {L.}~\bibnamefont {Tolos}},
  \bibinfo {author} {\bibfnamefont {M.}~\bibnamefont {Le~Delliou}}, \bibinfo
  {author} {\bibfnamefont {X.}~\bibnamefont {Lee}}, \ and\ \bibinfo {author}
  {\bibfnamefont {F.}~\bibnamefont {Burgio}},\ }\href {\doibase
  10.1103/PhysRevD.99.063015} {\bibfield  {journal} {\bibinfo  {journal} {Phys.
  Rev. D}\ }\textbf {\bibinfo {volume} {99}},\ \bibinfo {pages} {063015}
  (\bibinfo {year} {2019})},\ \Eprint {http://arxiv.org/abs/1903.01183}
  {arXiv:1903.01183 [gr-qc]} \BibitemShut {NoStop}%
\bibitem [{\citenamefont {Liebling}\ and\ \citenamefont
  {Palenzuela}(2017)}]{Liebling:2012fv}%
  \BibitemOpen
  \bibfield  {author} {\bibinfo {author} {\bibfnamefont {S.~L.}\ \bibnamefont
  {Liebling}}\ and\ \bibinfo {author} {\bibfnamefont {C.}~\bibnamefont
  {Palenzuela}},\ }\href {\doibase 10.12942/lrr-2012-6} {\bibfield  {journal}
  {\bibinfo  {journal} {Living Rev. Rel.}\ }\textbf {\bibinfo {volume} {20}},\
  \bibinfo {pages} {5} (\bibinfo {year} {2017})},\ \Eprint
  {http://arxiv.org/abs/1202.5809} {arXiv:1202.5809 [gr-qc]} \BibitemShut
  {NoStop}%
\bibitem [{\citenamefont {Giudice}\ \emph {et~al.}(2016)\citenamefont
  {Giudice}, \citenamefont {McCullough},\ and\ \citenamefont
  {Urbano}}]{Giudice:2016zpa}%
  \BibitemOpen
  \bibfield  {author} {\bibinfo {author} {\bibfnamefont {G.~F.}\ \bibnamefont
  {Giudice}}, \bibinfo {author} {\bibfnamefont {M.}~\bibnamefont {McCullough}},
  \ and\ \bibinfo {author} {\bibfnamefont {A.}~\bibnamefont {Urbano}},\ }\href
  {\doibase 10.1088/1475-7516/2016/10/001} {\bibfield  {journal} {\bibinfo
  {journal} {JCAP}\ }\textbf {\bibinfo {volume} {10}},\ \bibinfo {pages} {001}
  (\bibinfo {year} {2016})},\ \Eprint {http://arxiv.org/abs/1605.01209}
  {arXiv:1605.01209 [hep-ph]} \BibitemShut {NoStop}%
\bibitem [{\citenamefont {Sanchis-Gual}\ \emph {et~al.}(2019)\citenamefont
  {Sanchis-Gual}, \citenamefont {Di~Giovanni}, \citenamefont {Zilhão},
  \citenamefont {Herdeiro}, \citenamefont {Cerdá-Durán}, \citenamefont
  {Font},\ and\ \citenamefont {Radu}}]{Sanchis-Gual:2019ljs}%
  \BibitemOpen
  \bibfield  {author} {\bibinfo {author} {\bibfnamefont {N.}~\bibnamefont
  {Sanchis-Gual}}, \bibinfo {author} {\bibfnamefont {F.}~\bibnamefont
  {Di~Giovanni}}, \bibinfo {author} {\bibfnamefont {M.}~\bibnamefont
  {Zilhão}}, \bibinfo {author} {\bibfnamefont {C.}~\bibnamefont {Herdeiro}},
  \bibinfo {author} {\bibfnamefont {P.}~\bibnamefont {Cerdá-Durán}}, \bibinfo
  {author} {\bibfnamefont {J.}~\bibnamefont {Font}}, \ and\ \bibinfo {author}
  {\bibfnamefont {E.}~\bibnamefont {Radu}},\ }\href {\doibase
  10.1103/PhysRevLett.123.221101} {\bibfield  {journal} {\bibinfo  {journal}
  {Phys. Rev. Lett.}\ }\textbf {\bibinfo {volume} {123}},\ \bibinfo {pages}
  {221101} (\bibinfo {year} {2019})},\ \Eprint
  {http://arxiv.org/abs/1907.12565} {arXiv:1907.12565 [gr-qc]} \BibitemShut
  {NoStop}%
\bibitem [{\citenamefont {Kesden}\ \emph {et~al.}(2005)\citenamefont {Kesden},
  \citenamefont {Gair},\ and\ \citenamefont {Kamionkowski}}]{Kesden:2004qx}%
  \BibitemOpen
  \bibfield  {author} {\bibinfo {author} {\bibfnamefont {M.}~\bibnamefont
  {Kesden}}, \bibinfo {author} {\bibfnamefont {J.}~\bibnamefont {Gair}}, \ and\
  \bibinfo {author} {\bibfnamefont {M.}~\bibnamefont {Kamionkowski}},\ }\href
  {\doibase 10.1103/PhysRevD.71.044015} {\bibfield  {journal} {\bibinfo
  {journal} {Phys. Rev. D}\ }\textbf {\bibinfo {volume} {71}},\ \bibinfo
  {pages} {044015} (\bibinfo {year} {2005})},\ \Eprint
  {http://arxiv.org/abs/astro-ph/0411478} {arXiv:astro-ph/0411478} \BibitemShut
  {NoStop}%
\bibitem [{\citenamefont {Berti}\ \emph {et~al.}(2015)\citenamefont {Berti}
  \emph {et~al.}}]{Berti:2015itd}%
  \BibitemOpen
  \bibfield  {author} {\bibinfo {author} {\bibfnamefont {E.}~\bibnamefont
  {Berti}} \emph {et~al.},\ }\href {\doibase 10.1088/0264-9381/32/24/243001}
  {\bibfield  {journal} {\bibinfo  {journal} {Class. Quant. Grav.}\ }\textbf
  {\bibinfo {volume} {32}},\ \bibinfo {pages} {243001} (\bibinfo {year}
  {2015})},\ \Eprint {http://arxiv.org/abs/1501.07274} {arXiv:1501.07274
  [gr-qc]} \BibitemShut {NoStop}%
\bibitem [{\citenamefont {Bambi}(2017)}]{Bambi:2015kza}%
  \BibitemOpen
  \bibfield  {author} {\bibinfo {author} {\bibfnamefont {C.}~\bibnamefont
  {Bambi}},\ }\href {\doibase 10.1103/RevModPhys.89.025001} {\bibfield
  {journal} {\bibinfo  {journal} {Rev. Mod. Phys.}\ }\textbf {\bibinfo {volume}
  {89}},\ \bibinfo {pages} {025001} (\bibinfo {year} {2017})},\ \Eprint
  {http://arxiv.org/abs/1509.03884} {arXiv:1509.03884 [gr-qc]} \BibitemShut
  {NoStop}%
\bibitem [{\citenamefont {Sennett}\ \emph {et~al.}(2017)\citenamefont
  {Sennett}, \citenamefont {Hinderer}, \citenamefont {Steinhoff}, \citenamefont
  {Buonanno},\ and\ \citenamefont {Ossokine}}]{Sennett:2017etc}%
  \BibitemOpen
  \bibfield  {author} {\bibinfo {author} {\bibfnamefont {N.}~\bibnamefont
  {Sennett}}, \bibinfo {author} {\bibfnamefont {T.}~\bibnamefont {Hinderer}},
  \bibinfo {author} {\bibfnamefont {J.}~\bibnamefont {Steinhoff}}, \bibinfo
  {author} {\bibfnamefont {A.}~\bibnamefont {Buonanno}}, \ and\ \bibinfo
  {author} {\bibfnamefont {S.}~\bibnamefont {Ossokine}},\ }\href {\doibase
  10.1103/PhysRevD.96.024002} {\bibfield  {journal} {\bibinfo  {journal} {Phys.
  Rev. D}\ }\textbf {\bibinfo {volume} {96}},\ \bibinfo {pages} {024002}
  (\bibinfo {year} {2017})},\ \Eprint {http://arxiv.org/abs/1704.08651}
  {arXiv:1704.08651 [gr-qc]} \BibitemShut {NoStop}%
\bibitem [{\citenamefont {Macedo}\ \emph {et~al.}(2013)\citenamefont {Macedo},
  \citenamefont {Pani}, \citenamefont {Cardoso},\ and\ \citenamefont
  {Crispino}}]{Macedo:2013jja}%
  \BibitemOpen
  \bibfield  {author} {\bibinfo {author} {\bibfnamefont {C.~F.}\ \bibnamefont
  {Macedo}}, \bibinfo {author} {\bibfnamefont {P.}~\bibnamefont {Pani}},
  \bibinfo {author} {\bibfnamefont {V.}~\bibnamefont {Cardoso}}, \ and\
  \bibinfo {author} {\bibfnamefont {L.~C.~B.}\ \bibnamefont {Crispino}},\
  }\href {\doibase 10.1103/PhysRevD.88.064046} {\bibfield  {journal} {\bibinfo
  {journal} {Phys. Rev. D}\ }\textbf {\bibinfo {volume} {88}},\ \bibinfo
  {pages} {064046} (\bibinfo {year} {2013})},\ \Eprint
  {http://arxiv.org/abs/1307.4812} {arXiv:1307.4812 [gr-qc]} \BibitemShut
  {NoStop}%
\bibitem [{\citenamefont {Yoshida}\ \emph {et~al.}(1994)\citenamefont
  {Yoshida}, \citenamefont {Eriguchi},\ and\ \citenamefont
  {Futamase}}]{Yoshida:1994xi}%
  \BibitemOpen
  \bibfield  {author} {\bibinfo {author} {\bibfnamefont {S.}~\bibnamefont
  {Yoshida}}, \bibinfo {author} {\bibfnamefont {Y.}~\bibnamefont {Eriguchi}}, \
  and\ \bibinfo {author} {\bibfnamefont {T.}~\bibnamefont {Futamase}},\ }\href
  {\doibase 10.1103/PhysRevD.50.6235} {\bibfield  {journal} {\bibinfo
  {journal} {Phys. Rev. D}\ }\textbf {\bibinfo {volume} {50}},\ \bibinfo
  {pages} {6235} (\bibinfo {year} {1994})}\BibitemShut {NoStop}%
\bibitem [{\citenamefont {Guzman}\ and\ \citenamefont
  {Rueda-Becerril}(2009)}]{Guzman:2009zz}%
  \BibitemOpen
  \bibfield  {author} {\bibinfo {author} {\bibfnamefont {F.}~\bibnamefont
  {Guzman}}\ and\ \bibinfo {author} {\bibfnamefont {J.}~\bibnamefont
  {Rueda-Becerril}},\ }\href {\doibase 10.1103/PhysRevD.80.084023} {\bibfield
  {journal} {\bibinfo  {journal} {Phys. Rev. D}\ }\textbf {\bibinfo {volume}
  {80}},\ \bibinfo {pages} {084023} (\bibinfo {year} {2009})},\ \Eprint
  {http://arxiv.org/abs/1009.1250} {arXiv:1009.1250 [astro-ph.HE]} \BibitemShut
  {NoStop}%
\bibitem [{\citenamefont {Palenzuela}\ \emph {et~al.}(2008)\citenamefont
  {Palenzuela}, \citenamefont {Lehner},\ and\ \citenamefont
  {Liebling}}]{Palenzuela:2007dm}%
  \BibitemOpen
  \bibfield  {author} {\bibinfo {author} {\bibfnamefont {C.}~\bibnamefont
  {Palenzuela}}, \bibinfo {author} {\bibfnamefont {L.}~\bibnamefont {Lehner}},
  \ and\ \bibinfo {author} {\bibfnamefont {S.~L.}\ \bibnamefont {Liebling}},\
  }\href {\doibase 10.1103/PhysRevD.77.044036} {\bibfield  {journal} {\bibinfo
  {journal} {Phys. Rev. D}\ }\textbf {\bibinfo {volume} {77}},\ \bibinfo
  {pages} {044036} (\bibinfo {year} {2008})},\ \Eprint
  {http://arxiv.org/abs/0706.2435} {arXiv:0706.2435 [gr-qc]} \BibitemShut
  {NoStop}%
\bibitem [{\citenamefont {Palenzuela}\ \emph {et~al.}(2007)\citenamefont
  {Palenzuela}, \citenamefont {Olabarrieta}, \citenamefont {Lehner},\ and\
  \citenamefont {Liebling}}]{Palenzuela:2006wp}%
  \BibitemOpen
  \bibfield  {author} {\bibinfo {author} {\bibfnamefont {C.}~\bibnamefont
  {Palenzuela}}, \bibinfo {author} {\bibfnamefont {I.}~\bibnamefont
  {Olabarrieta}}, \bibinfo {author} {\bibfnamefont {L.}~\bibnamefont {Lehner}},
  \ and\ \bibinfo {author} {\bibfnamefont {S.~L.}\ \bibnamefont {Liebling}},\
  }\href {\doibase 10.1103/PhysRevD.75.064005} {\bibfield  {journal} {\bibinfo
  {journal} {Phys. Rev. D}\ }\textbf {\bibinfo {volume} {75}},\ \bibinfo
  {pages} {064005} (\bibinfo {year} {2007})},\ \Eprint
  {http://arxiv.org/abs/gr-qc/0612067} {arXiv:gr-qc/0612067} \BibitemShut
  {NoStop}%
\bibitem [{\citenamefont {Choptuik}\ and\ \citenamefont
  {Pretorius}(2010)}]{Choptuik:2009ww}%
  \BibitemOpen
  \bibfield  {author} {\bibinfo {author} {\bibfnamefont {M.~W.}\ \bibnamefont
  {Choptuik}}\ and\ \bibinfo {author} {\bibfnamefont {F.}~\bibnamefont
  {Pretorius}},\ }\href {\doibase 10.1103/PhysRevLett.104.111101} {\bibfield
  {journal} {\bibinfo  {journal} {Phys. Rev. Lett.}\ }\textbf {\bibinfo
  {volume} {104}},\ \bibinfo {pages} {111101} (\bibinfo {year} {2010})},\
  \Eprint {http://arxiv.org/abs/0908.1780} {arXiv:0908.1780 [gr-qc]}
  \BibitemShut {NoStop}%
\bibitem [{\citenamefont {Palenzuela}\ \emph {et~al.}(2017)\citenamefont
  {Palenzuela}, \citenamefont {Pani}, \citenamefont {Bezares}, \citenamefont
  {Cardoso}, \citenamefont {Lehner},\ and\ \citenamefont
  {Liebling}}]{Palenzuela:2017kcg}%
  \BibitemOpen
  \bibfield  {author} {\bibinfo {author} {\bibfnamefont {C.}~\bibnamefont
  {Palenzuela}}, \bibinfo {author} {\bibfnamefont {P.}~\bibnamefont {Pani}},
  \bibinfo {author} {\bibfnamefont {M.}~\bibnamefont {Bezares}}, \bibinfo
  {author} {\bibfnamefont {V.}~\bibnamefont {Cardoso}}, \bibinfo {author}
  {\bibfnamefont {L.}~\bibnamefont {Lehner}}, \ and\ \bibinfo {author}
  {\bibfnamefont {S.}~\bibnamefont {Liebling}},\ }\href {\doibase
  10.1103/PhysRevD.96.104058} {\bibfield  {journal} {\bibinfo  {journal} {Phys.
  Rev. D}\ }\textbf {\bibinfo {volume} {96}},\ \bibinfo {pages} {104058}
  (\bibinfo {year} {2017})},\ \Eprint {http://arxiv.org/abs/1710.09432}
  {arXiv:1710.09432 [gr-qc]} \BibitemShut {NoStop}%
\bibitem [{\citenamefont {Helfer}\ \emph {et~al.}(2017)\citenamefont {Helfer},
  \citenamefont {Marsh}, \citenamefont {Clough}, \citenamefont {Fairbairn},
  \citenamefont {Lim},\ and\ \citenamefont {Becerril}}]{Helfer:2016ljl}%
  \BibitemOpen
  \bibfield  {author} {\bibinfo {author} {\bibfnamefont {T.}~\bibnamefont
  {Helfer}}, \bibinfo {author} {\bibfnamefont {D.~J.~E.}\ \bibnamefont
  {Marsh}}, \bibinfo {author} {\bibfnamefont {K.}~\bibnamefont {Clough}},
  \bibinfo {author} {\bibfnamefont {M.}~\bibnamefont {Fairbairn}}, \bibinfo
  {author} {\bibfnamefont {E.~A.}\ \bibnamefont {Lim}}, \ and\ \bibinfo
  {author} {\bibfnamefont {R.}~\bibnamefont {Becerril}},\ }\href {\doibase
  10.1088/1475-7516/2017/03/055} {\bibfield  {journal} {\bibinfo  {journal}
  {JCAP}\ }\textbf {\bibinfo {volume} {03}},\ \bibinfo {pages} {055} (\bibinfo
  {year} {2017})},\ \Eprint {http://arxiv.org/abs/1609.04724} {arXiv:1609.04724
  [astro-ph.CO]} \BibitemShut {NoStop}%
\bibitem [{\citenamefont {Dietrich}\ \emph {et~al.}(2019)\citenamefont
  {Dietrich}, \citenamefont {Ossokine},\ and\ \citenamefont
  {Clough}}]{Dietrich:2018bvi}%
  \BibitemOpen
  \bibfield  {author} {\bibinfo {author} {\bibfnamefont {T.}~\bibnamefont
  {Dietrich}}, \bibinfo {author} {\bibfnamefont {S.}~\bibnamefont {Ossokine}},
  \ and\ \bibinfo {author} {\bibfnamefont {K.}~\bibnamefont {Clough}},\ }\href
  {\doibase 10.1088/1361-6382/aaf43e} {\bibfield  {journal} {\bibinfo
  {journal} {Class. Quant. Grav.}\ }\textbf {\bibinfo {volume} {36}},\ \bibinfo
  {pages} {025002} (\bibinfo {year} {2019})},\ \Eprint
  {http://arxiv.org/abs/1807.06959} {arXiv:1807.06959 [gr-qc]} \BibitemShut
  {NoStop}%
\bibitem [{\citenamefont {Olivares}\ \emph {et~al.}(2020)\citenamefont
  {Olivares}, \citenamefont {Younsi}, \citenamefont {Fromm}, \citenamefont
  {De~Laurentis}, \citenamefont {Porth}, \citenamefont {Mizuno}, \citenamefont
  {Falcke}, \citenamefont {Kramer},\ and\ \citenamefont
  {Rezzolla}}]{Olivares:2018abq}%
  \BibitemOpen
  \bibfield  {author} {\bibinfo {author} {\bibfnamefont {H.}~\bibnamefont
  {Olivares}}, \bibinfo {author} {\bibfnamefont {Z.}~\bibnamefont {Younsi}},
  \bibinfo {author} {\bibfnamefont {C.~M.}\ \bibnamefont {Fromm}}, \bibinfo
  {author} {\bibfnamefont {M.}~\bibnamefont {De~Laurentis}}, \bibinfo {author}
  {\bibfnamefont {O.}~\bibnamefont {Porth}}, \bibinfo {author} {\bibfnamefont
  {Y.}~\bibnamefont {Mizuno}}, \bibinfo {author} {\bibfnamefont
  {H.}~\bibnamefont {Falcke}}, \bibinfo {author} {\bibfnamefont
  {M.}~\bibnamefont {Kramer}}, \ and\ \bibinfo {author} {\bibfnamefont
  {L.}~\bibnamefont {Rezzolla}},\ }\href {\doibase 10.1093/mnras/staa1878}
  {\bibfield  {journal} {\bibinfo  {journal} {Mon. Not. Roy. Astron. Soc.}\
  }\textbf {\bibinfo {volume} {497}},\ \bibinfo {pages} {521} (\bibinfo {year}
  {2020})},\ \Eprint {http://arxiv.org/abs/1809.08682} {arXiv:1809.08682
  [gr-qc]} \BibitemShut {NoStop}%
\bibitem [{\citenamefont {Sanchis-Gual}\ \emph {et~al.}(2020)\citenamefont
  {Sanchis-Gual}, \citenamefont {Zilh\~ao}, \citenamefont {Herdeiro},
  \citenamefont {Di~Giovanni}, \citenamefont {Font},\ and\ \citenamefont
  {Radu}}]{Sanchis-Gual:2020mzb}%
  \BibitemOpen
  \bibfield  {author} {\bibinfo {author} {\bibfnamefont {N.}~\bibnamefont
  {Sanchis-Gual}}, \bibinfo {author} {\bibfnamefont {M.}~\bibnamefont
  {Zilh\~ao}}, \bibinfo {author} {\bibfnamefont {C.}~\bibnamefont {Herdeiro}},
  \bibinfo {author} {\bibfnamefont {F.}~\bibnamefont {Di~Giovanni}}, \bibinfo
  {author} {\bibfnamefont {J.~A.}\ \bibnamefont {Font}}, \ and\ \bibinfo
  {author} {\bibfnamefont {E.}~\bibnamefont {Radu}},\ }\href@noop {} {\
  (\bibinfo {year} {2020})},\ \Eprint {http://arxiv.org/abs/2007.11584}
  {arXiv:2007.11584 [gr-qc]} \BibitemShut {NoStop}%
\bibitem [{\citenamefont {Bezares}\ \emph {et~al.}(2017)\citenamefont
  {Bezares}, \citenamefont {Palenzuela},\ and\ \citenamefont
  {Bona}}]{Bezares:2017mzk}%
  \BibitemOpen
  \bibfield  {author} {\bibinfo {author} {\bibfnamefont {M.}~\bibnamefont
  {Bezares}}, \bibinfo {author} {\bibfnamefont {C.}~\bibnamefont {Palenzuela}},
  \ and\ \bibinfo {author} {\bibfnamefont {C.}~\bibnamefont {Bona}},\ }\href
  {\doibase 10.1103/PhysRevD.95.124005} {\bibfield  {journal} {\bibinfo
  {journal} {Phys. Rev. D}\ }\textbf {\bibinfo {volume} {95}},\ \bibinfo
  {pages} {124005} (\bibinfo {year} {2017})},\ \Eprint
  {http://arxiv.org/abs/1705.01071} {arXiv:1705.01071 [gr-qc]} \BibitemShut
  {NoStop}%
\bibitem [{\citenamefont {Di~Giovanni}\ \emph
  {et~al.}(2020{\natexlab{b}})\citenamefont {Di~Giovanni}, \citenamefont
  {Sanchis-Gual}, \citenamefont {Cerd\'a-Dur\'an}, \citenamefont {Zilh\~ao},
  \citenamefont {Herdeiro}, \citenamefont {Font},\ and\ \citenamefont
  {Radu}}]{DiGiovanni:2020ror}%
  \BibitemOpen
  \bibfield  {author} {\bibinfo {author} {\bibfnamefont {F.}~\bibnamefont
  {Di~Giovanni}}, \bibinfo {author} {\bibfnamefont {N.}~\bibnamefont
  {Sanchis-Gual}}, \bibinfo {author} {\bibfnamefont {P.}~\bibnamefont
  {Cerd\'a-Dur\'an}}, \bibinfo {author} {\bibfnamefont {M.}~\bibnamefont
  {Zilh\~ao}}, \bibinfo {author} {\bibfnamefont {C.}~\bibnamefont {Herdeiro}},
  \bibinfo {author} {\bibfnamefont {J.}~\bibnamefont {Font}}, \ and\ \bibinfo
  {author} {\bibfnamefont {E.}~\bibnamefont {Radu}},\ }\href@noop {} {\
  (\bibinfo {year} {2020}{\natexlab{b}})},\ \Eprint
  {http://arxiv.org/abs/2010.05845} {arXiv:2010.05845 [gr-qc]} \BibitemShut
  {NoStop}%
\bibitem [{\citenamefont {Friedberg}\ \emph {et~al.}(1987)\citenamefont
  {Friedberg}, \citenamefont {Lee},\ and\ \citenamefont
  {Pang}}]{Friedberg:1986tq}%
  \BibitemOpen
  \bibfield  {author} {\bibinfo {author} {\bibfnamefont {R.}~\bibnamefont
  {Friedberg}}, \bibinfo {author} {\bibfnamefont {T.}~\bibnamefont {Lee}}, \
  and\ \bibinfo {author} {\bibfnamefont {Y.}~\bibnamefont {Pang}},\ }\href
  {\doibase 10.1103/PhysRevD.35.3658} {\bibfield  {journal} {\bibinfo
  {journal} {Phys. Rev. D}\ }\textbf {\bibinfo {volume} {35}},\ \bibinfo
  {pages} {3658} (\bibinfo {year} {1987})}\BibitemShut {NoStop}%
\bibitem [{\citenamefont {Kleihaus}\ \emph {et~al.}(2005)\citenamefont
  {Kleihaus}, \citenamefont {Kunz},\ and\ \citenamefont
  {List}}]{Kleihaus:2005me}%
  \BibitemOpen
  \bibfield  {author} {\bibinfo {author} {\bibfnamefont {B.}~\bibnamefont
  {Kleihaus}}, \bibinfo {author} {\bibfnamefont {J.}~\bibnamefont {Kunz}}, \
  and\ \bibinfo {author} {\bibfnamefont {M.}~\bibnamefont {List}},\ }\href
  {\doibase 10.1103/PhysRevD.72.064002} {\bibfield  {journal} {\bibinfo
  {journal} {Phys. Rev. D}\ }\textbf {\bibinfo {volume} {72}},\ \bibinfo
  {pages} {064002} (\bibinfo {year} {2005})},\ \Eprint
  {http://arxiv.org/abs/gr-qc/0505143} {arXiv:gr-qc/0505143} \BibitemShut
  {NoStop}%
\bibitem [{\citenamefont {Kleihaus}\ \emph {et~al.}(2008)\citenamefont
  {Kleihaus}, \citenamefont {Kunz}, \citenamefont {List},\ and\ \citenamefont
  {Schaffer}}]{Kleihaus:2007vk}%
  \BibitemOpen
  \bibfield  {author} {\bibinfo {author} {\bibfnamefont {B.}~\bibnamefont
  {Kleihaus}}, \bibinfo {author} {\bibfnamefont {J.}~\bibnamefont {Kunz}},
  \bibinfo {author} {\bibfnamefont {M.}~\bibnamefont {List}}, \ and\ \bibinfo
  {author} {\bibfnamefont {I.}~\bibnamefont {Schaffer}},\ }\href {\doibase
  10.1103/PhysRevD.77.064025} {\bibfield  {journal} {\bibinfo  {journal} {Phys.
  Rev. D}\ }\textbf {\bibinfo {volume} {77}},\ \bibinfo {pages} {064025}
  (\bibinfo {year} {2008})},\ \Eprint {http://arxiv.org/abs/0712.3742}
  {arXiv:0712.3742 [gr-qc]} \BibitemShut {NoStop}%
\bibitem [{\citenamefont {Kleihaus}\ \emph {et~al.}(2012)\citenamefont
  {Kleihaus}, \citenamefont {Kunz},\ and\ \citenamefont
  {Schneider}}]{Kleihaus:2011sx}%
  \BibitemOpen
  \bibfield  {author} {\bibinfo {author} {\bibfnamefont {B.}~\bibnamefont
  {Kleihaus}}, \bibinfo {author} {\bibfnamefont {J.}~\bibnamefont {Kunz}}, \
  and\ \bibinfo {author} {\bibfnamefont {S.}~\bibnamefont {Schneider}},\ }\href
  {\doibase 10.1103/PhysRevD.85.024045} {\bibfield  {journal} {\bibinfo
  {journal} {Phys. Rev. D}\ }\textbf {\bibinfo {volume} {85}},\ \bibinfo
  {pages} {024045} (\bibinfo {year} {2012})},\ \Eprint
  {http://arxiv.org/abs/1109.5858} {arXiv:1109.5858 [gr-qc]} \BibitemShut
  {NoStop}%
\bibitem [{\citenamefont {Coleman}(1985)}]{Coleman:1985ki}%
  \BibitemOpen
  \bibfield  {author} {\bibinfo {author} {\bibfnamefont {S.~R.}\ \bibnamefont
  {Coleman}},\ }\href {\doibase 10.1016/0550-3213(86)90520-1} {\bibfield
  {journal} {\bibinfo  {journal} {Nucl. Phys. B}\ }\textbf {\bibinfo {volume}
  {262}},\ \bibinfo {pages} {263} (\bibinfo {year} {1985})},\ \bibinfo {note}
  {[Erratum: Nucl.Phys.B 269, 744 (1986)]}\BibitemShut {NoStop}%
\bibitem [{\citenamefont {Schunck}\ and\ \citenamefont
  {Torres}(2000)}]{Schunck:1999zu}%
  \BibitemOpen
  \bibfield  {author} {\bibinfo {author} {\bibfnamefont {F.~E.}\ \bibnamefont
  {Schunck}}\ and\ \bibinfo {author} {\bibfnamefont {D.~F.}\ \bibnamefont
  {Torres}},\ }\href {\doibase 10.1142/S0218271800000608} {\bibfield  {journal}
  {\bibinfo  {journal} {Int. J. Mod. Phys. D}\ }\textbf {\bibinfo {volume}
  {9}},\ \bibinfo {pages} {601} (\bibinfo {year} {2000})},\ \Eprint
  {http://arxiv.org/abs/gr-qc/9911038} {arXiv:gr-qc/9911038} \BibitemShut
  {NoStop}%
\bibitem [{\citenamefont {Choi}\ \emph {et~al.}(2019)\citenamefont {Choi},
  \citenamefont {He},\ and\ \citenamefont {Schiappacasse}}]{Choi:2019mva}%
  \BibitemOpen
  \bibfield  {author} {\bibinfo {author} {\bibfnamefont {G.}~\bibnamefont
  {Choi}}, \bibinfo {author} {\bibfnamefont {H.-J.}\ \bibnamefont {He}}, \ and\
  \bibinfo {author} {\bibfnamefont {E.~D.}\ \bibnamefont {Schiappacasse}},\
  }\href {\doibase 10.1088/1475-7516/2019/10/043} {\bibfield  {journal}
  {\bibinfo  {journal} {JCAP}\ }\textbf {\bibinfo {volume} {10}},\ \bibinfo
  {pages} {043} (\bibinfo {year} {2019})},\ \Eprint
  {http://arxiv.org/abs/1906.02094} {arXiv:1906.02094 [astro-ph.CO]}
  \BibitemShut {NoStop}%
\bibitem [{\citenamefont {Hertzberg}\ and\ \citenamefont
  {Schiappacasse}(2018)}]{Hertzberg:2018lmt}%
  \BibitemOpen
  \bibfield  {author} {\bibinfo {author} {\bibfnamefont {M.~P.}\ \bibnamefont
  {Hertzberg}}\ and\ \bibinfo {author} {\bibfnamefont {E.~D.}\ \bibnamefont
  {Schiappacasse}},\ }\href {\doibase 10.1088/1475-7516/2018/08/028} {\bibfield
   {journal} {\bibinfo  {journal} {JCAP}\ }\textbf {\bibinfo {volume} {08}},\
  \bibinfo {pages} {028} (\bibinfo {year} {2018})},\ \Eprint
  {http://arxiv.org/abs/1804.07255} {arXiv:1804.07255 [hep-ph]} \BibitemShut
  {NoStop}%
\bibitem [{\citenamefont {Davidson}\ and\ \citenamefont
  {Schwetz}(2016)}]{Davidson:2016uok}%
  \BibitemOpen
  \bibfield  {author} {\bibinfo {author} {\bibfnamefont {S.}~\bibnamefont
  {Davidson}}\ and\ \bibinfo {author} {\bibfnamefont {T.}~\bibnamefont
  {Schwetz}},\ }\href {\doibase 10.1103/PhysRevD.93.123509} {\bibfield
  {journal} {\bibinfo  {journal} {Phys. Rev. D}\ }\textbf {\bibinfo {volume}
  {93}},\ \bibinfo {pages} {123509} (\bibinfo {year} {2016})},\ \Eprint
  {http://arxiv.org/abs/1603.04249} {arXiv:1603.04249 [astro-ph.CO]}
  \BibitemShut {NoStop}%
\bibitem [{\citenamefont {Lee}\ and\ \citenamefont {Pang}(1989)}]{Lee:1988av}%
  \BibitemOpen
  \bibfield  {author} {\bibinfo {author} {\bibfnamefont {T.}~\bibnamefont
  {Lee}}\ and\ \bibinfo {author} {\bibfnamefont {Y.}~\bibnamefont {Pang}},\
  }\href {\doibase 10.1016/0550-3213(89)90365-9} {\bibfield  {journal}
  {\bibinfo  {journal} {Nucl. Phys. B}\ }\textbf {\bibinfo {volume} {315}},\
  \bibinfo {pages} {477} (\bibinfo {year} {1989})}\BibitemShut {NoStop}%
\bibitem [{\citenamefont {Gleiser}(1988)}]{Gleiser:1988rq}%
  \BibitemOpen
  \bibfield  {author} {\bibinfo {author} {\bibfnamefont {M.}~\bibnamefont
  {Gleiser}},\ }\href {\doibase 10.1103/PhysRevD.38.2376} {\bibfield  {journal}
  {\bibinfo  {journal} {Phys. Rev. D}\ }\textbf {\bibinfo {volume} {38}},\
  \bibinfo {pages} {2376} (\bibinfo {year} {1988})},\ \bibinfo {note}
  {[Erratum: Phys.Rev.D 39, 1257 (1989)]}\BibitemShut {NoStop}%
\bibitem [{\citenamefont {Gleiser}\ and\ \citenamefont
  {Watkins}(1989)}]{Gleiser:1988ih}%
  \BibitemOpen
  \bibfield  {author} {\bibinfo {author} {\bibfnamefont {M.}~\bibnamefont
  {Gleiser}}\ and\ \bibinfo {author} {\bibfnamefont {R.}~\bibnamefont
  {Watkins}},\ }\href {\doibase 10.1016/0550-3213(89)90627-5} {\bibfield
  {journal} {\bibinfo  {journal} {Nucl. Phys. B}\ }\textbf {\bibinfo {volume}
  {319}},\ \bibinfo {pages} {733} (\bibinfo {year} {1989})}\BibitemShut
  {NoStop}%
\bibitem [{\citenamefont {Kusmartsev}\ \emph {et~al.}(1991)\citenamefont
  {Kusmartsev}, \citenamefont {Mielke},\ and\ \citenamefont
  {Schunck}}]{Kusmartsev:1990cr}%
  \BibitemOpen
  \bibfield  {author} {\bibinfo {author} {\bibfnamefont {F.~V.}\ \bibnamefont
  {Kusmartsev}}, \bibinfo {author} {\bibfnamefont {E.~W.}\ \bibnamefont
  {Mielke}}, \ and\ \bibinfo {author} {\bibfnamefont {F.~E.}\ \bibnamefont
  {Schunck}},\ }\href {\doibase 10.1103/PhysRevD.43.3895} {\bibfield  {journal}
  {\bibinfo  {journal} {Phys. Rev. D}\ }\textbf {\bibinfo {volume} {43}},\
  \bibinfo {pages} {3895} (\bibinfo {year} {1991})},\ \Eprint
  {http://arxiv.org/abs/0810.0696} {arXiv:0810.0696 [astro-ph]} \BibitemShut
  {NoStop}%
\bibitem [{\citenamefont {Tamaki}\ and\ \citenamefont
  {Sakai}(2011)}]{Tamaki:2011zza}%
  \BibitemOpen
  \bibfield  {author} {\bibinfo {author} {\bibfnamefont {T.}~\bibnamefont
  {Tamaki}}\ and\ \bibinfo {author} {\bibfnamefont {N.}~\bibnamefont {Sakai}},\
  }\href {\doibase 10.1103/PhysRevD.83.044027} {\bibfield  {journal} {\bibinfo
  {journal} {Phys. Rev. D}\ }\textbf {\bibinfo {volume} {83}},\ \bibinfo
  {pages} {044027} (\bibinfo {year} {2011})},\ \Eprint
  {http://arxiv.org/abs/1105.2932} {arXiv:1105.2932 [gr-qc]} \BibitemShut
  {NoStop}%
\bibitem [{\citenamefont {Sorkin}(1981)}]{Sorkin:1981jc}%
  \BibitemOpen
  \bibfield  {author} {\bibinfo {author} {\bibfnamefont {R.}~\bibnamefont
  {Sorkin}},\ }\href {\doibase 10.1086/159282} {\bibfield  {journal} {\bibinfo
  {journal} {Astrophys. J.}\ }\textbf {\bibinfo {volume} {249}},\ \bibinfo
  {pages} {254} (\bibinfo {year} {1981})}\BibitemShut {NoStop}%
\bibitem [{\citenamefont {Schiffrin}\ and\ \citenamefont
  {Wald}(2014)}]{Schiffrin:2013zta}%
  \BibitemOpen
  \bibfield  {author} {\bibinfo {author} {\bibfnamefont {J.~S.}\ \bibnamefont
  {Schiffrin}}\ and\ \bibinfo {author} {\bibfnamefont {R.~M.}\ \bibnamefont
  {Wald}},\ }\href {\doibase 10.1088/0264-9381/31/3/035024} {\bibfield
  {journal} {\bibinfo  {journal} {Class. Quant. Grav.}\ }\textbf {\bibinfo
  {volume} {31}},\ \bibinfo {pages} {035024} (\bibinfo {year} {2014})},\
  \Eprint {http://arxiv.org/abs/1310.5117} {arXiv:1310.5117 [gr-qc]}
  \BibitemShut {NoStop}%
\bibitem [{\citenamefont {Seidel}\ and\ \citenamefont
  {Suen}(1990)}]{Seidel:1990jh}%
  \BibitemOpen
  \bibfield  {author} {\bibinfo {author} {\bibfnamefont {E.}~\bibnamefont
  {Seidel}}\ and\ \bibinfo {author} {\bibfnamefont {W.-M.}\ \bibnamefont
  {Suen}},\ }\href {\doibase 10.1103/PhysRevD.42.384} {\bibfield  {journal}
  {\bibinfo  {journal} {Phys. Rev. D}\ }\textbf {\bibinfo {volume} {42}},\
  \bibinfo {pages} {384} (\bibinfo {year} {1990})}\BibitemShut {NoStop}%
\bibitem [{\citenamefont {Balakrishna}\ \emph {et~al.}(1998)\citenamefont
  {Balakrishna}, \citenamefont {Seidel},\ and\ \citenamefont
  {Suen}}]{Balakrishna:1997ej}%
  \BibitemOpen
  \bibfield  {author} {\bibinfo {author} {\bibfnamefont {J.}~\bibnamefont
  {Balakrishna}}, \bibinfo {author} {\bibfnamefont {E.}~\bibnamefont {Seidel}},
  \ and\ \bibinfo {author} {\bibfnamefont {W.-M.}\ \bibnamefont {Suen}},\
  }\href {\doibase 10.1103/PhysRevD.58.104004} {\bibfield  {journal} {\bibinfo
  {journal} {Phys. Rev. D}\ }\textbf {\bibinfo {volume} {58}},\ \bibinfo
  {pages} {104004} (\bibinfo {year} {1998})},\ \Eprint
  {http://arxiv.org/abs/gr-qc/9712064} {arXiv:gr-qc/9712064} \BibitemShut
  {NoStop}%
\bibitem [{\citenamefont {Guzman}(2004)}]{Guzman:2004jw}%
  \BibitemOpen
  \bibfield  {author} {\bibinfo {author} {\bibfnamefont {F.}~\bibnamefont
  {Guzman}},\ }\href {\doibase 10.1103/PhysRevD.70.044033} {\bibfield
  {journal} {\bibinfo  {journal} {Phys. Rev. D}\ }\textbf {\bibinfo {volume}
  {70}},\ \bibinfo {pages} {044033} (\bibinfo {year} {2004})},\ \Eprint
  {http://arxiv.org/abs/gr-qc/0407054} {arXiv:gr-qc/0407054} \BibitemShut
  {NoStop}%
\bibitem [{\citenamefont {Valdez-Alvarado}\ \emph {et~al.}(2013)\citenamefont
  {Valdez-Alvarado}, \citenamefont {Palenzuela}, \citenamefont {Alic},\ and\
  \citenamefont {Ureña-López}}]{ValdezAlvarado:2012xc}%
  \BibitemOpen
  \bibfield  {author} {\bibinfo {author} {\bibfnamefont {S.}~\bibnamefont
  {Valdez-Alvarado}}, \bibinfo {author} {\bibfnamefont {C.}~\bibnamefont
  {Palenzuela}}, \bibinfo {author} {\bibfnamefont {D.}~\bibnamefont {Alic}}, \
  and\ \bibinfo {author} {\bibfnamefont {L.}~\bibnamefont {Ureña-López}},\
  }\href {\doibase 10.1103/PhysRevD.87.084040} {\bibfield  {journal} {\bibinfo
  {journal} {Phys. Rev. D}\ }\textbf {\bibinfo {volume} {87}},\ \bibinfo
  {pages} {084040} (\bibinfo {year} {2013})},\ \Eprint
  {http://arxiv.org/abs/1210.2299} {arXiv:1210.2299 [gr-qc]} \BibitemShut
  {NoStop}%
\bibitem [{\citenamefont {Friedman}(1978)}]{friedman1978}%
  \BibitemOpen
  \bibfield  {author} {\bibinfo {author} {\bibfnamefont {J.~L.}\ \bibnamefont
  {Friedman}},\ }\href@noop {} {\bibfield  {journal} {\bibinfo  {journal}
  {Comm. Math. Phys.}\ }\textbf {\bibinfo {volume} {63}},\ \bibinfo {pages}
  {243} (\bibinfo {year} {1978})}\BibitemShut {NoStop}%
\bibitem [{\citenamefont {Collodel}\ \emph {et~al.}(2017)\citenamefont
  {Collodel}, \citenamefont {Kleihaus},\ and\ \citenamefont
  {Kunz}}]{Collodel:2017biu}%
  \BibitemOpen
  \bibfield  {author} {\bibinfo {author} {\bibfnamefont {L.~G.}\ \bibnamefont
  {Collodel}}, \bibinfo {author} {\bibfnamefont {B.}~\bibnamefont {Kleihaus}},
  \ and\ \bibinfo {author} {\bibfnamefont {J.}~\bibnamefont {Kunz}},\ }\href
  {\doibase 10.1103/PhysRevD.96.084066} {\bibfield  {journal} {\bibinfo
  {journal} {Phys. Rev. D}\ }\textbf {\bibinfo {volume} {96}},\ \bibinfo
  {pages} {084066} (\bibinfo {year} {2017})},\ \Eprint
  {http://arxiv.org/abs/1708.02057} {arXiv:1708.02057 [gr-qc]} \BibitemShut
  {NoStop}%
\bibitem [{\citenamefont {Keir}(2016)}]{Keir:2014oka}%
  \BibitemOpen
  \bibfield  {author} {\bibinfo {author} {\bibfnamefont {J.}~\bibnamefont
  {Keir}},\ }\href {\doibase 10.1088/0264-9381/33/13/135009} {\bibfield
  {journal} {\bibinfo  {journal} {Class. Quant. Grav.}\ }\textbf {\bibinfo
  {volume} {33}},\ \bibinfo {pages} {135009} (\bibinfo {year} {2016})},\
  \Eprint {http://arxiv.org/abs/1404.7036} {arXiv:1404.7036 [gr-qc]}
  \BibitemShut {NoStop}%
\bibitem [{\citenamefont {Reisswig}\ \emph {et~al.}(2013)\citenamefont
  {Reisswig}, \citenamefont {Ott}, \citenamefont {Abdikamalov}, \citenamefont
  {Haas}, \citenamefont {Moesta},\ and\ \citenamefont
  {Schnetter}}]{Reisswig:2013sqa}%
  \BibitemOpen
  \bibfield  {author} {\bibinfo {author} {\bibfnamefont {C.}~\bibnamefont
  {Reisswig}}, \bibinfo {author} {\bibfnamefont {C.}~\bibnamefont {Ott}},
  \bibinfo {author} {\bibfnamefont {E.}~\bibnamefont {Abdikamalov}}, \bibinfo
  {author} {\bibfnamefont {R.}~\bibnamefont {Haas}}, \bibinfo {author}
  {\bibfnamefont {P.}~\bibnamefont {Moesta}}, \ and\ \bibinfo {author}
  {\bibfnamefont {E.}~\bibnamefont {Schnetter}},\ }\href {\doibase
  10.1103/PhysRevLett.111.151101} {\bibfield  {journal} {\bibinfo  {journal}
  {Phys. Rev. Lett.}\ }\textbf {\bibinfo {volume} {111}},\ \bibinfo {pages}
  {151101} (\bibinfo {year} {2013})},\ \Eprint {http://arxiv.org/abs/1304.7787}
  {arXiv:1304.7787 [astro-ph.CO]} \BibitemShut {NoStop}%
\bibitem [{\citenamefont {Pretorius}(2005)}]{Pretorius:2004jg}%
  \BibitemOpen
  \bibfield  {author} {\bibinfo {author} {\bibfnamefont {F.}~\bibnamefont
  {Pretorius}},\ }\href {\doibase 10.1088/0264-9381/22/2/014} {\bibfield
  {journal} {\bibinfo  {journal} {Class. Quant. Grav.}\ }\textbf {\bibinfo
  {volume} {22}},\ \bibinfo {pages} {425} (\bibinfo {year} {2005})},\ \Eprint
  {http://arxiv.org/abs/gr-qc/0407110} {arXiv:gr-qc/0407110 [gr-qc]}
  \BibitemShut {NoStop}%
\bibitem [{\citenamefont {East}\ \emph {et~al.}(2012)\citenamefont {East},
  \citenamefont {Pretorius},\ and\ \citenamefont {Stephens}}]{East:2011aa}%
  \BibitemOpen
  \bibfield  {author} {\bibinfo {author} {\bibfnamefont {W.~E.}\ \bibnamefont
  {East}}, \bibinfo {author} {\bibfnamefont {F.}~\bibnamefont {Pretorius}}, \
  and\ \bibinfo {author} {\bibfnamefont {B.~C.}\ \bibnamefont {Stephens}},\
  }\href {\doibase 10.1103/PhysRevD.85.124010} {\bibfield  {journal} {\bibinfo
  {journal} {Phys. Rev. D}\ }\textbf {\bibinfo {volume} {85}},\ \bibinfo
  {pages} {124010} (\bibinfo {year} {2012})},\ \Eprint
  {http://arxiv.org/abs/1112.3094} {arXiv:1112.3094 [gr-qc]} \BibitemShut
  {NoStop}%
\bibitem [{\citenamefont {Lindblom}\ and\ \citenamefont
  {Szilagyi}(2009)}]{Lindblom:2009tu}%
  \BibitemOpen
  \bibfield  {author} {\bibinfo {author} {\bibfnamefont {L.}~\bibnamefont
  {Lindblom}}\ and\ \bibinfo {author} {\bibfnamefont {B.}~\bibnamefont
  {Szilagyi}},\ }\href {\doibase 10.1103/PhysRevD.80.084019} {\bibfield
  {journal} {\bibinfo  {journal} {Phys. Rev.}\ }\textbf {\bibinfo {volume}
  {D80}},\ \bibinfo {pages} {084019} (\bibinfo {year} {2009})},\ \Eprint
  {http://arxiv.org/abs/0904.4873} {arXiv:0904.4873 [gr-qc]} \BibitemShut
  {NoStop}%
\end{thebibliography}%

\end{document}